\documentclass[12pt]{article}
\usepackage{a41}
\usepackage{cite}
\usepackage{graphicx} 
\usepackage{epsfig}
\usepackage{amsmath}
\usepackage{amssymb}
\usepackage{ulem}
\usepackage{mdwlist}                       
\usepackage{color}
\usepackage{float}
\usepackage[rflt]{floatflt}
\usepackage{slashed}
\usepackage{afterpage}

 \newcommand{\N}{\nonumber}

\newcommand{\bea}{\begin{eqnarray}}
\newcommand{\eea}{\end{eqnarray}}
\newcommand{\eq}{\end{equation}}

\newcommand\Mvec{\mbox{\bf M}}

\newcommand\be{\begin{eqnarray}}
\newcommand\ee{\end{eqnarray}}

\newcommand\HA{{\rm H}}

\usepackage{array}
\usepackage[english]{babel}
\usepackage[latin1]{inputenc}
\usepackage[T1]{fontenc}
\usepackage{ae}

\usepackage{url}

\usepackage{amsmath, amsthm, amssymb}

\usepackage{array}
\usepackage[english]{babel}
\usepackage[latin1]{inputenc}
\usepackage[T1]{fontenc}
\usepackage{ae}

\newcommand{\gsim}{\raisebox{-0.07cm   }
{$\, \stackrel{>}{{\scriptstyle\sim}}\, $}}
\newcommand{\GeV}{\rm GeV}

 \newcommand{\PS}{\hat{P}}
 \newcommand{\adag}{/\!\!\! }

\newcommand{\NN}{\nonumber}

\newcommand{\Li}{{\rm Li}}

\newcommand{\ep}{\varepsilon}

\usepackage{rotating}

\usepackage{graphicx}
\newcounter{mmacnt}
\def\restartmma{\setcounter{mmacnt}{0}}
\restartmma \catcode`|=\active
\def|#1|{\mathrm{#1}}
\catcode`|=12
\newenvironment{mma}{
 \par\smallskip
 \catcode`|=\active
 \parskip=0pt\parindent=0pt 
 \small
 \def\In##1\\{%
   \def\linebreak{\hfill\break\null\qquad}%
   \refstepcounter{mmacnt}
   \hangindent=2.5em\hangafter=0
   \leavevmode
   \llap{\tiny\sffamily In[\arabic{mmacnt}]:=\kern.5em}%
   \mathversion{bold}\footnotesize$\displaystyle##1$\normalsize
   \mathversion{normal}\par
 }%
 \def\Print##1\\{%
   \def\linebreak{\hfill\break}%
   \hangindent=2.5em\hangafter=0
   \leavevmode ##1\par}%
 \def\Out##1\\{%
   \def\linebreak{$\hfill\break\null\hfill$}%
   \kern\abovedisplayskip\par
   \hangindent=2.5em\hangafter=0
   \leavevmode
   \llap{\tiny\sffamily Out[\arabic{mmacnt}]=\kern.5em}
   \footnotesize$\displaystyle##1$\normalsize\hfill\null\par
   \kern\belowdisplayskip
 }%
 \def\Warning##1##2\\{%
   \def\linebreak{\hfill\break}%
   \hangindent=2.5em\hangafter=0
   \leavevmode
   {\scriptsize##1 : ##2}\par}
 \def\Warning##1##2\\{%
   \def\linebreak{\hfill\break}%
   \hangindent=2.5em\hangafter=0
   \leavevmode
   {\scriptsize##1 : ##2}\par}%
}{%
 \par\smallskip
}

\usepackage{color}
\newenvironment{fshaded}{%
\MakeFramed {\FrameRestore}
}%
{\endMakeFramed}

\usepackage{tikz} 
\usetikzlibrary{matrix}
   
\allowdisplaybreaks[4]

\begin{document}
\noindent
\sloppy
\thispagestyle{empty}
\begin{flushleft}
DESY 15--004 \hfill
\\
DO--TH 15/01\\
TTP 22--023\\
ZU-TH 57/22\\
SAGEX--20--11\\
November 2022
\end{flushleft}
%
\vspace*{\fill}
\begin{center}
{\LARGE\bf \boldmath{$O(\alpha_s^2$)} Polarized Heavy Flavor Corrections} 

\vspace{2mm}
{\LARGE\bf to Deep-Inelastic Scattering at \boldmath{$Q^2 \gg m^2$}}

\vspace{2cm}
\large
I.~Bierenbaum$^a$,
J.~Bl\"umlein$^a$, 
A.~De Freitas$^{a,b}$,
A.~Goedicke$^a$,\\
S.~Klein$^a$, 
and K.~Sch\"onwald$^{a,c,d}$
\\
\vspace{2em}
\normalsize
{\it $^a$~Deutsches Elektronen--Synchrotron DESY,\\
Platanenallee 6, 15738 Zeuthen, Germany}
\\

\vspace*{2mm}
{\it $^b$~
Johannes Kepler University Linz, Research Institute for Symbolic Computation (RISC),
Altenbergerstra\ss{}e 69, A-4040, Linz, Austria}
\\

\vspace*{2mm}
{\it $^c$~Institut f\"ur Theoretische Teilchenphysik,\\ 
Karlsruher Institut f\"ur Technologie (KIT) D-76128
Karlsruhe, Germany}
\\

\vspace*{2mm}
{\it $^d$~
Physik-Institut, Universit\"at Z\"urich, Winterthurerstrasse 190,\\ CH-8057 Z\"urich, Switzerland}
\\
\vspace{2em}
\end{center}
\vspace*{\fill}
%
\begin{abstract}
\noindent
We calculate the quarkonic $O(\alpha_s^2)$ massive operator matrix elements $\Delta A_{Qg}(N), 
\Delta A_{Qq}^{\rm PS}(N)$ and $\Delta A_{qq,Q}^{\rm NS}(N)$ for the twist--2 operators and 
the associated heavy flavor Wilson coefficients in polarized deeply inelastic scattering in 
the region $Q^2 \gg m^2$ to $O(\varepsilon)$ in the case of the inclusive heavy flavor 
contributions. The evaluation is performed in Mellin space, without applying the 
integration-by-parts method. The result is given in terms of harmonic sums. This leads to a 
significant compactification of the operator matrix elements and massive Wilson coefficients 
in the region $Q^2 \gg m^2$ derived previously in \cite{BUZA2}, which we partly confirm, and 
also partly correct. The results allow to determine the heavy flavor Wilson coefficients for 
$g_1(x,Q^2)$ to $O(\alpha_s^2)$ for all but the power suppressed terms $\propto (m^2/Q^2)^k, 
k \geq 1$. The results in momentum fraction $z$-space are also presented.  We also discuss the 
small $x$ effects in the polarized case. Numerical results are presented. We also compute the 
gluonic matching coefficients in the two--mass variable flavor number scheme to $O(\ep)$. 
\end{abstract}

\vspace*{4mm}
\begin{center}
{\sf Dedicated to the memory of Dieter Robaschik.}
\end{center}
\vspace*{\fill}
\newpage
\section{Introduction}
\label{sec:1}

\vspace{1mm}\noindent
The question of the composition of the nucleon spin in terms of partonic degrees of freedom has 
attracted much interest after the initial experimental finding \cite{SPUZ} that the polarizations 
of the three light quarks alone do not provide the required value of 1/2. Subsequently, the polarized 
nucleon structure functions have been measured in great detail by various experiments \cite{EXP1}.\footnote{
For theoretical surveys see \cite{LR,Blumlein:2012bf,Deur:2018roz}.} To determine the different contributions to the 
nucleon spin, both the flavor dependence as well as the contribution due to the gluons and angular excitations 
at virtualities $Q^2$ in the perturbative region have to be studied in more detail in the future \cite{Boer:2011fh}
experimentally.
Since the nucleon spin contributions are related to the first moments of the respective distribution 
functions, it is desirable to measure to very small values of $x$, i.e. to highest possible 
hadronic energies, cf.~\cite{PROC}. A detailed treatment 
of the flavor structure requires the inclusion of heavy flavor. As in the unpolarized case
\cite{UHEAV1,BUZA1,BBK1,BBK1a,BBK2,BBKS8a,Bierenbaum:2009mv,BBK-III,FL} this contribution 
is driven by the gluon and sea--quark densities, since the flavor non--singlet contributions 
contribute from $O(a_s^2)$ only, where $a_s = \alpha_s/(4\pi)$ is the renormalized coupling 
constant in the $\overline{\sf MS}$ scheme.
The Wilson coefficients are known to 
first order in the whole kinematic 
range \cite{WIL1a,WIL1b}.\footnote{For a fast, precise numerical implementation of the heavy corrections in Mellin 
space see \cite{AB}.} The photo--production cross section for polarized scattering has been calculated to 
next-to-leading order (NLO) in \cite{PHOTO}. Very recently also the NLO corrections for polarized 
deep--inelastic 
production of tagged heavy quarks have been computed in \cite{Hekhorn:2018ywm}, partly numerically, retaining 
also the power corrections. Previously, only the deep inelastic scattering cross section in the case 
$Q^2 \gg m^2$, 
with $m$ the heavy quark mass, had been calculated in Ref.~\cite{BUZA2}.
Exclusive data on charm--quark pair production in polarized deep--inelastic scattering are available only in the 
region of very low photon virtualities \cite{COMPASS} at present. However, the inclusive measurement of the structure 
functions $g_1(x,Q^2)$ and $g_2(x,Q^2)$ contains the heavy flavor contributions for hadronic masses 
$W^2 \geq (2 m + m_N)^2$, with $m_N$ the nucleon mass. The scaling violations of the heavy quark contributions to 
the structure functions are different from those of the light partons. Therefore one may not model these 
contributions in a simple manner changing the number of active massless flavors. Numerical illustrations for the leading 
order (LO) contributions were given in \cite{BRN1} using the parton densities \cite{BB}.\footnote{Other
polarized parton density parameterizations can be found in Refs.~\cite{OPDF}.} 
In Ref.~\cite{Blumlein:2010rn}
the LO heavy charm contributions were accounted for in the fit explicitly.

Quantitative comparisons between the results of \cite{UHEAV1} and \cite{BUZA1,BBK1} show that the approximation $Q^2 
\gg m^2$ is valid for heavy flavor contributions to the structure function $F_1(x,Q^2)$ for $Q^2/m^2 \gsim 10$, i.e. 
$Q^2 \gsim 22.5~\GeV^2$ in the case of charm. A similar approximation should hold in the case of the polarized structure 
function $g_1(x,Q^2)$. By comparing the pure singlet contributions in the full and asymptotic kinematics 
\cite{Blumlein:2019zux}, one finds e.g. $Q^2/m^2 > 10$ at $x = 10^{-4}$, $Q^2/m^2 > 50$ at $x = 10^{-2}$ and $Q^2/m^2 > 110$ 
at $x = 0.5$ allowing for a deviation from the exact two--loop result by 3\%.

In the present paper we re-calculate for the first time the heavy flavor contributions to the longitudinally 
polarized structure function $g_1(x,Q^2)$ analytically to $O(a_s^2)$ in the asymptotic region $Q^2 \gg m^2$ 
and provide various phenomenological illustrations. We will consider the case of inclusive heavy flavor 
corrections in the following.\footnote{Tagged heavy flavor corrections, Ref.~\cite{BUZA2}, can be considered 
up to two--loop order. Starting with three--loop order this separation is not possible in the inclusive 
case~\cite{Bierenbaum:2009mv,HQ3N,Blumlein:2014zxa,Ablinger:2014vwa}.} At the time when Ref.~\cite{BUZA2} 
was published, the understanding of polarized processes in $D$ dimensions has still been under development 
\cite{ZN,SP_PS,BUZA2,MAT1,Weinzierl:1999xb,Jegerlehner:2000dz} and results on the loop level need to be checked.

The contributions to the structure function $g_2(x,Q^2)$ can be obtained by using the Wandzura--Wilczek relation 
\cite{WW} at the level of twist--2 operators \cite{G2A,G2B}. The general validity of this relation was shown in 
Ref.~\cite{G2A} using the covariant parton model \cite{Landshoff:1971xb}. This also applies to the heavy flavor 
contributions \cite{BRN1}. The Wandzura--Wilczek relation holds also in the case of 
diffractive scattering \cite{DIFFR}, for 
the target mass corrections \cite{BT,TM} and for non--forward scattering \cite{BR1}. 

As has been outlined in Ref.~\cite{BUZA1} already in the case of the twist--2 contributions, the asymptotic heavy 
flavor corrections factorize into massive operator matrix elements (OMEs) and the light flavor Wilson 
coefficients \cite{ZN} as a consequence of the renormalization group equations.\footnote{This has been proven
analytically in the case of QCD to two--loops by calculating the complete mass dependence for the 
non--singlet \cite{BUZA2,BUZA1,Blumlein:2016xcy} and the pure singlet contributions 
\cite{Blumlein:2019zux,Blumlein:2019qze}. Furthermore, it has been proven for the two--loop QED corrections for 
$e^+ e^- \rightarrow \gamma^*/Z^*$ \cite{Blumlein:2011mi,Blumlein:2020jrf}, where the initial state consist of 
massive particles.} 

In calculating the polarized heavy flavor Wilson coefficients to $O(a_s^2)$, we proceed in the same way as 
was followed in the unpolarized case \cite{BBK1}. Furthermore, we calculate newly the $O(\varepsilon)$ terms at 
this order for the unrenormalized OMEs in the Larin scheme \cite{GAM5}. They contribute to the $O(a_s^3)$ 
corrections through renormalization.\footnote{For the calculation of the moments of the unpolarized 
massive OMEs, see \cite{Bierenbaum:2009mv,BBK-III}. Corresponding moments in the case of transversity 
were obtained in \cite{TRANSV} to $O(a_s^3)$, with complete results at $O(a_s^2)$.} 
Later on we will translate the two--loop results into the scheme used in \cite{MAT1,Moch:2014sna,Behring:2019tus,
Blumlein:2021ryt,Blumlein:2022ndg}.

The calculation was performed in Mellin space without applying the integration-by-parts (IBP) method \cite{IBP} 
for the Feynman diagrams. 
This leads to much more compact representations in terms of harmonic sums 
\cite{Vermaseren:1998uu,Blumlein:1998if} both for the individual diagrams and the final results. In the 
course of the calculation we use representations through Mellin--Barnes
integrals \cite{BBK1,BW} and generalized hypergeometric functions \cite{GHF}.\footnote{For 
a survey on other calculation methods see \cite{Blumlein:2018cms}.}

The flavor non--singlet and pure singlet results are known analytically to two--loop order, 
including the power corrections \cite{BUZA1,Blumlein:2016xcy,Blumlein:2019zux}, and the 
asymptotic non--singlet three--loop corrections  have been 
calculated in 
\cite{Ablinger:2014vwa}. Phenomenological applications were given in \cite{Behring:2015zaa}
in the non--singlet case.
Furthermore, the polarized three-loop operator matrix elements 
$\Delta A_{Qq}^{\rm PS,(3)}, \Delta A_{qq,Q}^{\rm PS,(3)}, 
\Delta A_{qg,Q}^{\rm S,(3)}$ and $\Delta A_{gq,Q}^{(3)}$ in the single mass case 
\cite{Ablinger:2019etw,Behring:2021asx,LOGPOL}
and the OMEs $\Delta A_{qq,Q}^{\rm NS,(3)}, \Delta A_{Qq}^{\rm PS,(3)}, 
\Delta A_{gg,Q}^{\rm (3)}$ and $\Delta A_{gq,Q}^{\rm (3)}$ in the two--mass case have been computed 
\cite{Ablinger:2017err,Ablinger:2019gpu,
Ablinger:2020snj,Schonwald:2019gmn,Behring:2021asx}.

In the present paper we deal with corrections of a single heavy quark and $N_F$ massless quarks and 
also consider the first two--mass contributions, which contribute starting with $O(a_s^2)$.
The paper is organized as follows. In Section~\ref{sec:2} we summarize main relations, such 
as 
the differential cross sections for polarized deeply inelastic scattering and
the leading order heavy flavor corrections, and give a brief outline on the
representation of the asymptotic heavy flavor corrections at next-to-leading order
(NLO).  In Section~\ref{sec:3} we summarize details of the renormalization of the massive
operator matrix elements. The polarized gluonic and quarkonic massive operator 
matrix elements at two--loop order are calculated in Section~\ref{sec:4} 
in the Larin scheme \cite{GAM5} (for other schemes 
see~\cite{HVBM}). Since the specific prescriptions of $\gamma_5$ and the Levi--Civita pseudo--tensor in $D = 4 
+ \varepsilon$ 
dimensions violate Ward--identities, a finite renormalization has to be performed to transform all related 
quantities, 
i.e. the massive operator matrix elements, the massless Wilson coefficients and the parton 
distribution functions into 
the {\sf M} scheme, cf.~\cite{MAT1,Moch:2014sna,Behring:2019tus,Blumlein:2021ryt}. We  describe in 
detail the different treatments in Refs.~\cite{ZN,SP_PS,BUZA2}, in which partly mixed concepts were used, to 
understand and correct the final result of the previous calculation \cite{BUZA2}. In particular, also a 
recalculation of the massless two--loop pure singlet Wilson coefficient is needed 
for this comparison, cf.~Ref.~\cite{Blumlein:2019zux}.\footnote{Very recently also the three--loop massless 
polarized Wilson coefficients have been calculated in the Larin scheme \cite{Blumlein:2022gpp}.}

We checked our results for a number of moments with the help of the Mellin--Barnes method numerically. The mathematical 
structure of the results is discussed. Again, it can be represented in terms of a few {\sf basic} 
harmonic sums in a more 
compact form if compared to the results given in Ref.~\cite{BUZA2}. There are no new sums appearing 
if compared to the unpolarized calculation given in \cite{BBKS8a}. 
We also specify the 1st moment of the heavy flavor Wilson coefficient. The small $x$ behaviour
of these quantities is of special interest. We discuss it in the context of other quantities with  
leading small $x$ singularity at $N = 0$ to clarify the present status. Numerical results 
are presented in Section~\ref{sec:6}. In Section~\ref{sec:7} we present also the gluonic 2--loop OMEs, which 
contribute in the variable flavor number scheme (VFNS), cf. e.g.~\cite{Blumlein:2018jfm}.  
Section~\ref{sec:8} contains 
the 
conclusions. In Appendix~\ref{sec:A}, the results for the individual Feynman diagrams are presented.
Technical details of the calculation are given in Appendix~\ref{sec:B}. The 
asymptotic polarized heavy quark Wilson coefficients are listed both in 
momentum fraction $z$--space and Mellin $N$ space in Appendix~\ref{sec:C}, where we 
also correct  results given in Ref.~\cite{BUZA2}. 
\section{Heavy flavor structure functions in polarized deep--inelastic scattering}
\label{sec:2}

\vspace{1mm}\noindent
The process of deeply inelastic longitudinally polarized charged lepton scattering
off longitudinally (L) or transversely (T) polarized nucleons in the case of single
photon exchange\footnote{For the scattering cross sections in the case of also electro--weak 
contributions
see Refs.~\cite{G2B,BT}.} is given by 
\begin{equation}
l^{\pm} N \rightarrow {l'}^{\pm} X~.
\end{equation}
The differential scattering cross sections read
\begin{equation}
\frac{d^3 \sigma}{dx dy d \theta} = \frac{y \alpha^2}{Q^4} L^{\mu\nu} W_{\mu\nu}~,
\end{equation}
cf.~\cite{G2B},
where $x= Q^2/2 P.q$ and $y = P.q /k.P$ are the Bjorken variables, $P$ and $k$ are the incoming 
nucleon and lepton 4--momenta, $q = k - k'$ is the 4--momentum transfer, $Q^2 = - q^2$,
$\theta$ the azimuthal angle of the final state lepton, and $L^{\mu\nu}$ and $W_{\mu\nu}$
denote the leptonic and hadronic tensors. We consider the asymmetries $A(x,y,\theta)_{L,T}$ between
the differential cross sections for opposite nucleon polarization both in the longitudinal and transverse
case 
\begin{equation}
A(x,y,\theta)_{L,T} = 
 \frac{d^3 \sigma_{L,T}^{\rightarrow}}{dx dy d \theta}
-\frac{d^3 \sigma_{L,T}^{\leftarrow}}{dx dy d \theta}~, 
\end{equation}
which projects onto the polarized parts of both the leptonic and hadronic tensors,
$L^A_{\mu\nu}$ and $W^A_{\mu\nu}$. The hadronic tensor at the level of the twist $\tau = 2$ contributions is then 
determined by two nucleon structure functions
\begin{equation}
W^A_{\mu\nu} = i \varepsilon_{\mu\nu\lambda\sigma} \left[\frac{q^\lambda S^\sigma}{P.q} g_1(x,Q^2)
+ \frac{q^\lambda (P.q S^\sigma - S.q P^\sigma)}{(P.q)^2} g_2(x,Q^2) \right]~. 
\end{equation}
Here $S$ denotes the nucleon spin vector 
\begin{eqnarray}
S_L &=& (0,0,0,m_N) \nonumber\\
S_T &=& m_N (0,\cos(\bar{\theta}),\sin(\bar{\theta}),0)~,
\end{eqnarray}
in the longitudinally and transversely polarized cases in the nucleon rest frame,
with $m_N$ the nucleon mass, $\bar{\theta}$ a fixed angle in the plane transverse to the 
proton beam direction, and 
$\varepsilon_{\mu\nu\lambda\sigma}$ is the Levi--Civita symbol.

One obtains \cite{G2B,BT}\footnote{The QED radiative corrections were calculated in Ref.~\cite{Bardin:1996ch} and 
are contained in the present release of the code {\tt HECTOR} \cite{Arbuzov:1995id}.}

\begin{eqnarray}
A(x,y)_{L} &=& 4 \lambda \frac{\alpha^2}{Q^2} \left[  \left(2 - y - \frac{2xy m_N^2}{s} \right)
 g_1(x,Q^2) + 4 \frac{yx  m_N^2}{s} g_2(x,Q^2) \right], \\
A(x,y,\bar{\theta},\theta)_{T} &=& -8 \lambda \frac{\alpha^2}{Q^2} 
\sqrt{\frac{m_N^2}{s}} \sqrt{\frac{x}{y} 
\left[1-y-\frac{xy m_N^2}{S}\right]}
\cos(\bar{\theta} - \theta) [ y g_1(x,Q^2) + 2 g_2(x,Q^2)],
\end{eqnarray}
where $\alpha = e^2/(4\pi)$ is the fine structure constant, 
$\lambda$ the degree of polarization,  and $s = (P+k)^2$. In the case of $A(x,y)_{L}$ 
the dependence on $\theta$ is trivial and has been integrated out.
The structure function $g_1(x,Q^2)$ has the following representation to $O(\alpha_s^2)$, 
\begin{eqnarray}
g_1^{\tau = 2}(x,Q^2) 
= 
g_1^{\tau = 2,\rm light}(x,Q^2) + g_1^{\tau = 2,\rm heavy}(x,Q^2), 
\end{eqnarray}
where we account for the heavy flavor contributions in the asymptotic region, with \cite{ZN}, 
Eqs.~(2.6-2.8),
\begin{eqnarray}
x g_1^{\tau = 2,\rm light}(x,Q^2) &=& 
\frac{x}{2} \Biggl\{
\frac{1}{N_F} \sum_{k=1}^{N_F} e_k^2 \Biggl[
  \Delta C_q^{\rm S} \otimes \Delta \Sigma
+ \Delta C_g^{\rm S} \otimes \Delta G
\Biggr] + \Delta C_q^{\rm NS} \Delta_{\rm NS}
\Biggr\}, 
\\
x g_1^{\tau = 2,\rm heavy}(x,Q^2) &=& 
 x g_1^{\tau = 2,\rm c}(x,Q^2) 
+  x g_1^{\tau = 2,\rm b}(x,Q^2) 
+ x g_1^{\tau = 2,\rm cb}(x,Q^2), 
\end{eqnarray}
with $N_F = 3$.\footnote{In the presence of a finite nucleon mass also the structure function $g_1(x,Q^2)$
has twist--3 contributions, cf.~\cite{BT}.} 
The heavy flavor corrections contain both one and two heavy flavor contributions.
The polarized singlet and non--singlet distribution functions are given by
\begin{eqnarray}
\Delta \Sigma   &=& \sum_{k=1}^{N_F} \left(\Delta f_k + \Delta f_{\bar{k}}\right),
\\
\Delta_{\rm NS} &=& \sum_{k=1}^{N_F} \left[e_k^2 - \frac{1}{N_F} \sum_{i=1}^{N_F} e_i^2\right]
\left(\Delta f_k + \Delta 
f_{\bar{k}}\right),
\end{eqnarray}
and $\Delta G$ denotes the polarized gluon distribution and all parton distributions depend 
on $x$, 
the  factorization scale $\mu^2$ and the number of massless flavors $N_F$, with $(1/N_F) 
\sum_{i=1}^{N_F} e_i^2 = 2/9$ for $N_F =3$. The massless quarkonic 
Wilson coefficient, $\Delta C_q^{\rm S}$, is given by
\begin{eqnarray}
\Delta C_q^{\rm S} &=& \Delta C_q^{\rm NS} +  \Delta C_q^{\rm PS},
\end{eqnarray}
where $\Delta C_q^{\rm PS}$ contributes from $O(a_s^2)$ onward; $\otimes$ denotes the Mellin convolution
\begin{eqnarray}
\label{eq:AconB}
[A \otimes B](z) = \int_0^1 \int_0^1 dx dy~\delta(z-xy)~A(x) B(y)~.
\end{eqnarray}
By using the Mellin transform
\begin{eqnarray}
\Mvec[F(z)](N) = \int_0^1 dz z^{N-1} F(z), 
\end{eqnarray}
one obtains 
\begin{eqnarray}
\Mvec[[A \otimes B](z)](N) = \Mvec[A(z)](N) \cdot \Mvec[B(z)](N).
\end{eqnarray}

The heavy quark contributions in the asymptotic region are given by the single heavy flavor corrections
to two--loop order, \cite{Buza:1996wv} Eq.~(2.29),\footnote{$\Delta L_q^{\rm PS}$ contributes only with
three--loop order.}
\begin{eqnarray}
\lefteqn{x g_1^{\tau = 2, Q}(x,Q^2) =}
\NN\\ &&
\frac{x}{2} \Biggl\{
     \left(\sum_{k=1}^{N_F}e_k^2\right) 
\Delta L_{g,1}^{\sf 
S}\left(x,\frac{Q^2}{m^2_Q}\right)
\otimes \Delta G 
                 +  \left[\sum_{k=1}^{N_F} e_k^2 \Delta L_{q,1}^{\sf NS}\left(x,\frac{Q^2}{\mu^2}
                                                ,\frac{m^2_Q}{\mu^2}\right)
\otimes [
\Delta f_k + \Delta f_{\bar{k}}] \right]
\NN\\ &&
       + e_Q^2\Biggl[
                   \Delta H_{q,1}^{\sf PS}\left(x,\frac{Q^2}{\mu^2},\frac{m^2_Q}{\mu^2}\right)
                \otimes
                   \Delta \Sigma
                  + \Delta H_{g,1}^{\sf S}\left(x,\frac{Q^2}{\mu^2}
                                           ,\frac{m^2_Q}{\mu^2}\right)
                \otimes
                   \Delta G
                                  \Biggr]\Biggr\},
\nonumber\\ 
\end{eqnarray}
with $Q = c,b$. The single mass heavy flavor Wilson coefficients are 
\begin{eqnarray}
\Delta L_{q,1}^{\sf NS} &=& a_s^2 \Delta L_{q,1}^{\sf NS,(2)},
\\
\Delta L_{g,1}^{\sf S} &=& a_s^2 \Delta L_{g,1}^{\sf S,(2)},
\\
\Delta H_{q,1}^{\sf PS} &=& a_s^2 \Delta H_{q,1}^{\sf PS,(2)},
\\
\Delta H_{g,1}^{\sf S} &=& 
  a_s \Delta H_{g,1}^{\sf S,(1)}
+ a_s^2 \Delta H_{g,1}^{\sf S,(2)}.
\end{eqnarray}
The two--mass corrections are given by \cite{Ablinger:2017err}
\begin{eqnarray}
\label{eqTWOm}
x g_1^{\tau = 2,\rm cb}(x,Q^2) 
&=& a_s^2 \frac{x}{2} \int_x^1 \frac{dz}{z} \Delta H_g^{\rm two-mass}(z,Q^2,m_c,m_b)(z)
\Delta G\left(\frac{x}{z},Q^2\right),
\end{eqnarray}
with 
\begin{eqnarray}
\label{eqTWO1}
\Delta H_g^{\rm two-mass}(z) &=& \frac{32}{3} T_F^2 \Biggl\{
(1-2z)(e_c^2 + e_b^2) 
\ln\left(\frac{Q^2}{m_c^2}\right)
\ln\left(\frac{Q^2}{m_b^2}\right)
- \left(
  e_c^2 \ln\left(\frac{Q^2}{m_b^2}\right) 
+ e_b^2 \ln\left(\frac{Q^2}{m_c^2}\right)\right) 
\nonumber\\ &&
\Biggl[(3-4z)-(1-2z)\ln\left(\frac{1-z}{z}\right)
\Biggr] \Biggr\},
\end{eqnarray}
and $e_Q$ the charge of the heavy quark.\footnote{The double-logarithmic two--mass correction to $F_2(x,Q^2)$
in \cite{Blumlein:2018jfm}, Eq.~(21), has to be corrected by the factor $-(e_c^2+e_b^2)$.}

The massless two--loop Wilson coefficients are given in 
\cite{Zijlstra:1992kj,Moch:1999eb,FORTRAN,Moch:2008fj,Vogt:2008yw,Blumlein:2022gpp}
and the massive Wilson coefficients $\Delta L_{q,1}^{\sf NS, (2)}, \Delta L_{g,1}^{\sf S, (2)},
\Delta H_{q,1}^{\sf PS, (2)}$ and $\Delta H_{g,1}^{\sf S, (2)}$ are 
given in Appendix~\ref{sec:C} and  $\Delta H_{g,1}^{\sf S, (1)}$ in (\ref{eq:Hg1}).

The twist--2 heavy flavor contributions to the structure function $g_1(x,Q^2)$ are 
calculated using the collinear parton model. This is not possible in the case of the 
structure function $g_2(x,Q^2)$. As shown in Ref.~\cite{BRN1}, for the 
gluonic contributions the Wandzura--Wilczek relation also holds for the heavy flavor contributions
\begin{eqnarray}
g_2^{\tau = 2}(x,Q^2) = - g_1^{\tau = 2}(x,Q^2) 
+  \int_x^1 \frac{dz}{z} g_1^{\tau = 2}(z,Q^2)~,
\end{eqnarray}
which can be proven in the covariant parton model and derived from the analytic continuation of the 
moments obtained in the light cone expansion \cite{G2A,G2B,BT,BRN1}. Here the twist expansion is 
necessary. 

At leading order the heavy flavor Wilson corrections are known in the whole 
kinematic region, \cite{WIL1a,WIL1b}
\begin{eqnarray}
\label{eq:Hg1}
\Delta H_{g,1}^{(1)}\left(z,\frac{m^2}{Q^2}\right) 
= 4 T_F \left[\beta (3- 4 z) - (1- 2z) \ln \left| 
\frac{1 + \beta}{1- \beta} \right|\right]~,
\end{eqnarray}
where $\beta$ denotes the center of mass (cms) velocity of the heavy quarks,
\begin{eqnarray}
\beta = \sqrt{1- \frac{4 m^2}{Q^2} \frac{z}{1-z}}~.
\end{eqnarray}
The support of $\Delta H_{g,1}^{(1)}\left(z,{m^2}/{Q^2}\right)$ is given by 
$z~\epsilon~[0,1/a]$,
where $a = 1 + 4 m^2/Q^2$ and $m$ denotes the heavy quark mass. As it is well know,
the first moment of the Wilson coefficient vanishes 
\begin{eqnarray}
\label{eq2}
\int_0^{1/a} dz \Delta H_{g,1}^{(1)}\left(z,\frac{m^2}{Q^2}\right) =  0~,
\end{eqnarray}
which has a phenomenological implication on the heavy flavor contributions to polarized structure 
functions, resulting into an oscillatory profile \cite{BRN1}. The 
unpolarized heavy flavor Wilson 
coefficients \cite{UHEAV1,BUZA1,BBK1} do not obey a 
relation like (\ref{eq2}) but exhibit a rising behaviour towards  small 
values of $x$. 

The massive contribution to the structure function $g_1(x,Q^2)$ at $O(a_s)$ is given by
\begin{eqnarray}
x g_1^{Q\overline{Q}}(x,Q^2) = \frac{x}{2} e_Q^2  a_s(Q^2) \int_{ax}^1 \frac{dz}{z} 
\Delta H_{g,1}^{(1)}\left(\frac{x}{z},
\frac{m^2}{Q^2}\right) \Delta G(z,Q^2)~.
\end{eqnarray}

At asymptotic values $Q^2 \gg m^2$ one obtains the leading order heavy flavor Wilson coefficient
\begin{eqnarray}
\label{eq1}
\Delta H_{g,1}^{{\rm as},(1)}\left(z,\frac{m^2}{Q^2}\right)
= 4 T_F \left[(3- 4 z) 
+ (2z - 1) \ln \left(
\frac{1-z}{z}\right)
+ (2z - 1) \ln \left(
\frac{Q^2}{m^2}\right)
\right] 
\end{eqnarray}
and $a = 1$.
The factor in front of the logarithmic term $\ln(Q^2/m^2)$ in (\ref{eq1}) is the leading
order polarized splitting function $P_{qg}^{(0)}(z)$
\begin{eqnarray}
P_{qg}^{(0)}(z) = 8 T_F \left[ z^2 - (1-z)^2\right] = 8 T_F \left[2z - 
1\right]~. 
\end{eqnarray}
The sum--rule (\ref{eq2}) also holds in the asymptotic case extending the range of
integration to $z~\epsilon~[0,1]$,
\begin{eqnarray}
\label{eq2as}
\int_0^{1} dz \Delta H_{g,1}^{{\rm as}, (1)}\left(z,\frac{m^2}{Q^2}\right) =  0~.
\end{eqnarray}
Note that $H_{g,1}^{(1)}$ does not depend on the factorization scale $\mu^2$ due to the absence 
of collinear singularities.\footnote{Sometimes it is assumed in the literature that the {\it phase space}
logarithm $\ln(Q^2/m^2)$ would be a collinear logarithm, and has to be resummed. This, however, is not the 
case for the differential scattering cross section. See, however, Section~\ref{sec:6}.}

As has been shown in Ref.~\cite{BUZA1} the asymptotic heavy flavor Wilson coefficients obey a factorized form given by certain 
Mellin--convolutions of the massive OMEs and the massless Wilson 
coefficients.
The expression at one-- and two--loop order in the tagged heavy flavor case were given in Ref.~\cite{BUZA1}.
In the inclusive case the general structure of the Wilson coefficients is \cite{Bierenbaum:2009mv}
\begin{eqnarray}
\label{eqH2g}
\Delta H_{g_1,g}^{\rm S, (1)}\left(\frac{Q^2}{m^2}\right)
&=&    \Delta \widehat{C}^{(1)}_{g_1,g}\left(\frac{Q^2}{\mu^2}\right)
               +  \Delta A_{Qg}^{(1)}\left(\frac{\mu^2}{m^2}\right), 
\\
\Delta H_{g_1,g}^{\rm S, (2)}\left(\frac{Q^2}{\mu^2}, \frac{m^2}{\mu^2}\right)
&=&   
                  \Delta \widehat{C}^{(2)}_{g_1,g}\left(\frac{Q^2}{\mu^2}\right)
+ \Delta A_{Qg}^{(1)}\left(\frac{\mu^2}{m^2}\right) \otimes
                  \Delta C^{(1)}_{g_1,q}\left(\frac{Q^2}{\mu^2}\right)
+ \Delta A_{Qg}^{(2)}\left(\frac{\mu^2}{m^2}\right)
\nonumber\\ && 
 + \Delta A_{gg,Q}^{(2)} \left(\frac{\mu^2}{m^2}\right) 
\Delta \widetilde{C}^{,(1)}_{g_1,g}\left(\frac{Q^2}{\mu^2}\right),
\\
\label{eqH2ps}
\Delta H_{g_1,q}^{\rm PS, (2)}\left(\frac{Q^2}{m^2}, \frac{m^2}{\mu^2}\right)
&=&  \Delta \widehat{C}^{{\rm PS},(2)}_{g_1,q}\left(\frac{Q^2}{\mu^2}\right)
 +               \Delta A_{Qq}^{{\rm PS},(2)}\left(\frac{\mu^2}{m^2}\right), 
\\
\label{eqH2ns}
\Delta L_{g_1,q}^{\rm NS, (2)}\left(\frac{Q^2}{m^2}, \frac{m^2}{\mu^2}\right)
&=&  \Delta \widehat{C}^{{\rm NS},(2)}_{g_1,q,Q}\left(\frac{Q^2}{\mu^2}\right)
 +              \Delta A_{qq,Q}^{{\rm NS},(2)}\left(\frac{\mu^2}{m^2}\right)
+ \Delta L_{g_1,q}^{\rm NS (2), massless}\left(\frac{Q^2}{m^2}, \frac{m^2}{\mu^2}\right),
\nonumber\\
\\
\label{eqL2g}
\Delta L_{g_1,g}^{\rm S, (2)}\left(\frac{Q^2}{m^2}, \frac{m^2}{\mu^2}\right)
&=&                \Delta A_{gg,Q}^{(2)}\left(\frac{\mu^2}{m^2}\right) N_F
\Delta \widetilde{C}^{,(1)}_{g_1,g}\left(\frac{Q^2}{\mu^2}\right),
\end{eqnarray}
with
\begin{eqnarray}
\Delta L_{g_1,q}^{\rm NS,(2),massless} &=& - \beta_{0,Q} \ln\left(\frac{m^2}{\mu^2}\right) \left[P_{qq}^{(0)}(z) 
\ln\left(\frac{Q^2}{\mu^2}\right) + c_{g_1,q}^{(1)}(z)\right]
\end{eqnarray}
and 
\begin{eqnarray}
\beta_{0,Q} = - \frac{4}{3} T_F,
\end{eqnarray}
and $T_F = 1/2$ in $SU(N_c)$. 
Note the difference between the definition of $\Delta L_{g_1,q}^{\rm NS (2)}$ in \cite{BUZA2} and (\ref{eqH2ns}), 
cf.~\cite{Blumlein:2016xcy}. Since $\Delta L_{g_1,q}^{\rm NS (2)}\left(\frac{Q^2}{m^2}, \frac{m^2}{\mu^2}
\right) - \Delta L_{g_1,q}^{\rm NS,(2),massless}$ is finite in $D=4$ dimensions 
\cite{BUZA2,Blumlein:2016xcy}, there is no finite renormalization for this quantity, and 
$\Delta L_{g_1,q}^{\rm NS,(2),massless}$ is a pure bubble correction of the massless one--loop
Wilson coefficient, which is known in the ${\overline{\sf MS}}$ scheme too.

The massless Wilson coefficient $\Delta \widehat{C}^{(1)}_{g_1,g}\left(\frac{Q^2}{\mu^2}\right)$ depends on the 
factorization scale $\mu^2$. This dependence cancels, however, against that of the massive OME in
$\Delta H_{g_1,g}^{\rm S (1)}$.

In measuring the structure functions $g_1(x,Q^2)$ and $g_2(x,Q^2)$ the inclusive relations apply.
Here also heavy flavor corrections with massless di-quark final states and virtual heavy
flavor corrections contribute.\footnote{In Ref.~\cite{Blumlein:2016xcy} it has been shown that,
otherwise, the polarized Bjorken sum-rule cannot be obtained.}
The massless coefficient functions, related to 
$N_H$ heavy quarks, are denoted by
\begin{eqnarray}
\label{EQ:hat1}
\Delta \widehat{C}_{g_1;k}\left(\frac{Q^2}{\mu^2}\right) =
\Delta C_{g_1;k}\left(\frac{Q^2}{\mu^2},N_L+N_H\right)
- \Delta C_{g_1;k}\left(\frac{Q^2}{\mu^2},N_L\right)~,
\end{eqnarray}
where $N_L$ is the number of light flavors. In the following we will consider the case of a single heavy
quark, i.e. $N_H = 1$. 

The representation of the polarized two--loop massless Wilson coefficients 
in Ref.~\cite{ZN} have been corrected several times. They are partly given
in the Larin scheme and partly in the {\sf M} scheme, see also the comment in 
\cite{BUZA2} on the calculation of $A_{Qg}^{(2)}$ there. For clear
reference we present the pure singlet and gluonic contributions in the {\sf M} scheme using harmonic
polylogarithms (or alternatively, harmonic sums) in Appendix~\ref{sec:C}.
The massless flavor non--singlet Wilson coefficient is the same as for the unpolarized structure function $xF_3$ and it 
has been 
calculated in \cite{Moch:2008fj,Blumlein:2022gpp} to three--loop order.\footnote{
This also holds to three--loop order for the usual case of $N_F = 3$ massless flavors, but not for more 
(or less) massless flavors, resulting into a different $d_{abc}$ term, despite the first moment of its 
contribution vanishes.} 
The two--loop results were obtained in 
\cite{Zijlstra:1992kj,Moch:1999eb,Blumlein:2022gpp}. In \cite{Vogt:2008yw} it has been mentioned that the 
final result on the two--loop massless Wilson coefficients of \cite{ZN} for $g_1$ have been confirmed in the 
{\sf M} scheme. We have checked that our results also agree with the corresponding {\tt FORTRAN} program by 
W.L.~van Neerven \cite{FORTRAN}. 

In the following, we will use the notation $\hat{f}$, Eq.~(\ref{EQ:hat1}), also for the splitting 
functions and anomalous dimensions
\begin{eqnarray}
\label{Thehat}
\hat{f} = f(N_F+1) - f(N_F).
\end{eqnarray}
The operator matrix elements $ A_{k,i}^{\rm S, NS}$ obey the expansion
\begin{eqnarray}
\label{op1}
\Delta A_{k,i}^{\rm S, NS} \left(\frac{m^2}{\mu^2}\right) = \langle i|O_k|i\rangle
= \delta_{k,i} + \sum_{l=1}^{\infty} a_s^l \Delta A_{k,i}^{{\rm S,NS},(l)},~~~~k,i=q,g.
\end{eqnarray}
The twist--2 operators $O_k$ form  the massive OMEs between {\sf partonic} states $|i\rangle$,
which are related by collinear factorization to the initial--state nucleon
states $|N\rangle$. 

The operator matrix elements  $A_{k,i}^{\rm S, NS} \left({m^2}/{\mu^2},x\right)$ are process 
independent quantities. They are calculated from the diagrams in 
Figures~2--5 of Ref.~\cite{BBK1}, 
for the polarized local non--singlet, singlet and gluon operators
\begin{eqnarray}
\label{eq3}
O_{q,r}^{\{\mu_1 \ldots \mu_N\}}(x) &=& i^{N-1} {\bf S}\left[ 
\overline{\psi}(x) \gamma_5 \gamma^{\mu_1} 
D^{\mu_2} \ldots D^{\mu_N} \frac{\lambda_r}{2} \psi(x) \right] -~~{\sf 
trace~terms}, \\
O_{q}^{\{\mu_1 \ldots \mu_N\}}(x) &=& i^{N-1} {\bf S}\left[ 
\overline{\psi}(x) \gamma_5 \gamma^{\mu_1} 
D^{\mu_2} \ldots D^{\mu_N}  \psi(x) \right] -~~{\sf trace~terms}, \\
\label{eq4}
O_{g}^{\{\mu_1 \ldots \mu_N\}}(x) &=& 2 i^{N-2} {\bf S} {\bf Sp} \left[ 
\varepsilon^{\mu_1 \alpha 
\beta \gamma} {\rm Tr} \left( F_{\beta\gamma}^a(x) 
D^{\mu_2} \ldots D^{\mu_{N-1}}  F_{\alpha, a}^{\mu_N}\right) \right] -~~{\sf 
trace~terms}~.
\end{eqnarray}
Here $\psi(x)$ denotes the quark field, $\lambda_r$ the $SU(3)$ (light) flavor matrix, 
$D^{\mu}$ the covariant derivative including the gluon fields, $F_{\alpha\beta}^a$ the gluon
field strength tensor, with $a$ the $SU(3)_c$ color index. The trace ({\bf Sp}) is over color space. The curly 
brackets $\{ \ldots \}$ in the 
l.h.s. of Eqs.~(\ref{eq3}--\ref{eq4}) and the symbol ${\bf S}$ in the r.h.s. denote symmetrization of all Lorentz
indices, which projects onto the twist--2 operators. The corresponding Feynman rules 
are obtained by replacing in the quark case
\begin{eqnarray}
\Delta \!\!\!\! /~~~\rightarrow~~~\Delta \!\!\!\! /
\gamma_5
\end{eqnarray}
and turning from the field strength tensor $F_{\mu\nu}^a$ to its dual by introducing the Levi--Civita 
symbol $\ep^{\mu \alpha \beta \gamma}$ in the gluonic case
in the unpolarized Feynman rules of the operator insertions given  
in Figure~1 of Ref.~\cite{BBK1} and Figures~8 and 9 of Ref.~\cite{Bierenbaum:2009mv}.

The expansion coefficients of the unrenormalized OMEs $A_{k,i}^{{\rm S,NS},(l)}$ have the
representation
\begin{eqnarray}
\Delta A_{k,i}^{{\rm S,NS},(l)} = \sum_{m = -l}^ \infty \ep^{l} \Delta a_{k,i}^{(l),m}.
\end{eqnarray}
In the present calculation we need the following coefficients 
\begin{eqnarray}
&& \Delta a_{Qg}^{(1),0}   = 0,~~~
\Delta a_{Qg}^{(1),1}   = \Delta \overline{a}_{Qg}^{(1)},~~~
\Delta a_{qq,Q}^{{\rm NS},(2),0} = \Delta a_{qq,Q}^{{\rm NS},(2)},~~~ 
\Delta a_{Qq}^{{\rm PS},(2),0} =  \Delta a_{Qq}^{{\rm PS},(2)},~~~ 
\nonumber\\ &&
\Delta a_{Qg}^{(2),0} = \Delta a_{Qg}^{(2)}. 
\end{eqnarray}
We also calculate the $O(\ep)$ terms at two--loop order, denoted by a bar, for use at the three--loop level.

The massless Wilson coefficients to $O(a_s^2)$ are given by, cf.~\cite{Zijlstra:1992qd},
\begin{eqnarray}
\Delta C^{(1)}_{g_1,g} &=& 
\frac{1}{2} \Delta P_{qg}^{(0)} 
\ln\left(\frac{Q^2}{\mu^2}\right) + c_{g_1,g}^{(1)},
\\
\Delta C^{(2)}_{g_1,g} &=& 
\Biggl[
\frac{1}{8} \Delta P_{qg}^{(0)} 
\left[\Delta P_{gg}^{(0)} + \Delta P_{qq}^{(0)}\right]
- \frac{1}{4} \beta_0 \Delta P_{qg}^{(0)} 
\Biggr] \ln^2\left(\frac{Q^2}{\mu^2}\right)
\nonumber\\ &&
+ \Biggl[
\frac{1}{2} \Delta P_{qg}^{(1)} + \left(\frac{1}{2} 
\Delta P_{gg}^{(0)} - \beta_0\right) c_{g_1,g}^{(1)}
+ \frac{1}{2} \Delta P_{qg}^{(0)} c_{g_1,q}^{(1)}
\Biggr] \ln\left(\frac{Q^2}{\mu^2}\right)
+ c_{g_1,g}^{(2)},
\\
C^{{\rm PS},(2)}_{g_1,q} &=& 
\frac{1}{8} \Delta P_{qg}^{(0)} \Delta P_{gq}^{(0)} 
\ln^2\left(\frac{Q^2}{\mu^2}\right)
+ \Biggl[
\frac{1}{2} \Delta P_{qg}^{{\rm PS},(1)}
+ \frac{1}{2} P_{gq}^{(0)} c_{g_1,g}^{(1)}
\Biggr] 
\ln\left(\frac{Q^2}{\mu^2}\right)
+ c_{g_1,g}^{{\rm PS},(2)},
\\
C^{{\rm NS},(2)}_{g_1,q} &=& 
\Biggl[
\frac{1}{8} {\Delta P_{qq}^{(0)}}^2 - \frac{1}{4} \beta_0 \Delta P_{qq}^{(0)}
\Biggr]\ln^2\left(\frac{Q^2}{\mu^2}\right)
\nonumber\\ &&
+ \Biggl[
\frac{1}{2} \Delta P_{qq}^{{\rm NS-},(1)} 
+ \Biggl( \frac{1}{2} \Delta P_{qq}^{(0)} - \beta_0 c_{g_1,q}^{(1)}
\Biggr) c_{g_1,q}^{(1)} \Biggr] \ln\left(\frac{Q^2}{\mu^2}\right)
+ c_{g_1,q}^{(2)}.
\end{eqnarray}
Here $c_i^{(k)}$ denotes the contribution to $\Delta C_i^{(k)}$ for $Q^2 = \mu^2$ and 
$\beta_0$ is the lowest order expansion coefficient of the QCD 
$\beta$--function,
\begin{eqnarray}
\beta_0 &=& \frac{11}{3} C_A -  \frac{4}{3} N_F T_F, 
\end{eqnarray}
with $C_A = N_c, C_F = (N_c^2-1)/(2 N_c)$ and $N_F$ denotes the number of light quark 
flavors and $N_c = 3$ for QCD. $\zeta_k = \sum_{l=1}^\infty (1/l^k),~k \in \mathbb{N}, k \geq 2$ denote
values of the Riemann $\zeta$--function at integer argument and $P_{ij}^{{\rm (PS,NS)},(k-1)}$,
$c_{g_1,i}^{(k)}$ are the $k$th order splitting and coefficient functions. 
For the different $N$--dependent functions we use the shorthand notation $F(N) \equiv F$.

The splitting functions $\Delta P_{ij}^{(k)}(N)$ are related to the anomalous dimensions 
$\Delta \gamma_{ij}^{(k)}$ by
\begin{eqnarray}
\Delta P_{ij}^{(k)} = - \Delta \gamma_{ij}^{(k)},
\end{eqnarray}
used in other representations. In the representation in Mellin $N$ space the corresponding 
quantities depend on nested harmonic sums, $S_{\vec{a}}$, \cite{Vermaseren:1998uu,Blumlein:1998if}, which 
are recursively 
defined by
\begin{eqnarray}
S_{a_1, ... , a_l}(N) = \sum_{k=1}^N \frac{({\rm sign}(a_1))^k}{k^{|a_1|}}
S_{a_2, ... , a_l}(k)~, S_\emptyset = 1, a_i \in \mathbb{Z} \backslash \{0\}.
\end{eqnarray}
At LO and NLO the splitting functions \cite{SP_PS,Vogelsang:1995vh,Vogelsang:1996im,Moch:2014sna,
Behring:2019tus,Blumlein:2021ryt,Blumlein:2022gpp} in the 
{\sf M} scheme are given by\footnote{Here and in the following we drop the factor 
$\tfrac{1}{2}[1-(-1)^N]$ and
the integer moments are taken at the odd integers $N \geq 1$. Note the partly different normalizations comparing 
the splitting functions given in Refs.~\cite{SP_PS,Vogelsang:1995vh,Vogelsang:1996im,Moch:2014sna,
Behring:2019tus,Blumlein:2021ryt,Blumlein:2022gpp}.}
\begin{eqnarray}
\Delta P_{qq}^{(0)} &=&  \textcolor{black}{C_F} \left[\frac{2 (2+3 N+3 N^2)}{N (N+1)} 
- 8 S_1\right],
\\
\Delta P_{qg}^{(0)} &=&  8 \textcolor{black}{T_F N_F} \frac{N-1}{N(N+1)},
\\
\Delta P_{gq}^{(0)} &=&  4 \textcolor{black}{C_F} \frac{N+2}{N(N+1)},
\\
\Delta P_{gg}^{(0)} &=&  \textcolor{black}{C_A} \left[
\frac{2 (24+11 N+11 N^2)}{3 N (N+1)}-8 S_1 \right]
-\frac{8}{3} \textcolor{black}{T_F N_F},
\\
\Delta \hat{P}_{qq}^{(1),\rm NS} &=&  \textcolor{black}{C_F T_F N_F} \frac{4}{3} \Biggl\{
-8 S_2 + \frac{40}{3} S_1 - \frac{3 N^4 + 6 N^3 +47 N^2 +20 N -12}{3 N^2 (N+1)^2}
\Biggr\},
\\
\Delta P_{qq}^{(1),\rm PS} &=&  -\textcolor{black}{C_F T_F N_F} \frac{16 (N+2) (1+2 N+N^3)}{N^3 
(N+1)^3},
\\
\Delta P_{qg}^{(1)} &=&  \textcolor{black}{C_F T_F N_F}  \Biggl[
        \frac{8 (N-1) \big(
                2-N+10 N^3+5 N^4\big)}{N^3 (N+1)^3}
        -\frac{32 (N-1)}{N^2 (N+1)} S_1
\nonumber\\ &&
        +\frac{16 (N-1)}{N (N+1)} [S_1^2 - S_2]
\Biggr]
+\textcolor{black}{C_A T_F N_F} \Biggl[
        \frac{16\big(  
N^5
+N^4
-4 N^3
+3 N^2
-7 N
-2
\big)}{N^3 (1+N)^3} 
\nonumber\\ &&
        +\frac{64}{N (1+N)^2} S_1
        -\frac{16 (N-1)}{N (1+N)} [S_1^2+ S_2 + 2 S_{-2}]
\Biggr].
\end{eqnarray}
The first order polarized Wilson coefficients 
$c_{g_1,q}$ and  $c_{g_1,g}$  for $Q^2 = \mu^2$ read  
\cite{FP,WIL_p1,WIL1b,ZN,Blumlein:2019zux,Moch:1999eb,Blumlein:2022gpp}
\begin{eqnarray}
\label{eq11a}
c_{g_1,q}^{(1)} &=& 
\textcolor{black}{C_F} \left [
        -\frac{(2+3 N) \big(
                3 N^2-1\big)}{N^2 (N+1)}
        +\frac{3 N^2+3 N-2}{N (N+1)} S_1
        +2 [S_1^2 - S_2]
\right],
\\
c_{g_1,g}^{(1)} &=& -4 \textcolor{black}{T_F N_F} \frac{N-1}{N(N+1)} \left[\frac{N-1}{N} + S_1\right]~.
\end{eqnarray}
The first moment of $c_{g_1,q}^{(1)}$ yields $- 3 C_F$ in accordance with the Bjorken sum rule in the 
massless case \cite{Bjorken:1969mm} and \cite{Kodaira:1978sh}. The 2nd order contributions were given in
\cite{ZN,FORTRAN,Vogt:2008yw,Blumlein:2022gpp,Blumlein:2022gpp}.
\section{Renormalization}
\label{sec:3}

\vspace{1mm}\noindent
In the following we briefly summarize the renormalization of the polarized  massive operator matrix 
elements and Wilson coefficients to $O(a_s^2)$. It has been given for the case of tagged heavy flavor 
in Ref.~\cite{BUZA2,BUZA1}. We will consider, however, the inclusive case since we deal with the structure 
function $g_1(x,Q^2)$  and follow Ref.~\cite{Bierenbaum:2009mv}\footnote{Note that the renormalization 
applied in \cite{UHEAV1,BUZA1,BUZA2} is not generally valid in the case of inclusive structure functions.},
where the renormalization has been performed in the unpolarized case. Since we use the 
Larin prescription \cite{GAM5},
we perform subsequently a finite renormalization to the {\sf M} scheme given in 
Ref.~\cite{MAT1,Moch:2014sna,Behring:2019tus,Blumlein:2021ryt}, which is the only additional 
renormalization step beyond those described in Ref.~\cite{Bierenbaum:2009mv} in the single heavy mass case.

The unrenormalized polarized OMEs obey the series expansion
\begin{eqnarray}
\label{Aunren}
\Delta \hat{\hat{A}}_{ij} &=& \delta_{ij} + \sum_{k=1}^\infty \hat{a}_s^k 
\Delta \hat{\hat{A}}_{ij}^{(k)}~,
\end{eqnarray}
In the 
renormalized case, the corresponding expansion reads
\begin{eqnarray}
\label{Aren}
\Delta {A}_{ij} &=& \delta_{ij} + \sum_{k=1}^\infty {a}_s^k \Delta {A}_{ij}^{(k)}~.
\end{eqnarray}
One performs {\it i)} the mass renormalization, {\it ii)} the coupling constant 
renormalization, {\it iii)} the renormalization of the local operators by ultraviolet 
$Z$--factors, and for the massless sub--sets of the diagrams {\it iv)} one removes the
collinear singularities. By this one obtains the renormalized OMEs given in 
\cite{Bierenbaum:2009mv}, Eqs.~(4.16, 4.22, 4.35) and the renormalized asymptotic
massive Wilson coefficients in Eqs.~(2.11, 2.14, 2.15). These expressions can be written
in terms of the anomalous dimensions, massless Wilson coefficients, the expansion coefficients
of the unrenormalized heavy quark mass, the QCD $\beta$-function and the expansion coefficients of
the massive OMEs up to two--loop order.

Yet these expressions are given in the Larin scheme used in the present calculation. 
The massless Wilson coefficients to two--loop order transform from the Larin scheme to the 
{\sf M} scheme \cite{MAT1} by
\begin{eqnarray}
\Delta C_{g_1,q}^{\rm (1), NS, M} &=& \Delta C_{g_1,g}^{\rm (1), NS, L} - z_{qq}^{(1)},
\\
\Delta C_{g_1,q}^{\rm (2), NS, M} &=& \Delta C_{g_1,g}^{\rm (2), NS, L} + {z_{qq}^{(1)}}^2 - 
z_{qq}^{(2),\rm NS} - z_{qq}^{(1)} 
\Delta C_{g_1,q}^{\rm (1), NS, L}, 
\\
\Delta C_{g_1,q}^{\rm (2), PS, M} &=& \Delta C_{g_1,g}^{\rm (2), PS, L} - z_{qq}^{(2),\rm 
PS},
\\
\Delta C_{g_1,g}^{\rm (1), M} &=& \Delta C_{g_1,g}^{\rm (1), L}, 
\\
\Delta C_{g_1,g}^{\rm (2), M} &=& \Delta C_{g_1,g}^{\rm (2), L}~.
\end{eqnarray}
The relations can also be determined considering the massless physical evolution coefficients associated to the 
pair of observables $\{F_A, F_B\} = \{g_1(x,Q^2), dg_1(x,Q^2)/d \ln(Q^2)\}$, 
cf.~Ref.~\cite{Blumlein:2004xs,Blumlein:2021lmf}\footnote{We corrected typos in \cite{Blumlein:2004xs}.}.
One considers the evolution equation
\begin{eqnarray}
\frac{d}{dt} \left(\begin{array}{c} F_A \\ F_B \end{array}\right) = - \frac{1}{4} 
\left(\begin{array}{cc} 
K_{AA} & K_{AB} \\
K_{BA} & K_{BB} 
\end{array}\right) \left(\begin{array}{c} F_A \\ F_B \end{array}\right),
\end{eqnarray}
with $t = - (2/\beta_0) \ln(a_s(Q^2)/a_s(Q_0^2))$.
The scheme--invariant singlet evolution coefficients in the massless case read \footnote{To 1-- and 2--loop order 
they were given in Refs. \cite{FP,Blumlein:2000wh}. Here we drop the $\Delta$ in front of 
the $\gamma_{ij}$ and $c_l$.}

\begin{eqnarray}
    K_{d1}^{(0)} &=& \frac{1}{4}\left[
    \gamma_{qq}^{(0)}
    \gamma_{gg}^{(0)} -
    \gamma_{qg}^{(0)} 
    \gamma_{gq}^{(0)} \right],
    \\
    K_{dd}^{(0)} &=& \gamma_{qq}^{(0)} + \gamma_{gg}^{(0)},
    \\
    K_{d1}^{(1)} &=& 
    \frac{\beta_0}{\gamma_{qg}^{(0)}} \Biggl[
            c_{1,g}^{(1)} \Biggl(
                    \beta_0 \gamma_{qq}^{(0)}
                    +\frac{1}{2}  (\gamma_{gg}^{(0)}
                    -\gamma_{qq}^{(0)}
                    ) \gamma_{qq}^{(0)}
            \Biggr)
            +\frac{1}{2}  \gamma_{qg}^{(1)} \gamma_{qq}^{(0)}
    \Biggr]
    + \beta_0 c_{1,q}^{(1)} \Biggl[
            -\beta_0
            +\frac{1}{2}  (\gamma_{gg}^{(0)}
            +\gamma_{qq}^{(0)}
            )
    \Biggr]
    \nonumber\\ &&
    +\frac{1}{4} (
      \gamma_{gg}^{(1)} \gamma_{qq}^{(0)}
    + \gamma_{qq}^{(1)} \gamma_{gg}^{(0)} 
    - \gamma_{qg}^{(1)} \gamma_{gq}^{(0)} 
    -\gamma_{gq}^{(1)}  \gamma_{qg}^{(0)}
    )
    -\beta_0 c_{1,g}^{(1)} \gamma_{gq}^{(0)}
    + \frac{\beta_1}{2 \beta_0} \left(\gamma_{qg}^{(0)}  \gamma_{gq}^{(0)} - \gamma_{gg}^{(0)} \gamma_{qq}^{(0)}
    \right)
    \nonumber\\ &&
    -\frac{1}{2} \beta_0 \gamma_{qq}^{(1)},
    \\
    K_{dd}^{(1)} &=& 4 \beta_0 c_{1,q}^{(1)} 
    + \gamma_{qq}^{(1)}
    + \gamma_{gg}^{(1)} + 2 \beta_0 \frac{\gamma_{qg}^{(1)}}{\gamma_{qg}^{(0)}}
    - \frac{\beta_1}{\beta_0} \left(2 \beta_0 + \gamma_{qq}^{(0)} + \gamma_{gg}^{(0)}\right)
    + \frac{2 c_{1,g}^{(1)} \beta_0}{\gamma_{qg}^{(0)}} 
    \left(2 \beta_0 + \gamma_{gg}^{(0)} - \gamma_{qq}^{(0)}\right),
    \nonumber\\
    \\
    K_{d1}^{(2)} &=& -
    \frac{{\gamma_{gq}^{(2)}} {\gamma_{qg}^{(0)}}}{4}
    -\frac{{\gamma_{gq}^{(1)}} {\gamma_{qg}^{(1)}}}{4}
    -\frac{{\gamma_{gq}^{(0)}} {\gamma_{qg}^{(2)}}}{4}
    +\frac{{\gamma_{gg}^{(2)}} {\gamma_{qq}^{(0)}}}{4}
    +\frac{{\gamma_{gg}^{(1)}} {\gamma_{qq}^{(1)}}}{4}
    +\frac{{\gamma_{gg}^{(0)}} {\gamma_{qq}^{(2)}}}{4}
    \nonumber\\ &&
    +{\beta_0} \Biggl(
            -\frac{{c_{1,q}^{(1)}}^2 {\gamma_{gg}^{(0)}}}{2}
            +{c_{1,q}^{(2)}} {\gamma_{gg}^{(0)}}
            +\frac{{c_{1,q}^{(1)}} {\gamma_{gg}^{(1)}}}{2}
            -2 {c_{1,g}^{(2)}} {\gamma_{gq}^{(0)}}
            +2 {c_{1,g}^{(1)}} {c_{1,q}^{(1)}} {\gamma_{gq}^{(0)}}
            -\frac{3 {c_{1,g}^{(1)}} {\gamma_{gq}^{(1)}}}{2}
            -\frac{{c_{1,q}^{(1)}}^2 {\gamma_{qq}^{(0)}}}{2}
    \nonumber\\ && 
            +{c_{1,q}^{(2)}} {\gamma_{qq}^{(0)}}
            +\frac{{c_{1,q}^{(1)}} {\gamma_{qq}^{(1)}}}{2}
            -{\gamma_{qq}^{(2)}}
            +\frac{1}{{\gamma_{qg}^{(0)}}^2} \Biggl(
                    -\frac{1}{2} {c_{1,g}^{(1)}}^2 {\gamma_{gg}^{(0)}}^2 {\gamma_{qq}^{(0)}}
                    -{c_{1,g}^{(1)}} {\gamma_{gg}^{(0)}} {\gamma_{qg}^{(1)}} {\gamma_{qq}^{(0)}}
         -\frac{{\gamma_{qg}^{(1)}}^2 {\gamma_{qq}^{(0)}}}{2}
    \nonumber\\ &&            
                    +{c_{1,g}^{(1)}}^2 {\gamma_{gg}^{(0)}} {\gamma_{qq}^{(0)}}^2
                    +{c_{1,g}^{(1)}} {\gamma_{qg}^{(1)}} {\gamma_{qq}^{(0)}}^2
                    -\frac{1}{2} {c_{1,g}^{(1)}}^2 {\gamma_{qq}^{(0)}}^3
            \Biggr)
            +\frac{1}{{\gamma_{qg}^{(0)}}} \Biggl(
                    \frac{1}{2} {c_{1,g}^{(1)}}^2 {\gamma_{gg}^{(0)}} {\gamma_{gq}^{(0)}}
    +\frac{{c_{1,g}^{(1)}} {\gamma_{gq}^{(0)}} {\gamma_{qg}^{(1)}}}{2}
    \nonumber\\ &&                 
                    +{c_{1,g}^{(2)}} {\gamma_{gg}^{(0)}} {\gamma_{qq}^{(0)}}
                    +{c_{1,g}^{(1)}} {\gamma_{gg}^{(1)}} {\gamma_{qq}^{(0)}}
                    -\frac{3}{2} {c_{1,g}^{(1)}}^2 {\gamma_{gq}^{(0)}} {\gamma_{qq}^{(0)}}
                    +{\gamma_{qg}^{(2)}} {\gamma_{qq}^{(0)}}
                    -{c_{1,g}^{(2)}} {\gamma_{qq}^{(0)}}^2
                    +
                    \frac{{c_{1,g}^{(1)}} {\gamma_{gg}^{(0)}} {\gamma_{qq}^{(1)}}}{2}
    \nonumber\\ &&             
        +\frac{{\gamma_{qg}^{(1)}} {\gamma_{qq}^{(1)}}}{2}
                    -\frac{3}{2} {c_{1,g}^{(1)}} {\gamma_{qq}^{(0)}} {\gamma_{qq}^{(1)}}
                    +{c_{1,g}^{(1)}} {c_{1,q}^{(1)}} {\gamma_{qq}^{(0)}} (-{\gamma_{gg}^{(0)}}
                    +{\gamma_{qq}^{(0)}}
                    )
            \Biggr)
            +{\beta_1} \Biggl(
                    -{c_{1,q}^{(1)}}
                    +\frac{{c_{1,g}^{(1)}} {\gamma_{qq}^{(0)}}}{{\gamma_{qg}^{(0)}}}
            \Biggr)
    \Biggr)
    \nonumber\\ && 
    +{\beta_1} \Biggl(
            {c_{1,g}^{(1)}} {\gamma_{gq}^{(0)}}
            +\frac{{\gamma_{qq}^{(1)}}}{2}
            -\frac{1}{2} {c_{1,q}^{(1)}} ({\gamma_{gg}^{(0)}}
            +{\gamma_{qq}^{(0)}}
            )
            +\frac{
            -{c_{1,g}^{(1)}} 
             {\gamma_{gg}^{(0)}} 
             {\gamma_{qq}^{(0)}}
            -{\gamma_{qg}^{(1)}} 
             {\gamma_{qq}^{(0)}}
            +{c_{1,g}^{(1)}} 
             {\gamma_{qq}^{(0)}}^2}{2 \gamma_{qg}^{(0)}}
    \Biggr)
    \nonumber\\ && 
    +\frac{1}{{\beta_0}} \Biggl(
            {\beta_2} \big(
                    \frac{{\gamma_{gq}^{(0)}} {\gamma_{qg}^{(0)}}}{2}
                    -\frac{{\gamma_{gg}^{(0)}} {\gamma_{qq}^{(0)}}}{2}
            \big)
            +{\beta_1} \Biggl(
                    \frac{{\gamma_{gq}^{(1)}} {\gamma_{qg}^{(0)}}}{2}
                    +\frac{{\gamma_{gq}^{(0)}} {\gamma_{qg}^{(1)}}}{2}
                    -\frac{{\gamma_{gg}^{(1)}} {\gamma_{qq}^{(0)}}}{2}
                    -\frac{{\gamma_{gg}^{(0)}} {\gamma_{qq}^{(1)}}}{2}
            \Biggr)
    \Biggr)
    \nonumber\\ && 
    +{\beta_0^2} \Biggl(
            3 {c_{1,q}^{(1)}}^2
            -4 {c_{1,q}^{(2)}}
            +\frac{1}{\gamma_{qg}^{(0)}}
      \left({c_{1,g}^{(1)}}^2   {\gamma_{gq}^{(0)}}
            +{c_{1,q}^{(1)}}   {\gamma_{qg}^{(1)}}
            +4 {c_{1,g}^{(2)}} {\gamma_{qq}^{(0)}}
            +{c_{1,g}^{(1)}} {\gamma_{qq}^{(1)}}
            +{c_{1,g}^{(1)}} {c_{1,q}^{(1)}} ({\gamma_{gg}^{(0)}}
    \right.
    \nonumber\\ && \left. 
           -5 {\gamma_{qq}^{(0)}} )
      \right)
            +\frac{1}{{\gamma_{qg}^{(0)}}^2}\left(
             -2 {c_{1,g}^{(1)}} 
                {\gamma_{qg}^{(1)}} 
                {\gamma_{qq}^{(0)}}
             +2 {c_{1,g}^{(1)}}^2 {\gamma_{qq}^{(0)}} (-{\gamma_{gg}^{(0)}}
            +{\gamma_{qq}^{(0)}} ) \right)
    \Biggr)
    -\frac{3 \beta_1^2}{4\beta_0^2} \left(
             {\gamma_{gq}^{(0)}} {\gamma_{qg}^{(0)}}
            -{\gamma_{gg}^{(0)}} {\gamma_{qq}^{(0)}} \right)
    \nonumber\\ && 
    +{\beta_0^3} \left(
            \frac{2 {c_{1,g}^{(1)}} {c_{1,q}^{(1)}}}{{\gamma_{qg}^{(0)}}}
            -\frac{2 {c_{1,g}^{(1)}}^2 {\gamma_{qq}^{(0)}}}{{\gamma_{qg}^{(0)}}^2}
    \right),
    \\
    K_{dd}^{(2)} &=& -4 \beta_0 (c_{1,q}^{(1)})^2
    +8 \beta_0 c_{1,q}^{(2)}
    +\frac{\beta_0}{(\gamma_{qg}^{(0)})^2} \Biggl\{
            \left(c_{1,g}^{(1)}\right)^2 \Biggl[
                    -8 \beta_0^2
                    -8 \beta_0(\gamma_{gg}^{(0)} - \gamma_{qq}^{(0)})
                    -2 (\gamma_{gg}^{(0)} - \gamma_{qq}^{(0)})^2
            \Biggr]
    \nonumber\\ &&
            +c_{1,g}^{(1)} \Biggr[
                     -4 \gamma_{gg}^{(0)} \gamma_{qg}^{(1)}
                     +4 \gamma_{qg}^{(1)} \gamma_{qq}^{(0)}
                    -8 \beta_0 \gamma_{qg}^{(1)}
            \Biggr]
            -2  (\gamma_{qg}^{(1)})^2
    \Biggr\}
    +\frac{\beta_0}{\gamma_{qg}^{(0)}} \Biggl\{
            c_{1,g}^{(2)} \big(
                    16 \beta_0
                    + 4 (\gamma_{gg}^{(0)}
    \nonumber\\ &&               
     - \gamma_{qq}^{(0)}
                    )
            \big)
            +c_{1,g}^{(1)} c_{1,q}^{(1)} \big(
                    -16 \beta_0
                    +(-4 \gamma_{gg}^{(0)}
                    +4 \gamma_{qq}^{(0)}
                    )
            \big)
            +c_{1,g}^{(1)} (8 \beta_1
            +(4 \gamma_{gg}^{(1)}
            -4 \gamma_{qq}^{(1)}
            )
            )
    \nonumber\\ &&
            -4 \left(c_{1,g}^{(1)}\right)^2 \gamma_{gq}^{(0)}
            +4  \gamma_{qg}^{(2)}
    \Biggr\}
    +\frac{\beta_2}{\beta_0} \Biggl[
            -4 \beta_0
            +(-\gamma_{gg}^{(0)}
            -\gamma_{qq}^{(0)}
            )
    \Biggr]
    +\frac{\beta_1^2}{\beta_0^2} (2 \beta_0
    +\gamma_{gg}^{(0)}
    +\gamma_{qq}^{(0)}
    )
    \nonumber\\ &&
    - \frac{\beta_1}{\beta_0} (\gamma_{gg}^{(1)}
    + \gamma_{qq}^{(1)}
    )
    +\gamma_{gg}^{(2)}
    +\gamma_{qq}^{(2)}.
\end{eqnarray}
The transformation relations for the anomalous dimensions up to three--loop order are given e.g. in 
\cite{Blumlein:2021ryt}, Eqs.~(19--29). Since the scheme--invariant 
evolution equations do not affect phase space 
logarithms, such as $\ln(Q^2/m^2)$, which occur additionally in the heavy flavor Wilson 
coefficients, the massless case is extended to the single mass case by
\begin{eqnarray}
c_{1,g}^{(1)} & \rightarrow & c_{1,g}^{(1)} + \Delta H_{1,g}^{(1)},  
\\  
c_{1,g}^{(2)} & \rightarrow & c_{1,g}^{(2)} + \Delta H_{1,g}^{(2)} + \Delta L_{1,g}^{(2)},
\\  
c_{1,q}^{(2)} & \rightarrow & c_{1,q}^{(2)} + \Delta H_{1,q}^{(2),\rm PS},
\end{eqnarray}
with \cite{Bierenbaum:2009mv}
\begin{eqnarray}
\Delta H_{1,g}^{(1)} &=& - \frac{\Delta \hat{\gamma}_{qg}^{(0)}}{2} \ln \left(\frac{Q^2}{m^2}\right) 
+ 
\tilde{c}_{1,g}^{(1)}, 
\\  
\Delta H_{1,g}^{(2)} &=& -\frac{\Delta \hat{\gamma}_{qg}^{(0)}}{8}\Bigl[\Delta \gamma_{gg}^{(0)} - 
\Delta \gamma_{qq}^{(0)} + 2 \beta_0 
+ 4 \beta_{0,Q} \Bigr] \ln^2\left(\frac{Q^2}{m^2}\right) - \frac{\Delta \hat{\gamma}_{qg}^{(1)}}{2} 
\ln\left(\frac{Q^2}{m^2}\right) 
\nonumber\\ &&
+ \frac{\zeta_2}{8} \Delta \hat{\gamma}_{qg}^{(0)} \left( \Delta \gamma_{gg}^{(0)} - 
\Delta \gamma_{qq}^{(0)} + 2 \beta_0\right) + \Delta a_{Qg}^{(2)} +  \tilde{c}_{1,g}^{(2)} - 
\frac{\Delta \hat{\gamma}_{qg}^{(0)}}{2}
\ln\left(\frac{Q^2}{m^2}\right) c_{1,q}^{(1)} 
\nonumber\\ &&
+ \beta_{0,Q} \ln\left(\frac{Q^2}{m^2}\right) \tilde{c}_{1,g}^{(1)},
\\  
\Delta L_{1,g}^{(2)} &=& \beta_{0,Q} \ln\left(\frac{Q^2}{m^2}\right) \tilde{c}_{1,g}^{(1)},
\\
\Delta H_{1,q}^{(2),\rm PS} &=& - \frac{1}{8} \Delta \hat{\gamma}_{qg}^{(0)} \Delta \gamma_{gq}^{(0)} 
\ln^2\left(\frac{Q^2}{m^2}\right) - \frac{1}{2} \Delta \hat{\gamma}_{qq}^{(1),\rm PS} 
\ln\left(\frac{Q^2}{m^2}\right)+ \frac{1}{8} \Delta \hat{\gamma}_{qg}^{(0)} \Delta \gamma_{gq}^{(0)} 
\zeta_2 + a_{Qq}^{(2),\rm PS} + \tilde{c}_{1,q}^{(2)}.
\nonumber\\ 
\end{eqnarray}
The double--mass corrections to $O(a_s^2)$ are scheme--invariant, as are $\Delta H_{1,g}^{(1)}$ and $\Delta 
L_{1,g}^{(2)}$  and the following relations are implied, 
\begin{eqnarray}
\Delta a_{Qq}^{(2),\rm PS,M} &=& \Delta a_{Qq}^{(2),\rm PS,L} + z_{qq}^{(2),\rm PS}
\\
\Delta a_{Qg}^{(2),\rm M} &=& \Delta a_{Qg}^{(2),L},
\end{eqnarray}
Here the functions determining the finite renormalization are, cf.~\cite{Moch:2014sna},
\begin{eqnarray}
z_{qq}^{(1)} &=& - \frac{8 \textcolor{black}{C_F}}{N(N+1)}, 
\\
z_{qq}^{(2),\rm NS} &=& -\textcolor{black}{C_F T_F N_F}
 \frac{16(3 + N - 5 N^2)}{9 N^2 (N+1)^2} 
+\textcolor{black}{C_A C_F} \Biggl(
- \frac{4 V_2}{9 N^3 (N+1)^3} 
\nonumber\\ &&
- \frac{16}{N (N+1)} S_{-2}\Biggr) 
+\textcolor{black}{C_F^2} \Biggl(\frac{8 V_1}{N^3 (N+1)^3} 
       + 16 \frac{1 + 2N}{N^2 (N+1)^2} S_1 
\nonumber\\ &&     
  + \frac{16}{N(N+1)} S_2
       + \frac{32}{N (N+1)} S_{-2} \Biggr),
\\
z_{qq}^{(2),\rm PS} &=& - \textcolor{black}{C_F T_F N_F} \frac{8(2+N)(N^2-N-1)}{N^3(N+1)^3},
\end{eqnarray}
with
\begin{eqnarray}
V_1 &=& 2 N^4+N^3+8 N^2+5 N+2,
\\
V_2 &=& 103 N^4+140 N^3+58 N^2+21 N+36,
\end{eqnarray}
cf.~\cite{MAT1}.

Because of the Ward--Takahashi identity in the flavor non--singlet case, 
which implies to use anticommuting $\gamma_5$ along 
the external massless quark line, one obtains $\Delta L_{g_1,g}^{\rm (2),NS, M}$ directly. It can 
also be extracted from the inclusive 
full phase space calculation in Ref.~\cite{Blumlein:2016xcy}. In the pure singlet case the asymptotic expression can be 
obtained in a similar manner from a result in \cite{Blumlein:2019zux}. In both cases only very few Feynman diagrams 
contribute, unlike the case for $A_{Qg}^{(2)}$ and $H_{g_1,g}^{\rm (2)}$.
\section{The polarized operator matrix elements}
\label{sec:4}

\vspace{1mm}\noindent
The massless QCD Wilson coefficients for polarized deeply inelastic scattering were 
calculated to $O(a_s^2)$ in Ref.~\cite{ZN,FORTRAN,Vogt:2008yw,Blumlein:2022gpp} 
in the {\sf M} scheme. 
To derive the corresponding heavy flavor Wilson coefficients we calculate the 
corresponding massive operator matrix elements. We use first the Larin prescription for 
$\gamma_5$,~\cite{GAM5}, which has been applied in the calculation of the massless
Wilson coefficients in \cite{ZN,FORTRAN,Vogt:2008yw,Blumlein:2022gpp}.\footnote{See also footnote 5  in 
\cite{Vogelsang:1996im}, in which the calculation is perform using the CFP method 
\cite{Curci:1980uw}.} The Dirac-matrix $\gamma_5$ is represented in $D$ dimensions by
  \begin{eqnarray}
   \gamma^5 &=& \frac{i}{24}\ep_{\mu\nu\rho\sigma}\gamma^{\mu}
            \gamma^{\nu}\gamma^{\rho}\gamma^{\sigma}, \\
   \adag \hspace*{-0.5mm} \Delta \gamma^5 &=& \frac{i}{6}\ep_{\mu\nu\rho\sigma}\Delta^{\mu}
            \gamma^{\nu}\gamma^{\rho}\gamma^{\sigma}~. \label{gamma5}
  \end{eqnarray}
The Levi--Civita symbol will be contracted later with a second Levi--Civita symbol
emerging in the general expression for the Green's functions 
\begin{eqnarray}
\hat{G}^{ab}_{Q,\mu\nu} &=& \Delta \hat{A}^{(N)}_{Qg}\delta^{ab}(\Delta.p)^{N-1}
                      \ep_{\mu\nu\alpha\beta}\Delta^{\alpha}p^{\beta}~,
                      \label{greensing}
\\
\hat{G}_l^{ij} &=& \Delta \hat{A}^{(N)}_{lq} \delta^{ij} (\Delta.p)^{N-1} \Delta \hspace*{-3mm} \slash 
~\gamma_5.
\end{eqnarray}
In $D$ dimensions we apply 
the  following relation, \cite{zuber},
\begin{eqnarray}
   \ep_{\mu\nu\rho\sigma}\ep^{\alpha\lambda\tau\gamma}&=&
            -{\sf Det} \left[g^{\beta}_{\omega}\right]~,
              ~~~~~\beta=\alpha,\lambda,\tau,\gamma~,\quad
               \omega=\mu,\nu,\rho,\sigma~. \N
\label{levidet} 
\end{eqnarray}
The projectors for the quarkonic and the gluonic OMEs in the Larin scheme read
\begin{eqnarray}
P_q \hat{G}_l^{ij} &=& -\delta_{ij} \frac{i (\Delta.p)^{-N-1}}{4 N_c (D-2) (D-3)} \ep_{\mu\nu p\Delta} {\text tr}\left[
p \hspace*{-2mm} \slash \gamma^{\mu} \gamma^{\nu} \hat{G}_l^{ij} \right]
\\
P_g \hat{G}_{\mu\nu}^{ab} &=&
                   \frac{\delta^{ab}}{N_c^2-1}
                   \frac{1}{(D-2)(D-3)}
                   (\Delta.p)^{-N-1}\ep^{\mu\nu\rho\sigma}
                    \Delta_{\rho}p_{\sigma}
                    \hat{G}^{ab}_{\mu\nu}
\label{projecsing}.
\end{eqnarray}
In its practical application there are further requirements which we will describe in Section~\ref{sec:44}.

In combining the massless Wilson coefficients with the  massive operator matrix elements, (\ref{eqH2g}--\ref{eqH2ns}), 
and the parton densities, we obtain the scheme--invariant structure functions provided that all 
definitions 
are carried out in the {\it same} scheme.   

In the following we will first present the results for the operator matrix elements obtained in the 
Larin scheme and then perform the finite renormalization to the {\sf M} scheme. We will first  
derive the 
unrenormalized operator matrix elements, after the mass renormalization has been carried out. 
\subsection{The \boldmath{$O(a_s)$} operator matrix element}

\vspace{1mm}\noindent
The polarized leading order massive operator matrix element is obtained from
diagram in Figure~2a of Ref.~\cite{BBK1}, using the Feynman rules \cite{SP_PS,Behring:2019tus}.
Diagram 2b vanishes. 
Due to the crossing relations of the forward Compton amplitude \cite{G2B} corresponding to
the present process the overall factor
\begin{eqnarray}
\frac{1}{2} \left[1 - (-1)^N\right],~~N~\epsilon~\mathbb{N}, N \geq 1,
\end{eqnarray}
is implied, which we drop in the operator matrix elements in the following. 
To obtain the results in $z$--space, the analytic continuation to complex values
of $N$ is performed from the odd integers. For the unrenormalized operator matrix element
one obtains to $O(\varepsilon^2)$\footnote{Note a misprint in Eq.~(51) of Ref.~\cite{BBK1} which 
needs to be  corrected. 
There the exponents of $m^2/\mu^2$ should be $\varepsilon/2$ in all places.}
\begin{eqnarray}
\label{eq:A1}
\Delta \hat{A}^{(1)}_{Qg} &=& \frac{1}{a_s} A_a^{Qg} = - S_\varepsilon T_F 
\left(\frac{m^2}{\mu^2}\right)^{\varepsilon/2}
\frac{1}{\varepsilon} \exp\left\{\sum_{l=2}^\infty \frac{\zeta_l}{l} \left(\frac{\varepsilon}{2}\right)^l\right\}
\frac{8(N-1)}{N(N+1)} \nonumber\\
&=& S_\varepsilon T_F \left(\frac{m^2}{\mu^2}\right)^{\varepsilon/2}
\left[-\frac{1}{\varepsilon} - \frac{\zeta_2}{8} \varepsilon - \frac{\zeta_3}{24} \varepsilon^2\right]
~ \frac{8(N-1)}{N(N+1)} + O(\varepsilon^3)
\nonumber\\ 
&=& S_\varepsilon  \left(\frac{m^2}{\mu^2}\right)^{\varepsilon/2}
\left[-\frac{1}{\varepsilon} \Delta \widehat{P}^{(0)}_{qg}(N) 
+ \Delta a_{Qg}^{(1)} 
+ \varepsilon \Delta \overline{a}_{Qg}^{(1)} 
+ \varepsilon^2 \Delta \overline{\overline{a}}_{Qg}^{(1)} 
\right] 
+ O(\varepsilon^3)~,
\end{eqnarray}
with
\begin{eqnarray}
S_\ep = \exp\left[\frac{\ep}{2}\left(\gamma_E - \ln(4\pi)\right)\right]
\end{eqnarray}
and $\gamma_E$ the Euler--Mascheroni constant.\footnote{At the end of 
the calculation $S_\ep$ is set to one, as part of the 
renormalization in the $\overline{\rm MS}$ scheme.} The matrix element 
(\ref{eq:A1}) is proportional to the leading order splitting 
function $\widehat{P}_{qg}^{(0)}$ and one has
\begin{eqnarray}
\label{eq11}
\Delta a_{qg}^{(1)}(N)            &=& 0, \\
\Delta \overline{a}_{qg}^{(1)} &=& - \frac{\zeta_2}{8} \Delta 
\widehat{P}_{qg}^{(0)}, \\
\Delta \overline{\overline{a}}_{qg}^{(1)} &=& - \frac{\zeta_3}{24} \Delta 
\widehat{P}_{qg}^{(0)}~.
\end{eqnarray}
The renormalized one--loop operator matrix element is given by
\begin{eqnarray}
\Delta A_{Qg}^{(1)} = \Delta \hat{A}_{Qg}^{(1)} + (Z^{-1})_{qg}^{1}
= - \frac{1}{2} \Delta \PS_{qg}^{(0)}
\ln\left(\frac{m^2}{\mu^2}\right)~,
\end{eqnarray}
with
\begin{eqnarray}
(Z^{-1})_{qg} = S_\varepsilon \frac{1}{\varepsilon} \Delta 
\hat{P}_{qg}^{(0)}~. 
\end{eqnarray}
Eq.~(\ref{eq1}) yields then the corresponding expression
of $H_{Qg}^{(1)}(z,Q^2)$ 
\begin{eqnarray}
\Delta H_{Qg, g_1}^{(1),\rm as}(z) = 
\left[ \frac{1}{2} \Delta P_{qg}^{(0)}(z)~ \ln\left(\frac{Q^2}{m^2}\right)
+ c_{g1}^{(1)}(z)\right]~.
\end{eqnarray}
At $O(a_s)$ there is no finite renormalization due to the treatment of $\gamma_5$.
In Mellin space one has
\begin{eqnarray}
\Delta H_{Qg, g_1}^{(1),\rm as}(N)  \propto (N-1)~, 
\end{eqnarray}
cf.~(\ref{eq11}, \ref{eq11a}), and the first moment 
vanishes. 
\subsection{The \boldmath{$O(a_s^2)$} operator matrix element 
\boldmath{$\Delta A_{Qg}^{(2)}$}}
We express the unrenormalized operator matrix element  $\Delta A_{Qg}^{(2)}$,
after mass renormalization, in terms of splitting functions and the contributions 
of $O(\varepsilon^0, \varepsilon)$, cf. \cite{Bierenbaum:2009mv}, by 
 \begin{eqnarray}
\label{eq30}
  \Delta \hat A_{Qg}^{(2)} &=&S_\ep^2 \Bigg( \frac{m^2}{\mu^2}\Bigg)^\ep
                       \Bigg[ \frac{1}{\ep^2} 
\Big\{\frac{1}{2} 
                              \Delta \hat{\gamma}^{(0)}_{qg}  ( \Delta \gamma^{(0)}_{qq} - 
                              \Delta \gamma^{(0)}_{gg} -2 \beta_0 -4 \beta_{0,Q}) \Big\}
\nonumber\\ &&
+ \frac{1}{2 \ep} \Big\{\Delta \hat{\gamma}_{qg}^{(1)} -  2 \delta m_1^{(-1)} \Delta 
\hat{\gamma}_{qg}^{(0)} 
                               \Big\} 
+ \Delta {a}^{(2)}_{Qg}
- \delta m_1^{(0)} \Delta \hat{\gamma}_{qg}^{(0)} - \frac{1}{2} \Delta \hat{\gamma}_{qg}^{(0)} 
\beta_{0,Q} \zeta_2 
\nonumber\\ &&
+ \ep \Biggl(
 \Delta \bar{a}^{(2)}_{Qg}
- \delta m_1^{(1)} \Delta \hat{\gamma}_{qg}^{(0)} - \frac{1}{12} \Delta \hat{\gamma}_{qg}^{(0)} 
\beta_{0,Q} \zeta_2 
\Biggr)\Biggr]
\label{AQg2f}
 \end{eqnarray}
or the corresponding expression in $z$ space. Here the expansion coefficients of the unrenormalized 
mass $\hat{m}$ are given by
\begin{eqnarray}
\hat{m} &=&  m\left[1 
+ \hat{a}_s \left(\frac{m^2}{\mu^2}\right)^{\ep/2} \delta m_1 \right] + O(\hat{a}_s^2)
\\
\delta m_1 &=& \frac{1}{\ep} \delta m_1^{(1)} 
+ \delta m_1^{(0)} 
+ \ep \delta m_1^{(1)} + O(\ep^2).
\end{eqnarray}

After performing charge-- and operator renormalization and subtracting the 
collinear singularities one obtains
\begin{eqnarray}
  \Delta \hat A_{Qg}^{(2)} &=&
  \Big\{\frac{1}{8} 
                              \Delta \hat{\gamma}^{(0)}_{qg}  ( \Delta \gamma^{(0)}_{qq} - 
                              \Delta \gamma^{(0)}_{gg} -2 \beta_0 -4 \beta_{0,Q}) \Big\} 
\ln^2\left(\frac{m^2}{\mu^2}\right) 
+ \frac{\Delta \hat{\gamma}^{(1)}_{qg}}{2} \ln\left(\frac{m^2}{\mu^2}\right)
\nonumber\\ &&
+ \Delta {a}^{(2)}_{Qg} + \frac{1}{8} \Delta \hat{\gamma}^{(0)}_{qg}
\left(\Delta \gamma_{gg} - \Delta \gamma_{qq} + 2 \beta_0\right).
\label{eq:102}
\end{eqnarray}
While the leading order anomalous dimensions are scheme--independent, at NLO 
$\Delta \hat{\gamma}^{(1)}_{qg}$ is different in the Larin and {\sf M} scheme.

In an earlier version of Ref.~\cite{ZN}, $\Delta \hat{\widetilde{P}}_{qg}^{(1)}(N)$
was used as anomalous dimension departing from the {\sf M} scheme. Therefore,
in Ref.~\cite{BUZA2} the finite renormalization \cite{Moch:2014sna,Behring:2019tus,Blumlein:2021ryt} as a 
corresponding one in 
$c_{g_1}^{(2)}(z)$, \cite{ZN}, was not used calculating $\Delta A_{Qg}^{(2)}$, and 
analogously, $\Delta A_{Qq}^{(2), \rm PS}$. We refer to the final version of \cite{ZN} for the
two--loop Wilson coefficients in the {\sf M} scheme and apply the finite renormalizations
to $\Delta A_{Qg}^{(2)}$ and $\Delta A_{Qq}^{(2), \rm PS}$.

Comparing to (\ref{eq:102}) the unrenormalized two--loop OME $A_{Qg}^{(2)}(N)$ is given in the 
Larin scheme by
\begin{eqnarray}
 \Delta \hat{A}^{(2)}_{Qg}(N) &=&S^2_{\ep}\Bigl(\frac{m^2}{\mu^2}\Bigr)^{\ep}\Biggl\{
\frac{1}{\ep^2}\Biggl[
                          T_FC_F\Biggl(
                                     32\frac{N-1}
                                            {N(N+1)}S_1
                                     -8\frac{(N-1)(3N^2+3N+2)}
                                            {N^2(N+1)^2}
                                \Biggr) \N
\\
&&                       +T_FC_A\Biggl(
                                     -32\frac{N-1}
                                             {N(N+1)}S_1
                                     +64\frac{N-1}
                                             {N^2(N+1)^2}
                                \Biggr)
                          \Biggr] \N\\
&&        +\frac{1}{\ep}\Biggl[
                          T_FC_F\Biggl(
                                     -8\frac{N-1}
                                            {N(N+1)}S_2
                                     +8\frac{N-1}
                                           {N(N+1)}S^2_1
                                     -16\frac{N-1}
                                             {N^2(N+1)}S_1 \N\\
&&                                   +4\frac{(N-1)(5N^4+10N^3+8N^2+7N+2)}
                                            {N^3(N+1)^3}
                                \Biggr)
                         +T_FC_A\Biggl(
                                     -16\frac{N-1}
                                             {N(N+1)}\beta' 
\nonumber\\ & &
-8\frac{N-1}
                                            {N(N+1)}S_2
                                     -8\frac{N-1}
                                            {N(N+1)}S^2_1
                                     +8\frac{N-1}
                                            {N(N+1)}\zeta_2
\N\\ & &                                     
+\frac{32}
                                           {N(N+1)^2}S_1
                                  +8\frac{N^5+N^4-4N^3+3N^2-7N-2}
                                            {N^3(N+1)^3}
                                \Biggr)
                          \Biggr] \N\\
&&                         + \Delta a_{Qg}^{(2)}
                         + \Delta \overline{a}_{Qg}^{(2)}~\varepsilon 
\Biggr\}~, \label{Aqg2}
\end{eqnarray}
with the constant term in $\varepsilon$
\begin{eqnarray}
\label{eq:DaQg2}
 \Delta a_{Qg}^{(2)}&=&
           C_F T_F\Biggl\{
                        4\frac{N-1}{3N(N+1)}\Bigl(-4S_3
                                                  +S^3_1
                                                  +3S_1 S_2
                                                  +6S_1 \zeta_2 
                                            \Bigr) \N\\
&&                       -4\frac{N^4+17N^3+43N^2+33N+2}
                                {N^2(N+1)^2(N+2)}S_2
                         -4\frac{3N^2+3N-2}
                                {N^2(N+1)(N+2)}S^2_1 \N\\
&&                       -2\frac{(N-1)(3N^2+3N+2)}
                                {N^2(N+1)^2}\zeta_2
                         -4\frac{N^3-2N^2-22N-36}
                                {N^2(N+1)(N+2)}S_1 \N\\ & &
                         -\frac{2P_1}
                                {N^4(N+1)^4(N+2)}
                  \Biggr\} \N\\
&&         +T_FC_A\Biggl\{
                         4\frac{N-1}{3N(N+1)}\Biggl[
                             12\Mvec\left[\frac{\Li_2(x)}{1+x}\right](N+1)
                                    +3\beta''
                                    -8S_3
\N\\ &&
                                    -S^3_1 
                                  -9S_1 S_2
                                    -12S_1 \beta'
                                    -12\beta(N+1)\zeta_2
                                    -3\zeta_3
                                            \Biggr]\N\\ & &
                         -16\frac{N-1}
                                 {N(N+1)^2}\beta' 
                      +4\frac{N^2+4N+5}
                                {N(N+1)^2(N+2)}S^2_1 \N\\ & & 
                         +4\frac{7N^3+24N^2+15N-16}
                                {N^2(N+1)^2(N+2)}S_2 
                         +8\frac{(N-1)(N+2)}
                                {N^2(N+1)^2}\zeta_2 \N\\ & & 
                      +4\frac{N^4+4N^3-N^2-10N+2}
                                {N(N+1)^3(N+2)}S_1
                         -\frac{4P_2}
                                {N^4(N+1)^4(N+2)}
                  \Biggr\}~. \label{aqg2} 
\end{eqnarray}
At two--loop order single harmonic sums have to be calculated at $N=0$. This is done
expressing them first in terms of $S_{\pm k}(N)$, for which then the analytic
continuation
\begin{eqnarray}
S_1(N)    &=& \psi(N+1) + \gamma_E, \\
S_k(N)    &=& \frac{(-1)^{k-1}}{(k-1)!} \psi^{(k-1)}(N+1) + \zeta_k,~~~k 
\geq 2, \\
S_{-1}(N) &=& (-1)^{N} \beta(N+1) - \ln(2), \\
S_{-k}(N) &=& \frac{(-1)^{k+1}}{(k-1)!} \beta^{(k-1)}(N+1) - \left(1 -  
\frac{1}{2^{k-1}}\right)\zeta_k,~~~k \geq 2
\end{eqnarray}
is used, which suggests the following definition
\begin{eqnarray}
S_{\pm k}(0)    := 0~.
\end{eqnarray}
Here, the function $\beta(N)$ is related to the $\psi$--function $\psi(z) = d \ln(\Gamma(z))/dz$ by
\begin{eqnarray}
\beta(N) &=& \frac{1}{2} \left[\psi\left(\frac{N+1}{2}\right) - 
\psi\left(\frac{N}{2}\right)\right] 
\end{eqnarray}
and we use the short-hand notation 
\begin{eqnarray}
\beta^{(k)} \equiv \beta^{(k)}(N+1),~~~~k \in \mathbb{N},~~k \geq 0,
\end{eqnarray}
above and in the following.
The polynomials are
\begin{eqnarray}
P_1 &=& 12N^8+52N^7+60N^6-25N^4-2N^3+3N^2+8N+4~, \\
P_2 &=& 2N^8+10N^7+22N^6+36N^5+29N^4+4N^3+33N^2+12N+4~. 
\end{eqnarray}
The corresponding expression in Eq.~(A.2) of Ref.~\cite{BUZA2} differs by a global minus sign 
compared to (\ref{eq:DaQg2}),
which has to be corrected. The linear term in $\varepsilon$ reads
\begin{eqnarray}
\Delta \overline{a}_{Qg}^{(2)}&=&
       T_F C_F   \Biggl\{
                            \frac{N-1}
                                 {N(N+1)}
                             \Bigl(
                               16S_{2,1,1}
                              -8S_{3,1}
                              -8S_{2,1}S_1
                              +3S_4
                              -\frac{4}{3}S_3S_1
                              -\frac{1}{2}S^2_2
                              -\frac{1}{6}S^4_1
                              -\frac{8}{3}S_1\zeta_3 
\N\\
&& 
                           -S_2S^2_1
                              +2S_2\zeta_2
                              -2S^2_1\zeta_2
                             \Bigr)
                           -8\frac{S_{2,1}}{N^2}
                           +\frac{3N^2+3N-2}
                                 {N^2(N+1)(N+2)}
                             \Bigl(
                              2S_2S_1
                              +\frac{2}{3}S^3_1
                             \Bigl) \N\\
&&                         +2\frac{3N^4+48N^3+123N^2+98N+8}
                                 {3N^2(N+1)^2(2+N)}S_3
                           +\frac{4(N-1)}
                                  {N^2(N+1)}S_1\zeta_2
\N\\
&&                         
                           +\frac{2}{3}
                            \frac{(N-1)(3N^2+3N+2)}
                                 {N^2(N+1)^2}\zeta_3
+\frac{P_3}
                                 {N^3(N+1)^3(N+2)} S_2
                           +\frac{N^3-6N^2-22N-36}
                                 {N^2(N+1)(N+2)}S^2_1
\N\\
&&
                           +\frac{P_4\zeta_2}
                                 {N^3(N+1)^3} 
                         -2\frac{2N^4-4N^3-3N^2+20N+12}
                                  {N^2(N+1)^2(N+2)}S_1
                           +\frac{P_5}
                                 {N^5(N+1)^5(N+2)}
                   \Biggr\} \N
\\
%
%
&&   + T_F C_A   \Biggl\{
                            \frac{N-1}
                                 {N(N+1)}
                             \Bigl(
                               16S_{-2,1,1}
                              -4S_{2,1,1}
                              -8S_{-3,1}
                              -8S_{-2,2}
                               -4S_{3,1}
                              +\frac{2}{3}\beta'''
                              -16S_{-2,1} S_1 \N\\
&&                            -4\beta'' S_1
                              +8\beta' S_2
                              +8 \beta' S^2_1
                              +9S_4 
                              +\frac{40}{3} S_3 S_1
                              +\frac{1}{2}S^2_2
                              +5 S_2 S^2_1
                              +\frac{1}{6} S^4_1
                              +4\zeta_2 \beta'
                              -2\zeta_2 S_2 \N\\
&&                            -2\zeta_2  S^2_1
                              -\frac{10}{3} S_1\zeta_3
                              -\frac{17}{5}\zeta_2^2
                             \Bigr)
                           -\frac{N-1}
                                 {N(N+1)^2}
                             \Bigl(
                              16 S_{-2,1}
                              +4\beta''
                              -16 \beta' S_1
                             \Bigr) \N\\
&&                         -\frac{16}{3}\frac{N^3+7N^2+8N-6}
                                             {N^2(N+1)^2(N+2)}S_3
                           +\frac{2(3N^2-13)}
                                  {N(N+1)^2(N+2)} S_1 S_2
                           -\frac{2(N^2+4N+5)}
                                 {3N(N+1)^2(N+2)}S^3_1 \N\\
&&                         -\frac{8}
                                  {(N+1)^2} \zeta_2 S_1
                           -\frac{2}{3} 
                            \frac{(N-1)(9N+8)}
                                 {N^2(N+1)^2}\zeta_3
                           -\frac{8(N^2+3)}
                                  {N(N+1)^3}\beta'
                           -\frac{P_6}
                                 {N^3(N+1)^3(N+2)} S_2 \N\\
&&                         -\frac{N^4+2N^3-5N^2-12N+2}
                                 {N(N+1)^3(N+2)}S^2_1
                           -\frac{2P_7}
                                  {N^3(N+1)^3} \zeta_2
                           +\frac{2P_8}
                                  {N(N+1)^4(N+2)} S_1 \N\\ 
&&                         -\frac{2P_9}
                                  {N^5(N+1)^5(N+2)}
                   \Biggr\}~, 
\end{eqnarray}
and
\begin{eqnarray}
P_3 &=&3N^6+30N^5+107N^4+124N^3+48N^2+20N+8~, \\
P_4 &=&(N-1)(N^4+2N^3-2N^2-7N-2)~, \\
P_5 &=&8N^{10}+24N^9-11N^8-160N^7-311N^6-275N^5
       -111N^4-7N^3
\nonumber\\ &&
+11N^2+12N+4~,
\\
P_6 &=&N^6+18N^5+63N^4+84N^3+30N^2-64N-16~, \\
P_7 &=&N^5-N^4-4N^3-3N^2-7N-2~, \\
P_8 &=&2N^5+10N^4+29N^3+64N^2+67N+8~, \\
P_9 &=&4N^{10}+22N^9+45N^8+36N^7-11N^6
      -15N^5+25N^4-41N^3
\nonumber\\ &&
-21N^2-16N-4~. 
\end{eqnarray}
It is useful for the analytic continuation of the respective expressions to the complex plane to express 
harmonic sums containing also negative indices by their associated Mellin transforms referring to 
Ref.~\cite{Blumlein:1998if}, see also \cite{FL}. This allows to get rid of factors of $(-1)^N$, which would 
occur otherwise.

The calculation has been performed using {\tt FORM}~\cite{FORM}. Further 
mathematical simplifications were done with the help of {\tt MAPLE}.
The contributions due to the individual diagrams are given in Appendix~\ref{sec:A}. In the 
calculation,
extensive use was made of the representation of the Feynman--parameter integrals in terms
of generalized hypergeometric functions \cite{GHF}. Examples are given in Appendix~\ref{sec:B}. 
The infinite nested harmonic sums, partly weighted with Beta-functions and binomials, 
which occur in the present calculation, are similar to those in Ref.~\cite{BBK1}. 

We use
\begin{eqnarray}
      \Mvec\left[\frac{\Li_2(x)}{1+x}\right](N+1)  - \zeta_2 \beta(N+1)
      = (-1)^{N+1} \left[S_{-2,1}  + \frac{5}{8} \zeta_3\right]
      ~\label{hhhh}
\end{eqnarray}
to provide a proper representation for the analytic continuation.
The structural relations between the finite harmonic sums \cite{STRUCT} allow 
to express $\Delta a_{Qg}^{(2)}$ in terms of just two basic Mellin transforms, which 
are 
meromorphic functions in the complex $N$--plane with poles at the non--positive 
integers. They are related to the harmonic sums $S_1$ and $S_{-2,1}$. 
In the present calculation we refrain from using IBP reduction for 
the individual diagrams. Due to this and the consequent use of Mellin--space 
representations in terms of polynomial--weighted harmonic sums we obtain 
very compact results even 
for the individual diagrams. As in \cite{BBK1}, only one more harmonic sum, 
$S_{2,1}$, occurs, which cancels in the final result.  None of the harmonic 
sums containing the index $\{-1\}$ contributes, which has been observed in the 
case of all known space 
and time--like single scale processes up to three--loop orders \cite{MATH,STRUCT,MVV2,
Behring:2019tus,Blumlein:2021ryt,Moch:2004pa,Blumlein:2021enk,HQ3N,Ablinger:2014vwa,Ablinger:2017tan},
which can be written in terms of harmonic sums only. 
All other terms can be expressed by half--integer relations and 
derivatives w.r.t. $N$, cf.~\cite{STRUCT}.
\subsection{The \boldmath{$O(a_s^2)$} operator matrix element 
\boldmath{$A_{qq,Q}^{{\rm NS},(2)}$}}

\label{sec:44}
The diagrams for the non--singlet operator matrix element $A_{qq,Q}^{{\rm NS},(2)}$ are shown 
in Figure~5  of Ref.~\cite{BBK1}. Due to the Ward--Takahashi identity, it has to be the same as in 
the 
unpolarized case, i.e. one may treat $\gamma_5$ as anticommuting in the present case to obtain 
the OME in the {\sf M} scheme, using the quarkonic projector given in \cite{BUZA1}. 
The asymptotic Wilson coefficient is, however, different from the unpolarized one, cf.~Appendix~\ref{sec:C}.

The OME reads, \cite{BUZA2,BBK1,BBKS8a},
\begin{eqnarray}
\Delta \hat{A}^{{\rm NS},(2)}_{qq,Q}\Biggl(\frac{m^2}{\mu^2},\ep\Biggr)
                         &=&S_\ep^2 \Bigg( \frac{m^2}{\mu^2}
                            \Bigg)^\ep \Bigg[ \frac{1}{\ep^2}
                            {\beta}_{0,Q} \Delta \gamma^{(0)}_{qq} 
                            + \frac{1}{2\ep}
                            \Delta \hat{\gamma}^{{\rm NS},(1)}_{qq,Q}
                           +\Delta a^{{\rm NS},(2)}_{qq,Q}
                       +\Delta \overline{a}^{{\rm NS},(2)}_{qq,Q}~\ep
\Big] \,. \label{AqqQ2f}
\nonumber\\
\end{eqnarray}
The renormalized OME is given by
\begin{eqnarray}
\label{eq:NSOME}
\Delta {A}^{{\rm NS},(2)}_{qq,Q}\Biggl(\frac{m^2}{\mu^2}\Biggr)
&=& \frac{1}{4} \beta_{0,Q} \Delta \gamma_{qq}^{(0)} \ln^2\left(\frac{m^2}{\mu^2}\right)
+\frac{1}{2} \Delta \hat{\gamma}_{qq}^{(1),\rm NS} \ln\left(\frac{m^2}{\mu^2}\right) + \Delta 
a_{qq}^{(2),\rm NS} 
- \frac{1}{4} 
\beta_{0,Q} \Delta \gamma_{qq}^{(0)} \zeta_2. 
\nonumber\\
\end{eqnarray}
The constant term is given by
     \begin{eqnarray}
      \Delta a_{qq,Q}^{{\rm NS},(2)}(N)&=& T_F C_F\Biggl\{
                        -\frac{8}{3}S_3
                        -\frac{8}{3}\zeta_2S_1
                        +\frac{40}{9}S_2
                        +2\frac{3N^2+3N+2}
                               {3N(N+1)}\zeta_2
                        -\frac{224}{27}S_1 \N\\
                     && +\frac{219N^6+657N^5+1193N^4+763N^3-40N^2-48N+72}
                              {54N^3(N+1)^3}\Biggr\} \label{aNS2qqQ}.
     \end{eqnarray}
The corresponding expression in \cite{BUZA2} is defined without the color factor $C_F T_F$.
It agrees to the related quantity in the unpolarized case \cite{BUZA1,BBK1}.
The linear term in $\ep$ is given by
\begin{eqnarray}
 \Delta \overline{a}_{qq,Q}^{{\rm NS},(2)}&=&
                      T_FC_F\Biggl\{
                        \frac{4}{3}S_4
                        +\frac{4}{3}S_2\zeta_2
                        -\frac{8}{9}S_1\zeta_3
                        -\frac{20}{9}S_3
                        -\frac{20}{9}S_1\zeta_2
                        +2\frac{3N^2+3N+2}
                                 {9N(N+1)}\zeta_3
                        +\frac{112}{27}S_2\N\\
&&                      +\frac{3N^4+6N^3+47N^2+20N-12}
                              {18N^2(N+1)^2}\zeta_2
                        -\frac{656}{81}S_1
                        +\frac{P_{10}}{648N^4(N+1)^4}
                               \Biggr\}
                          \label{aNS2qqQep}~, 
\end{eqnarray}
with
\begin{eqnarray}
P_{10}&=&1551N^8+6204N^7+15338N^6+17868N^5+8319N^4
+944N^3+528N^2-144N-432~,
\N\\
\end{eqnarray}
again the same as in the unpolarized case \cite{BBKS8a}.
The OME (\ref{eq:NSOME}) in the Larin scheme is given in \cite{LOGPOL}. The part of the 
asymptotic heavy flavor Wilson coefficient corresponding to final heavy flavor states
is, however, the same in the Larin and the {\sf M} scheme, while that of the massless quark
final state has a finite renormalization, cf.~Eq.~(323) in  Ref.~\cite{LOGPOL}. 
\subsection{The \boldmath{$O(a_s^2)$} operator matrix element 
\boldmath{$A_{Qq}^{{\rm PS},(2)}$}}
\label{sec:44a}

\vspace{1mm}\noindent
The operator matrix element $A_{Qq}^{{\rm PS},(2)}$ is obtained from diagrams 
Figure~4 of Ref.~\cite{BBK1}. Here the contribution due to diagram~$b$ vanishes. 
The unrenormalized OME is given by
\begin{eqnarray}
\Delta \hat{A}_{Qq}^{(2),\rm PS, L} = S_\ep^2 \left(\frac{m^2}{\mu^2}\right)^\ep \left[
-\frac{1}{2 \ep^2} \Delta \hat{\gamma}_{qg}^{(0)} \Delta {\gamma}_{gq}^{(0)}
+ \frac{1}{2} \Delta \hat{\gamma}_{qq}^{(1),\rm PS} 
+ \Delta a_{Qq}^{(2),\rm PS, L}
+ \Delta \overline{a}_{Qq}^{(2),\rm PS, L} \ep \right] + O(\ep^2) 
\nonumber\\
\end{eqnarray}
and the renormalized OME reads
\begin{eqnarray}
\Delta {A}_{Qq}^{(2),\rm PS, L} = -\frac{\Delta \hat{\gamma}_{qg}^{(0)} \Delta \gamma_{gq}^{(0)}}{8} 
\ln^2\left(\frac{m^2}{\mu^2}\right) + \frac{1}{2} \Delta \hat{\gamma}_{qq}^{(1),\rm PS} 
\ln\left(\frac{m^2}{\mu^2}\right) 
+ \Delta a_{Qq}^{(2),\rm PS, L} + \frac{\Delta \hat{\gamma}_{qg}^{(0)} 
\Delta {\gamma}_{gq}^{(0)}}{8} \zeta_2.
\nonumber\\
\end{eqnarray}
The calculation is first performed in the Larin scheme, using the projector (11), Ref.~\cite{Behring:2019tus},
for diagrams with external massless quark lines in the polarized case.\footnote{Before \cite{Behring:2019tus} 
there was still some ambiguity in calculating the polarized pure singlet OME, cf.~Section~8.2.3 
of \cite{KLEINphd}.}

The next-to-leading order pure singlet anomalous dimension $\Delta \hat{\gamma}_{qq}^{(1),\rm PS}$ is 
the same in the Larin and in the {\sf M} scheme 
\cite{BUZA2,Moch:2014sna,Behring:2019tus,Blumlein:2021ryt}, as well as the asymptotic pure singlet 
Wilson coefficient. Because of
\begin{eqnarray}
\Delta H^{\rm PS,(2)}\left(N, \frac{Q^2}{m^2}, \frac{\mu^2}{m^2}\right) 
= 
\Delta A_{Qq}^{(2),\rm PS}\left(N,\frac{\mu^2}{m^2}\right) 
+ \Delta \tilde{C}_q^{(2),\rm PS}
\end{eqnarray}
and 
\begin{eqnarray}
\Delta \tilde{C}_q^{(2),\rm PS, M} = 
\Delta \tilde{C}_q^{(2),\rm PS, L} - z_{qq}^{(2), \rm PS}
\end{eqnarray}
one has
\begin{eqnarray}
\Delta A_{Qq}^{(2),\rm PS, M}\left(N,\frac{\mu^2}{m^2}\right) 
= \Delta A_{Qq}^{(2),\rm PS, L}\left(N,\frac{\mu^2}{m^2}\right) 
+ z_{qq}^{(2), \rm PS}.
\end{eqnarray}

One obtains the constant term $\Delta a^{(2), {\rm PS, L}}_{Qq}(N)$\footnote{Note a typographical error in
\cite{Blumlein:2019zux}, Eq.~(81). There the $\zeta_2$ term shall read 
$-[20(1-z)+8(1+z)\HA_0]\zeta_2$, switching one sign.} 
\begin{eqnarray}
\label{Da2PS}
\Delta a^{{(2), \rm PS, L} }_{Qq}(N) &=& - 4 T_F C_F \frac{N+2}{N^2 (N+1)^2} \Biggl[
(N-1) \left[2 S_2 + \zeta_2\right] - \frac{4 N^3 - 4 N^2 - 3N -1 }{N^2 (N+1)^2}
\Biggr].
\nonumber\\
\end{eqnarray}
The corresponding quantity in Eq. (A.4) of  Ref.~\cite{BUZA2} agrees with (\ref{Da2PS}) defined 
without the color factor $C_F T_F$ there. The term $\Delta \overline{a}^{(2), {\rm PS, L}}_{Qq}(N)$
is given by 
\begin{eqnarray}
\Delta \overline{a}_{Qq}^{(2),\rm PS, L} &=& 8 C_F T_F (N+2)
\Biggl[             \frac{N^3+2N+1}{4 N^3(N+1)^3}\Bigl(2S_2+\zeta_2\Bigr)
                   -\frac{N-1}{6 N^2(N+1)^2}\Bigl(3S_3+\zeta_3\Bigr)
\nonumber\\ &&            
       +\frac{N^5-7N^4+6N^3+7N^2+4N+1}{4 N^5(N+1)^5}
\Biggr].
\end{eqnarray}
\subsection{Discussion}

\vspace{1mm}\noindent
Our results for the massive operator matrix elements agree with those found in Ref.~\cite{BUZA2}.
There the calculation was performed in $z$ space and the integration-by-parts method was 
applied.
In Table \ref{table:results1} all functions contributing 
to (\ref{Aqg2}) in $z$ space are listed. These are 24 functions.
     \begin{table}[h]
\[
\renewcommand{\arraystretch}{1.0}
\begin{array}{ccccc}
 \delta(1-z)                   &
 \ln(z)              &
 \ln^2(z)            &
 \ln^3(z)            &
 \ln(1-z)            \\ 
  & & & & \\
 \ln^2(1-z)          &
 \ln^3(1-z)          &
 \ln(z)\ln(1-z)      &
 \ln(z)\ln^2(1-z)    &
 \ln^2(z)\ln(1-z)    \\ 
  & & & & \\
 \ln(1+z)            & 
 \ln(z) \ln(1+z)     & 
 \ln^2(z) \ln(1+z)   &  
 \ln(z) \ln^2(1+z)   &  
 \Li_2(1-z)          \\
  & & & & \\
 \Li_2(-z)           &
 \ln(z)\Li_2(1-z)    &
 \ln(1-z) \Li_2(1-z) &
 \Li_3(1-z)          & 
 S_{1,2}(1-z)        \\ 
& & & & \\
 \Li_3(-z)           &
 S_{1,2}(-z)         &
 \ln(z) \Li_2(-z)    &
 \ln(1+z) \Li_2(-z)  &
                     \\
\end{array}
\]
\renewcommand{\arraystretch}{1.0}
      \caption{\small \sf Functions contributing to the results in $z$--space}
      \label{table:results1}
     \end{table}
     \noindent
\noindent
The $O(\varepsilon^0)$ term depends on six harmonic sums. Since the single 
harmonic sums form one equivalence class, cf.~\cite{STRUCT}, the result
can be expressed by the two sums $S_1, S_{-2,1}$ only, by applying 
structural relations. Compared to the 24 functions needed in \cite{BUZA2}, we
reached a more compact representation.
The $O(\varepsilon)$ term depends on
the six sums  $S_1, S_{\pm 2,1}, S_{-3,1}, S_{\pm 2,1,1}$. The 
other sums can be expressed by structural relations. The $O(\varepsilon^0)$
terms have thus the same complexity as the two--loop anomalous dimensions,
while that of the $O(\varepsilon^0)$ terms corresponds to the level
observed for two--loop Wilson coefficients and other hard scattering processes
which depend on a single scale, cf.~\cite{MATH}.

Let us consider the first moment of the polarized heavy flavor operator matrix 
elements 
and Wilson coefficients in the region $Q^2 \gg m^2$. The splitting functions obey 
\begin{eqnarray}
\label{eq33}
\Delta P_{qq}^{(0)}(N=1) &=& 0,                     \\
\Delta P_{gg}^{(0)}(N=1) &=& 2 \beta_0,             \\  
\Delta P_{qg}^{(0)}(N=1) &=& 0,                     \\  
\Delta P_{gq}^{(0)}(N=1) &=& 6 C_F,                 \\
\Delta P_{qg}^{(1)}(N=1) &=& 0,                     \\  
\Delta P_{qq}^{{\rm PS},(1)}(N=1) &=& -24 T_F C_F,  \\  
\label{eq34}
\Delta P_{qq}^{{\rm NS},(1)}(N=1) &=& 0~.  
\end{eqnarray}

In Table~2 we illustrate the complexity of our  results in 
Mellin--space quoting the harmonic sums, which contribute to the individual Feynman diagrams, 
cf.~Appendix~\ref{sec:A}.
{\begin{table}[H]
\footnotesize
 \label{table:results2}
 \begin{center}
 \renewcommand{\arraystretch}{1.1}
 \begin{tabular}{||l|c|c|c|c|c|c|c|c|c|c|c|c|c|c||}
 \hline \hline
Diagram  & $S_1$ & $S_2$ & $S_3$ & $S_4$
& $S_{-2}$ & $S_{-3}$ & $S_{-4}$ & $S_{2,1}$ & $S_{-2,1}$
& $S_{3,1}$ & $S_{-3,1}$ & $S_{-2,2}$ &$S_{2,1,1}$ & $S_{-2,1,1}$
 \\
\hline \hline
A  &  &++&$-+$&    &  &  &    &    &   &    &    &    &    &     \\            
B  &++&++& ++ &$-+$&  &  &    &++  &   &$-+$&    &    &$-+$&     \\            
C  &  &++&$-+$&    &  &  &    &    &   &    &    &    &    &     \\            
D  &++&++&$-+$&    &  &  &    &    &   &    &    &    &    &     \\            
E  &++&++&$-+$&    &  &  &    &$-+$&   &    &    &    &    &     \\            
F  &++&++& ++ &$-+$&  &  &    & ++ &   &    &    &    &$-+$&     \\            
J  &  &++&$-+$&    &  &  &    &    &   &    &    &    &    &     \\            
L  &++&++& ++ &$-+$&  &  &    & ++ &   &$-+$&    &    &$-+$&     \\            
M  &  &++&$-+$&    &  &  &    &    &   &    &    &    &    &     \\            
N  &++&++& ++ &$-+$&++&++&$-+$& ++ &++ &$-+$&$-+$&$-+$&$-+$&$-+$ \\            
\hline
\hline
PS &  &++&$-+$& & & & & & & & & & & \\            
\hline
NS &++&++&++&$-+$& & & & & & & & & & \\            
\hline
\hline
$\Sigma$  &++&++&++&$-+$&++&++&$-+$&$-+$&++&$-+$&$-+$&$-+$&$-+$&$-+$ \\            
\hline
\hline
      \end{tabular}
      \renewcommand{\arraystretch}{1.0}
      \end{center}
 \caption{\small \sf Complexity of the results in Mellin space. The first $+$ 
denotes the contribution of the sum to in $O(\varepsilon^0)$, the second $+$ 
for the $O(\varepsilon)$ term and $-$ its absence.} 
     \end{table}
     \noindent
}

Furthermore one has
\begin{eqnarray}
\Delta \overline{a}_{Qg}^{(1)}(N=1)                 &=& 0,                   \\
\Delta \overline{\overline{a}}_{Qg}^{(1)}(N=1)      &=& 0,                   \\
\Delta {a}_{Qg}^{(2)}(N=1)                          &=& 0,                   \\
\Delta \overline{a}_{Qg}^{(2)}(N=1)                 &=& 0,                   \\
\Delta {a}_{Qq}^{{\rm PS},(2),\rm L}(N=1)           &=& 3 C_F T_F,                  \\
\Delta \overline{a}_{Qq}^{{\rm PS},(2) \rm L}(N=1)  &=& \frac{3}{4} C_F T_F (11 + 4 \zeta_2),                 
\\
\label{axv1}
\Delta {a}_{qq,Q}^{{\rm NS},(2)}(N=1)               &=& 0,                   \\
\label{axv2}
\Delta {\overline{a}}_{qq,Q}^{{\rm NS},(2)}(N=1)    &=& 0.                   
\end{eqnarray}
Relations (\ref{eq33}, \ref{eq34},\ref{axv1},\ref{axv2}) hold due to conservation of the axial 
vector current.

Since 
\begin{eqnarray}
\Delta C_{g_1}^{g,(2)}(Q^2/\mu^2,N=1)   &=& 0                   
\end{eqnarray}
holds, cf.~\cite{ZN}, one also obtains
\begin{eqnarray}
\Delta H_{g_1}^{Qg,(2)}(Q^2/m^2,N=1)   &=& 0
\end{eqnarray}
and the first moment of the gluonic contributions to the structure function 
$g_1(x,Q^2)$ both for the heavy and light flavor contributions vanishes, 
if calculated in the collinear parton model.
A related sum--rule for the  gluonic contribution to the photon structure 
function holds~\cite{PHO_G1}.

The first moment of the pure singlet contribution $H_{g_1,q}^{{\rm PS}, (2)}(Q^2/m^2,x)$ 
is given by 
\begin{eqnarray}
H_{g_1,q}^{{\rm PS}, (2)}(Q^2/m^2,N=1)   &=& -12 \ln\left(\frac{Q^2}{m^2}\right) + \frac{20}{3} + 16 \zeta_3.
\end{eqnarray}

We finally consider the small $x$ behaviour of the corrections calculated 
in the present paper.\footnote{For the unpolarized case see 
\cite{BUZA1,FL}.} The leading order small $x$ resummation for the 
polarized flavor non--singlet and singlet contributions were studied 
in \cite{Kirschner:1983di,Blumlein:1995jp,Blumlein:1996hb,Bartels:1995iu,Bartels:1996wc,SX2}. 
Unlike the unpolarized case where the most singular
contributions have poles at $N = 1$ in the perturbative expansion, the
leading poles are situated at $N = 0$ in the polarized case.
From a theoretical point of view, it is interesting to see to which series of a 
formal small $x$ expansion the different coefficients belong, in order to compare with ab initio 
calculations of these
terms, even though the resummation of these terms alone does not describe the small $x$
behaviour of the polarized structure functions, since sub--leading terms turn out 
to be as important,~cf.~\cite{Blumlein:1995jp,Blumlein:1996hb}.

In the polarized case  leading small $x$ terms are of the form
\begin{eqnarray}
a_s \left[a_s \ln^2(x)\right]^k~,
\end{eqnarray}
for the splitting functions in the {\sf M} scheme. Only for this case 
the all order resummation in $a_s$ has been derived so far. As has been pointed out in 
\cite{SX2} the small $x$ behaviour of the massless Wilson coefficient is found 
to be less singular by one power in $\ln(x)$, i.e. the $O(a_s^k)$ coefficient 
functions behave at most like  
\begin{eqnarray}
c_k(x) \propto a_s^k \ln^{2k-1}(x)~.
\end{eqnarray}
At leading order in $a_s$ the small $x$ asymptotic behaviour of the
polarized heavy flavor Wilson coefficient is given by
\begin{eqnarray}
\Delta H_{g_1}^{Qg,(1), x \rightarrow 0}(Q^2/m^2, N)   &=&         
a_s(Q^2) T_F \left[\frac{1}{N^2} + 
O\left(
\frac{1}{N}\right) 
\right], 
\\ 
\Delta H_{g_1}^{Qg,(1), x \rightarrow 0}(Q^2/m^2, x)   &\propto&   a_s(Q^2) T_F \ln(z)~. 
\end{eqnarray}
The leading singularity results from the massless one--loop Wilson coefficient, while
the massive operator matrix element behaves like
\begin{eqnarray}
\Delta A_{g_1}^{Qg,(1), x \rightarrow 0}(Q^2/m^2, N)   &=&         T_F \left[- \frac{1}{N} + 
O\left(1\right) \right] \ln\left(\frac{Q^2}{m^2}\right),\\ 
\Delta A_{g_1}^{Qg,(1), x \rightarrow 0}(Q^2/m^2, x)   &\propto&   - T_F 
\ln\left(\frac{Q^2}{m^2}\right)~.  
\end{eqnarray}
The logarithmic term $\ln(Q^2/m^2)$ thus belongs to the less singular series at small $x$.

As it is the case  at $O(a_s)$, the most singular terms at small $x$ for the asymptotic
heavy flavor Wilson coefficient
$\Delta H_{Qi}^{\rm S,(2)}(x,Q^2)$ at $O(a_s^2)$ are due to the constant terms 
in $Q^2$. Here the constant term in the massive operator matrix element, which is vanishing  
at $O(a_s)$, contains a term of same singularity as the massless Wilson coefficients \cite{ZN},
\begin{eqnarray}
\label{eq35}
\Delta H^{{\rm S, (2)}, N \rightarrow 0}_{g_1,g}\left(\frac{Q^2}{m^2}, N \right) &=& 
a_s^2(Q^2) \left\{-8 \frac{1}{N^4} 
T_F (4 C_A + 3 C_F) +O\left(\frac{1}{N^3}\right)\right\}, 
\nonumber\\
\Delta H^{{\rm S, (2)}, x \rightarrow 0}_{g_1,g}\left(\frac{Q^2}{m^2}, x \right) &\propto& 
a_s^2(Q^2)\left\{ \frac{4}{3} 
T_F (4 C_A + 3 C_F)\ln^3(x)\right\}, \\
\label{eq36}
\Delta H^{{\rm PS, (2)}, x \rightarrow 0}_{g_1,q}\left(\frac{Q^2}{m^2}, N \right) &=& a_s^2(Q^2) 
\left\{- \frac{32}{N^4} T_F 
C_F +  O\left(\frac{1}{N^3}\right) \right\},\nonumber\\
\Delta H^{{\rm PS, (2)}, x \rightarrow 0}_{g_1,q}\left(\frac{Q^2}{m^2}, x \right) & \propto & 
a_s^2(Q^2) \left\{\frac{16}{3} T_F C_F \ln^3(x)\right\}~.
\end{eqnarray}
The two--loop Wilson coefficients are by one power in $\ln(x)$ less singular at small $x$
in the non--singlet case if compared to the singlet case, 
\begin{eqnarray}
\label{eq37}
\Delta L^{{\rm NS, (2)}, x \rightarrow 0}_{g_1,q}\left(\frac{Q^2}{m^2}, N \right) &=& a_s^2(Q^2) \Biggl\{
\frac{4}{3} C_F T_F
\frac{8}{N^3} \Biggr\}
+ O\left(\frac{1}{N^2}\right),
\nonumber\\
\Delta L^{{\rm NS, (2)}, x \rightarrow 0}_{g_1,q}\left(\frac{Q^2}{m^2}, x \right) & \propto & -4 a_s^2(Q^2) 
C_F T_F \ln^2(x)~.
\end{eqnarray}
Furthermore, one has
\begin{eqnarray}
\Delta L^{(2), x \rightarrow 0}_{g_1,g}\left(\frac{Q^2}{m^2}, N \right) &=& a_s^2(Q^2) \Biggl\{\frac{16}{3} 
T_F^2 N_F 
\frac{1}{N^2}\Biggr\} 
+ O\left(\frac{1}{N}\right),
\nonumber\\
\Delta L^{(2), x \rightarrow 0}_{g_1,g}\left(\frac{Q^2}{m^2}, x \right) & \propto & - 
a_s^2(Q^2) \frac{16}{3} T_F^2 
N_F \ln(x)~.  
\end{eqnarray}
\section{Numerical results}
\label{sec:6}

\vspace{1mm}\noindent
In the following we illustrate the heavy flavor contributions to the twist--2 contributions of the 
polarized structure functions $g_{1,(2)}(x,Q^2)$ numerically.\footnote{To accelerate the 
numerical calculation we use splines over fine grids in very few cases.} 
\begin{figure}[H]
\centering
\includegraphics[width=0.8 \linewidth]{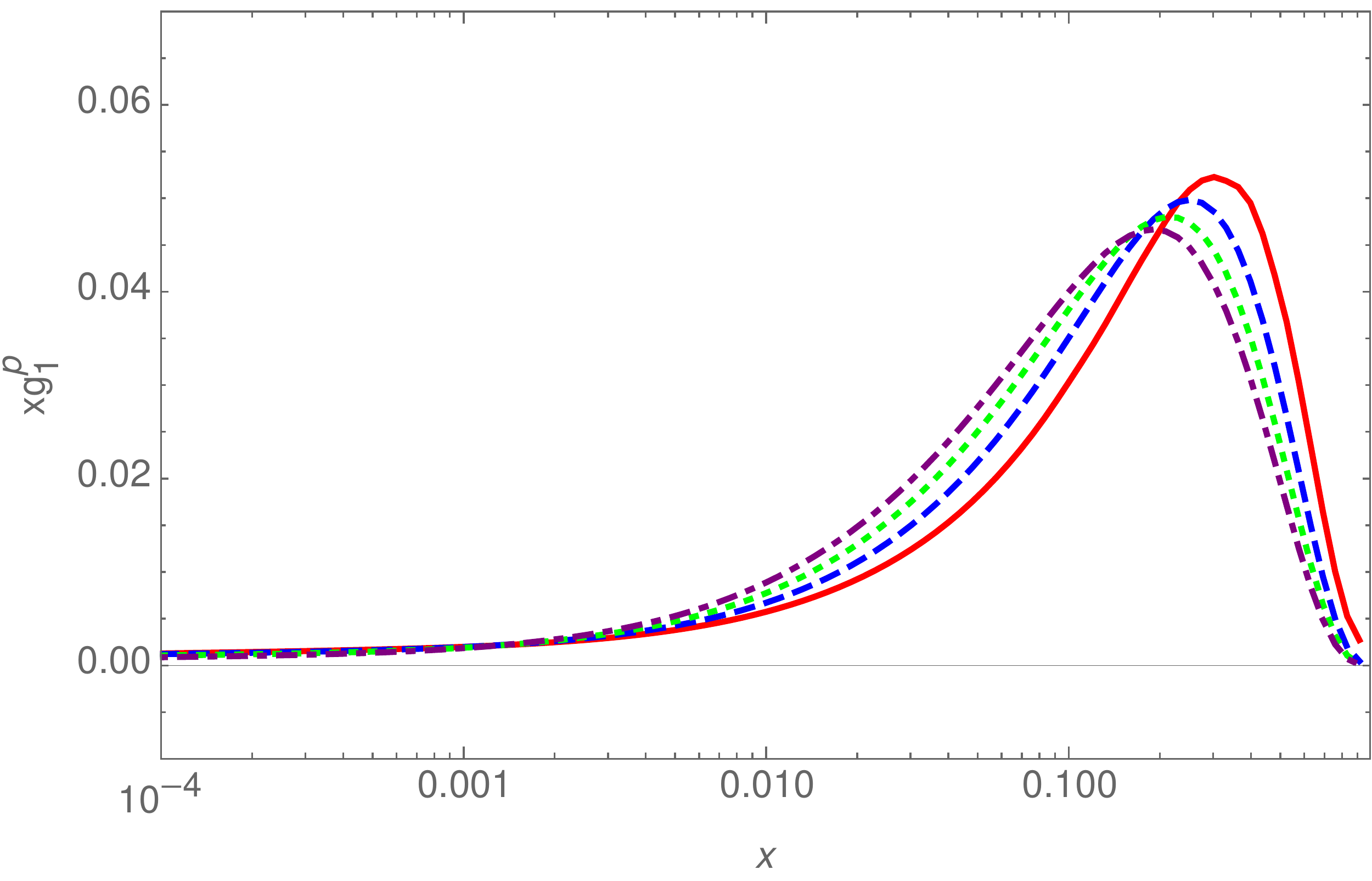}
\caption[]{\small \sf The NLO massless structure function $xg_1(x,Q^2)$ as a 
function of $x$ for $Q^2 = 10~\GeV^2$ (full line); $100~\GeV^2$ 
(dashed line); $1000~\GeV^2$ (dotted line), and  $10000~\GeV^2$ (dash--dotted line), using
the parton distribution functions of \cite{Blumlein:2010rn}. 
}
\label{fig:FIG1} 
\end{figure}

\begin{figure}[H]
\centering
\includegraphics[width=0.8 \linewidth]{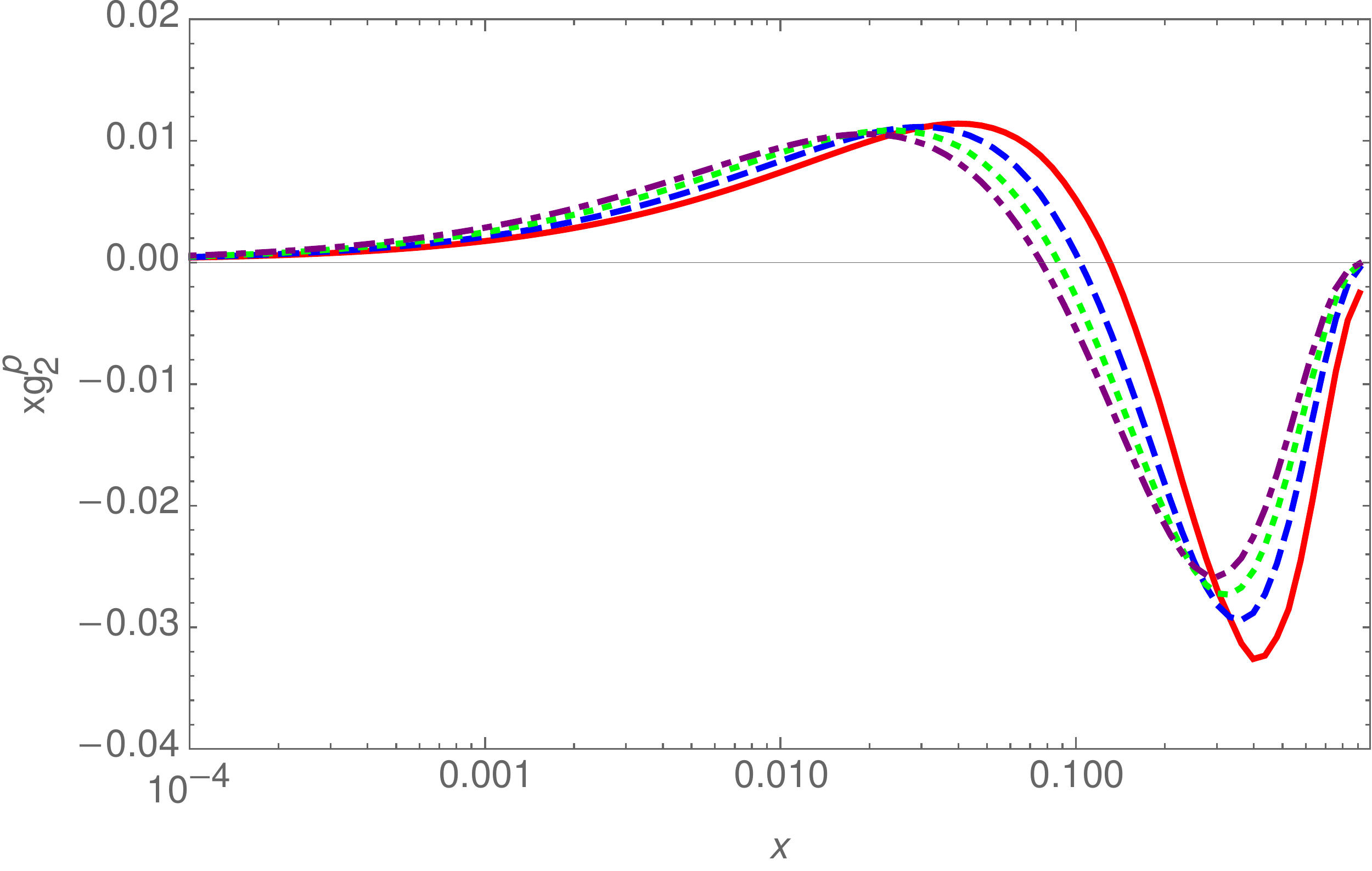}
\caption[]{\small \sf The NLO massless structure function $xg_2(x,Q^2)$ as a 
function of $x$ for $Q^2 = 10~\GeV^2$ (full line); $100~\GeV^2$ 
(dashed line); $1000~\GeV^2$ (dotted line), and  $10000~\GeV^2$ (dash--dotted line), using
the parton distribution functions of \cite{Blumlein:2010rn}. 
}
\label{fig:FIG2} 
\end{figure}

\noindent
The massless contributions to $xg_1^p(x,Q^2)$ and $xg_2^p(x,Q^2)$ are shown in Figures~\ref{fig:FIG1} 
and \ref{fig:FIG2} to NLO. In all illustrations we use the parton distribution functions of 
Ref.~\cite{Blumlein:2010rn} and $a_s(Q^2)$ at NLO and they are made for contributions to the proton 
structure functions $xg_{1,(2)}^p(x,Q^2)$. 
\begin{figure}[H]
\centering
\includegraphics[width=0.8 \linewidth]{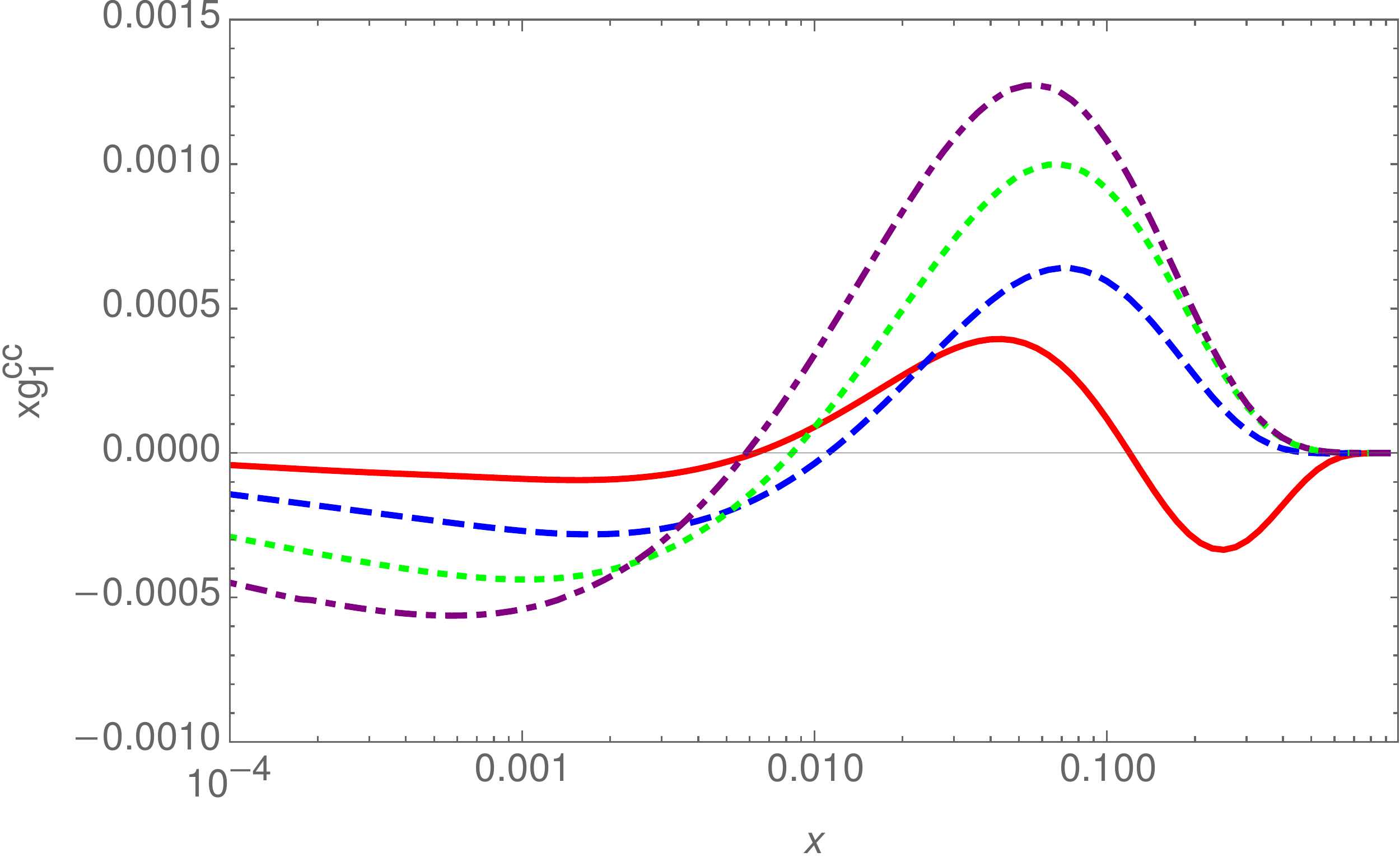}
\caption[]{\small \sf The $O(a_s)$ charm contribution the polarized structure function $xg_1(x,Q^2)$ as a 
function of $x$ for $Q^2 = 10~\GeV^2$ (full line); $100~\GeV^2$ 
(dashed line); $1000~\GeV^2$ (dotted line), and  $10000~\GeV^2$ (dash--dotted line) for $m_c = 1.59~\GeV$ and the 
parton distribution functions \cite{Blumlein:2010rn}. 
}
\label{fig:FIG3} 
\end{figure}
\begin{figure}[H]
\centering
\includegraphics[width=0.8 \linewidth]{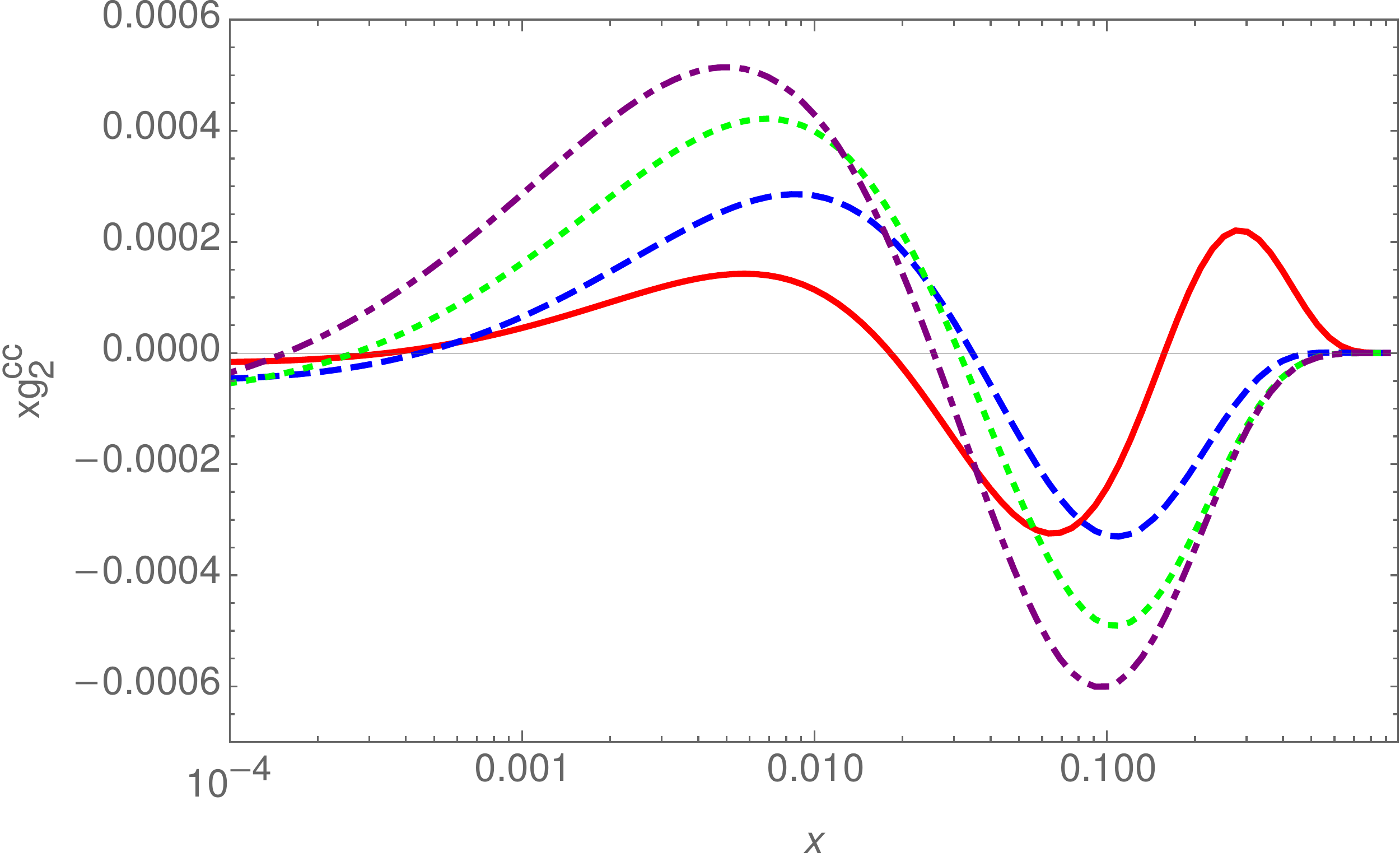}
\caption[]{\small \sf The $O(a_s)$ charm contribution the polarized structure function $xg_2(x,Q^2)$ as a 
function of $x$ for $Q^2 = 10~\GeV^2$ (full line); $100~\GeV^2$ 
(dashed line); $1000~\GeV^2$ (dotted line), and  $10000~\GeV^2$ (dash--dotted line) for $m_c = 1.59~\GeV$ and the 
parton distribution functions \cite{Blumlein:2010rn}. 
}
\label{fig:FIG4} 
\end{figure}

\noindent
In the small $x$ region both  structure functions tend to zero because of their principle shapes, which 
are similar to the unpolarized non--singlet structure functions. The change of sign in $xg_2(x,Q^2)$ is due 
to the Wandzura--Wilczek relation. In Figure~\ref{fig:FIG3} we illustrate the charm contributions to the 
structure function $xg_1(x,Q^2)$ at $O(a_s)$ for $Q^2 = 10, 100, 1000$ and $10000~\GeV^2$. 
\begin{figure}[H]
\centering
\includegraphics[width=0.7 \linewidth]{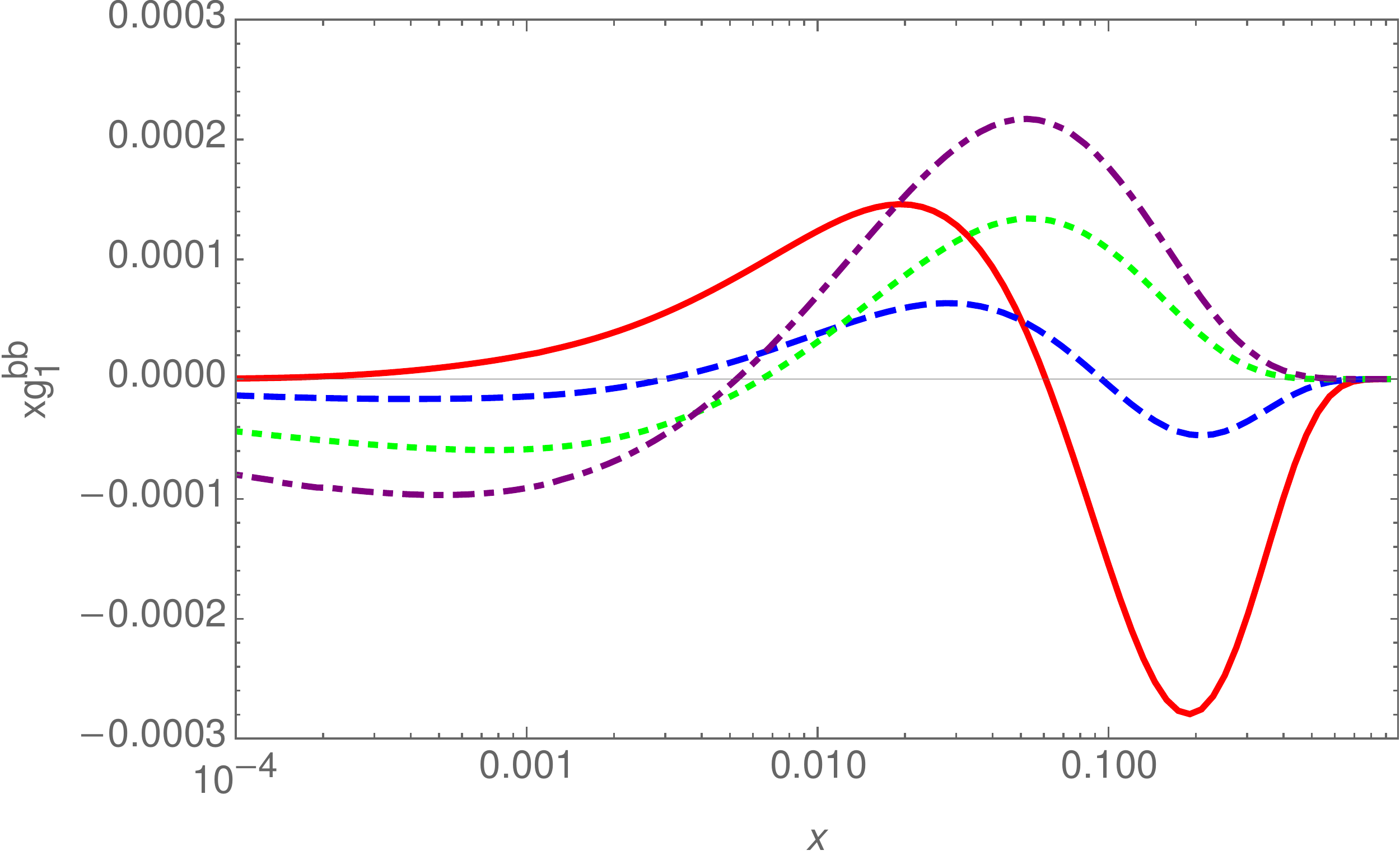}
\caption[]{\small \sf The $O(a_s)$ bottom contribution the polarized structure function $xg_1(x,Q^2)$ 
as a function of $x$ for $Q^2 = 10~\GeV^2$ (full line); $100~\GeV^2$ (dashed line); $1000~\GeV^2$ 
(dotted line), and  $10000~\GeV^2$ (dash--dotted line) for $m_b = 4.78~\GeV$ and the parton distribution 
functions \cite{Blumlein:2010rn}. 
}
\label{fig:FIG5} 
\end{figure}
\begin{figure}[H]
\centering
\includegraphics[width=0.8 \linewidth]{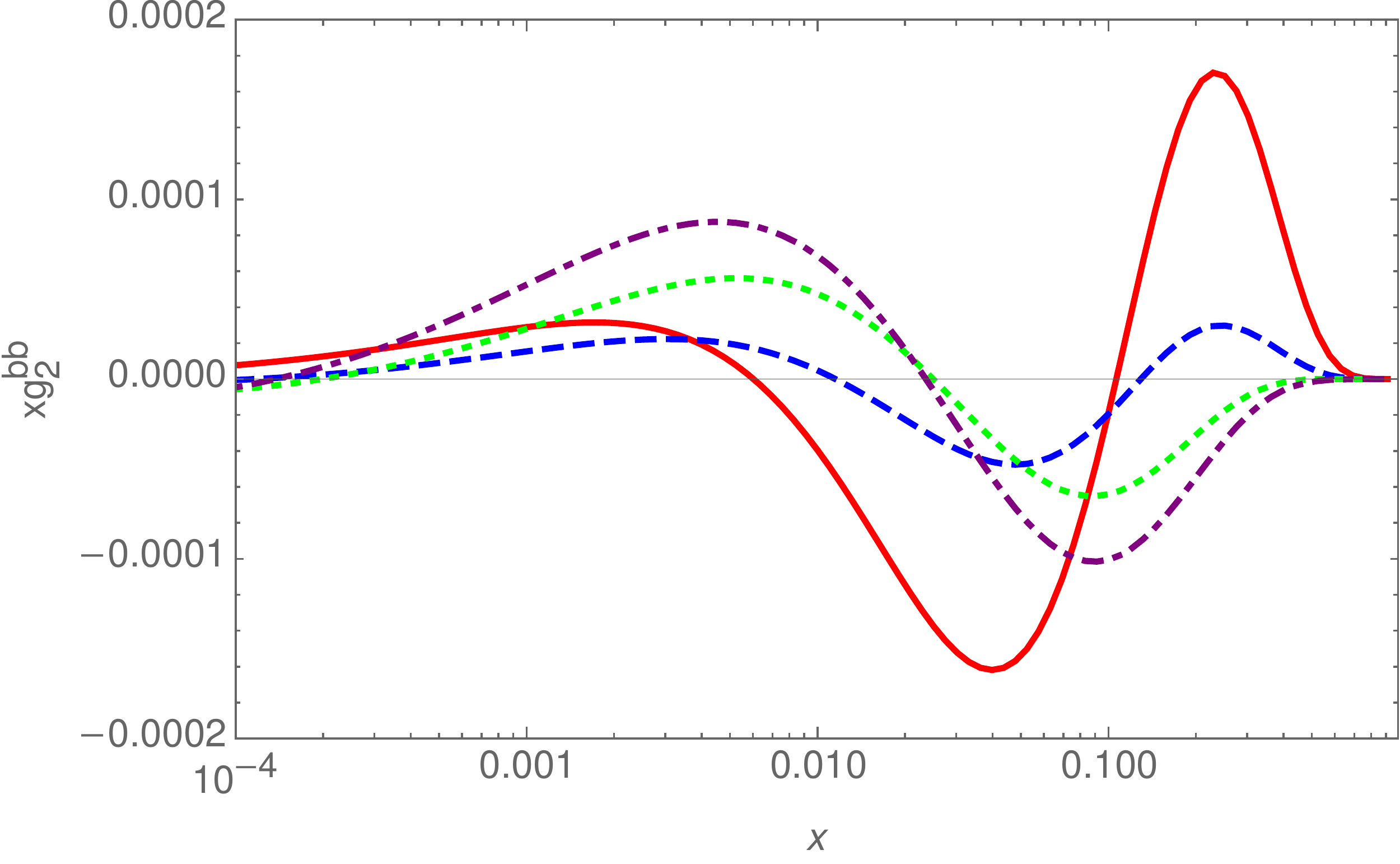}
\caption[]{\small \sf The $O(a_s)$ bottom contribution the polarized structure function $xg_2(x,Q^2)$ 
as a 
function of $x$ for $Q^2 = 10~\GeV^2$ (full line); $100~\GeV^2$ 
(dashed line); $1000~\GeV^2$ (dotted line), and  $10000~\GeV^2$ (dash--dotted line) for $m_b = 
4.78~\GeV$ and the 
parton distribution functions \cite{Blumlein:2010rn}. 
}
\label{fig:FIG6} 
\end{figure}
\begin{figure}[H]
\centering
\includegraphics[width=0.8 \linewidth]{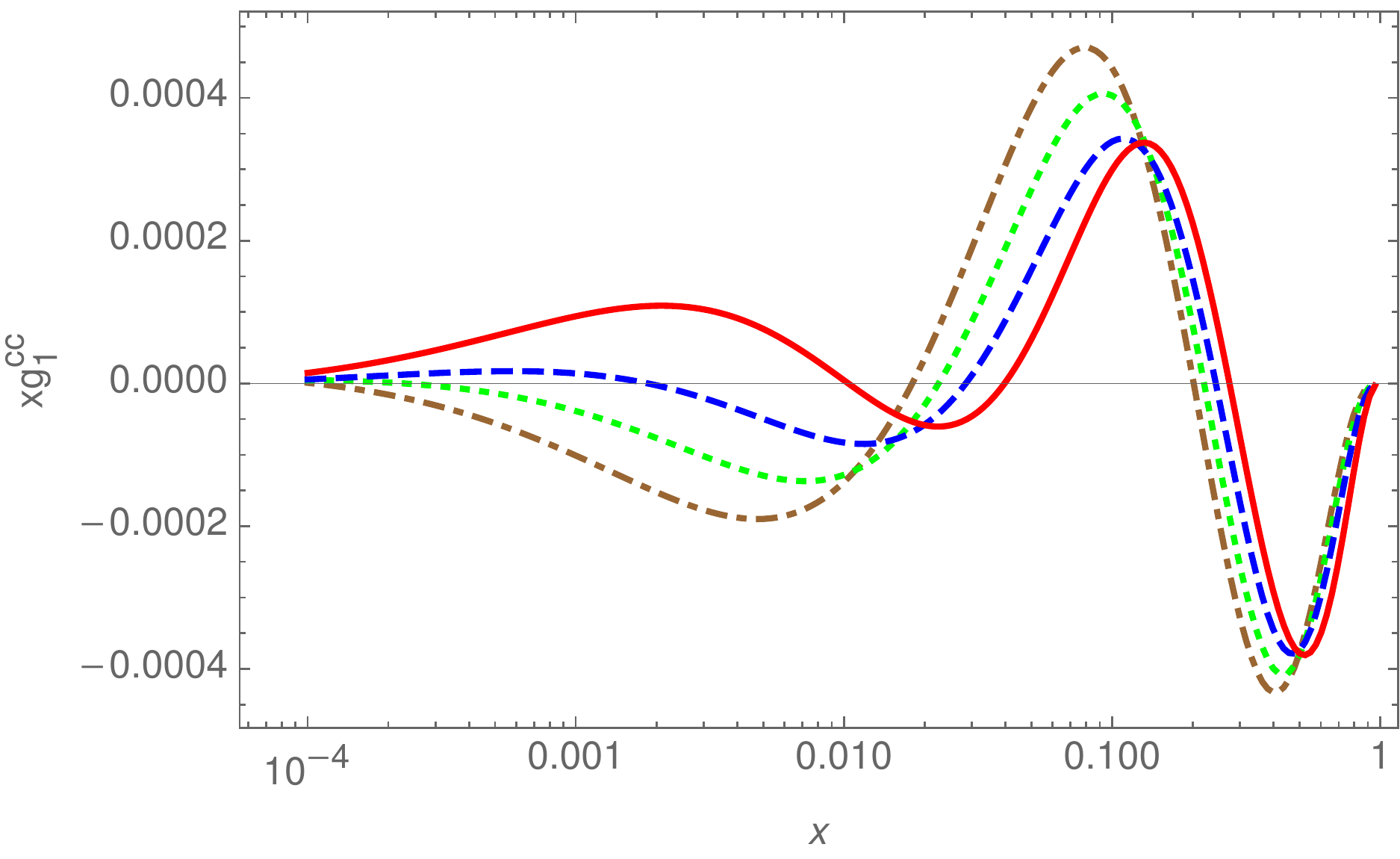}
\caption[]{\small \sf The $O(a_s^2)$ charm contributions to the structure function $xg_1(x,Q^2)$
for $m_c = 1.59~\GeV$ as a  function of $x$ for $Q^2 = 10~\GeV^2$ (full line); $100~\GeV^2$ 
(dashed line); $1000~\GeV^2$ (dotted line), and  $10000~\GeV^2$ (dash--dotted line), 
using the parton distribution functions of \cite{Blumlein:2010rn}. 
}
\label{FIG7}
\end{figure}

\noindent
The values of 
the charm and bottom quark masses are used in the on--shell scheme with $m_c = 1.59~\GeV$,~\cite{Alekhin:2012vu}, 
and $m_b = 4.78~\GeV$,~\cite{PDG}. 
\begin{figure}[H]
\centering
\includegraphics[width=0.8 \linewidth]{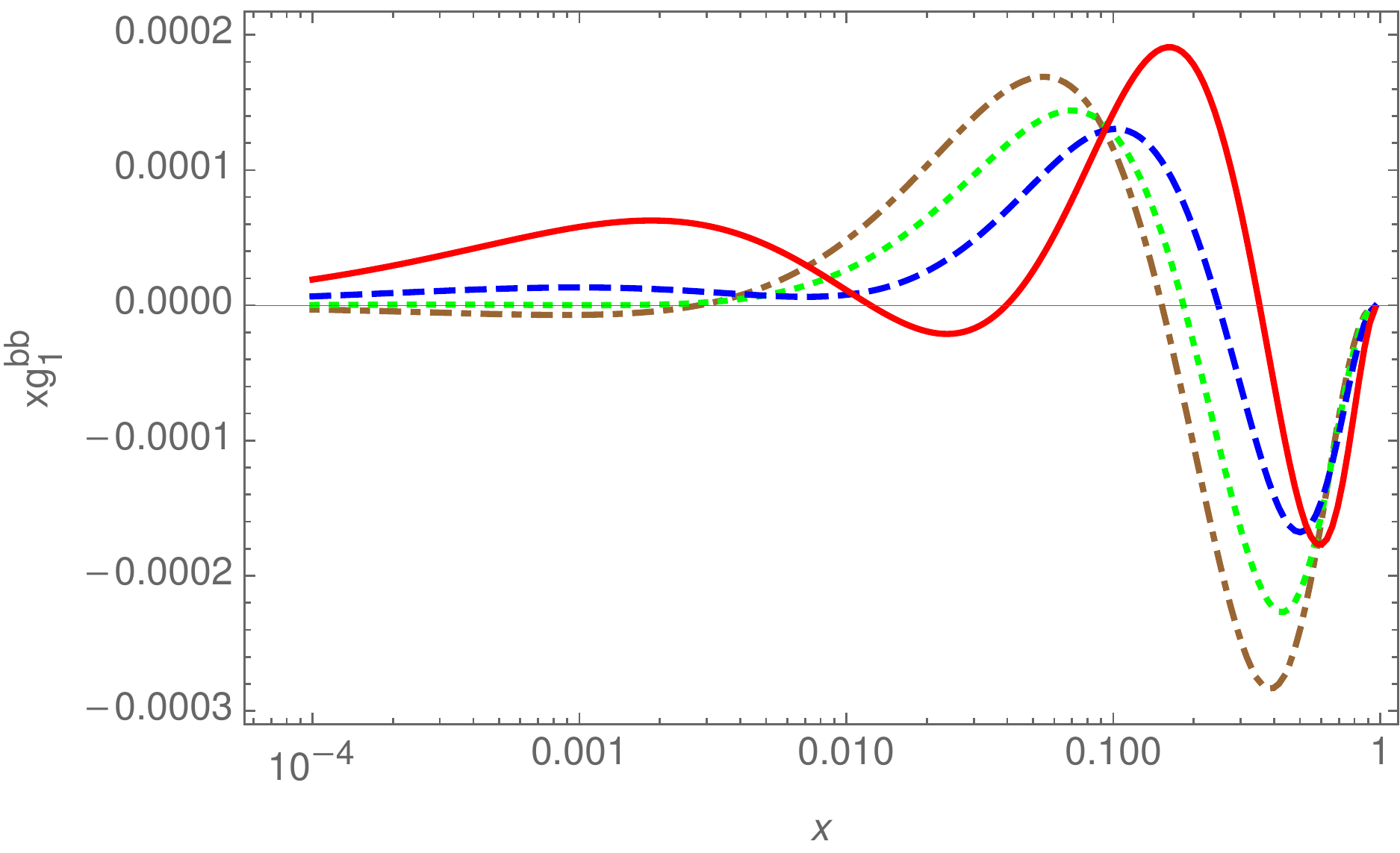}
\caption[]{\small \sf The $O(a_s^2)$ bottom contributions to the structure function $xg_1(x,Q^2)$
for $m_b = 4.78~\GeV$ as a  function of $x$ for $Q^2 = 10~\GeV^2$ (full line); $100~\GeV^2$ 
(dashed line); $1000~\GeV^2$ (dotted line), and  $10000~\GeV^2$ (dash--dotted line), 
using the parton distribution functions of
\cite{Blumlein:2010rn}.
}
\label{FIG8}
\end{figure}

\noindent
\begin{figure}[H]
\centering
\includegraphics[width=0.8 \linewidth]{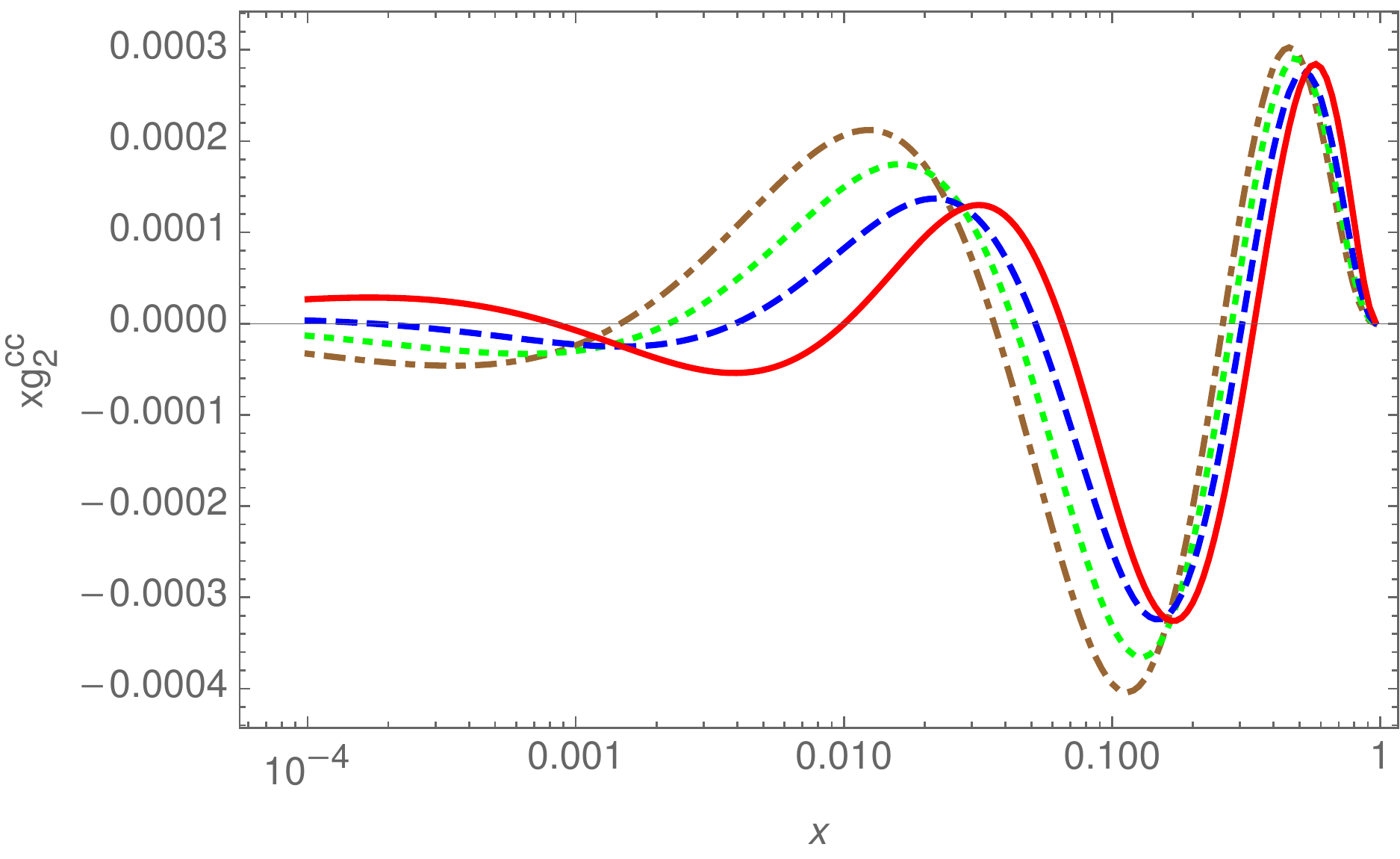}
\caption[]{\small \sf The $O(a_s^2)$ charm contributions to the structure function $xg_2(x,Q^2)$
for $m_c = 1.59~\GeV$ as a  function of $x$ for $Q^2 = 10~\GeV^2$ (full line); $100~\GeV^2$ 
(dashed line); $1000~\GeV^2$ (dotted line), and  $10000~\GeV^2$ (dash--dotted line), 
using the parton distribution functions of \cite{Blumlein:2010rn}. 
}
\label{FIG9}
\end{figure}
\begin{figure}[H]
\centering
\includegraphics[width=0.8 \linewidth]{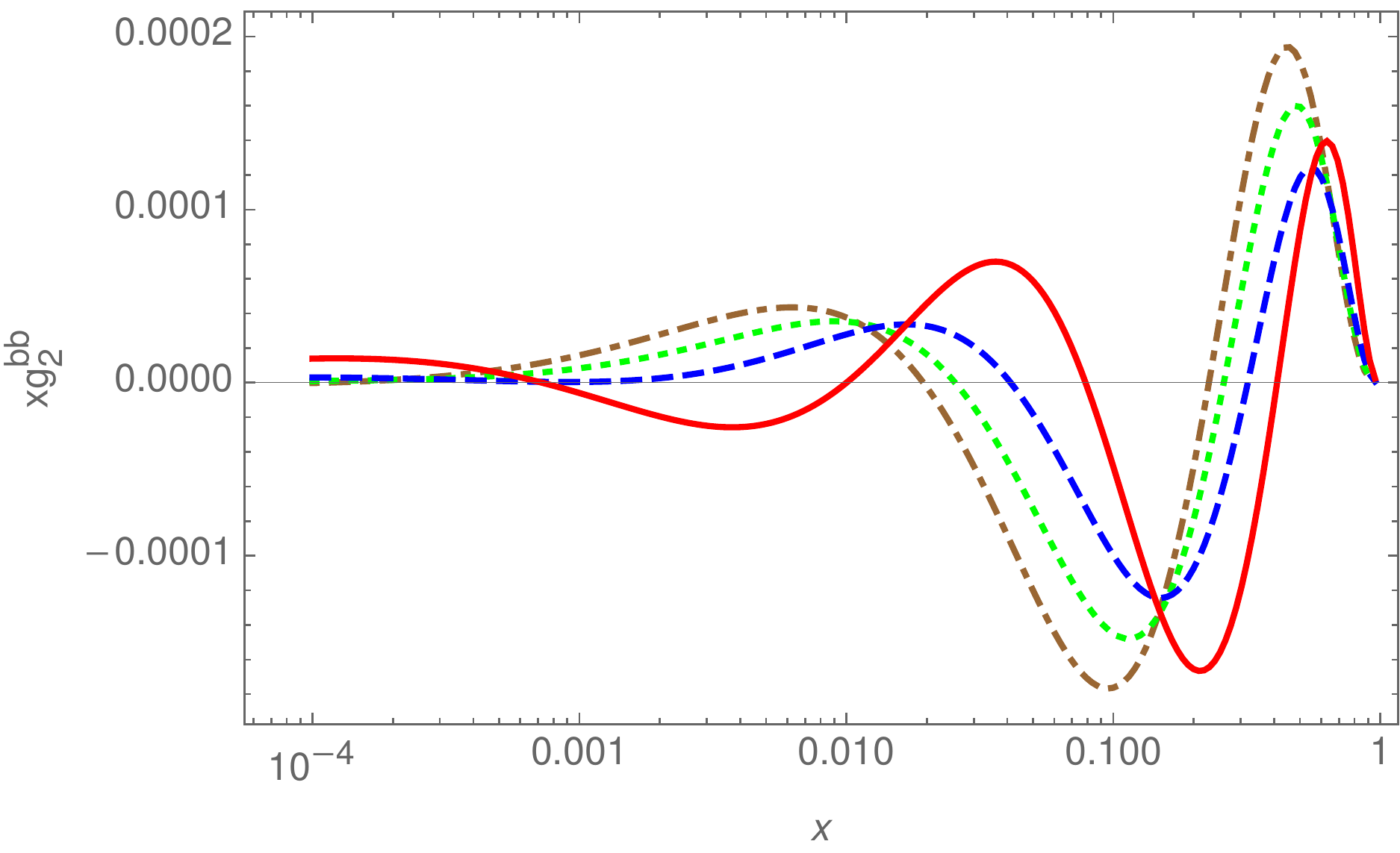}
\caption[]{\small \sf The $O(a_s^2)$ bottom contributions to the structure function $xg_2(x,Q^2)$
for $m_b = 4.78~\GeV$ as a  function of $x$ for $Q^2 = 10~\GeV^2$ (full line); $100~\GeV^2$ 
(dashed line); $1000~\GeV^2$ (dotted line), and  $10000~\GeV^2$ (dash--dotted line), 
using the parton distribution functions of \cite{Blumlein:2010rn}. 
}
\label{FIG10}
\end{figure}
\begin{figure}[H]
\centering
\includegraphics[width=0.8 \linewidth]{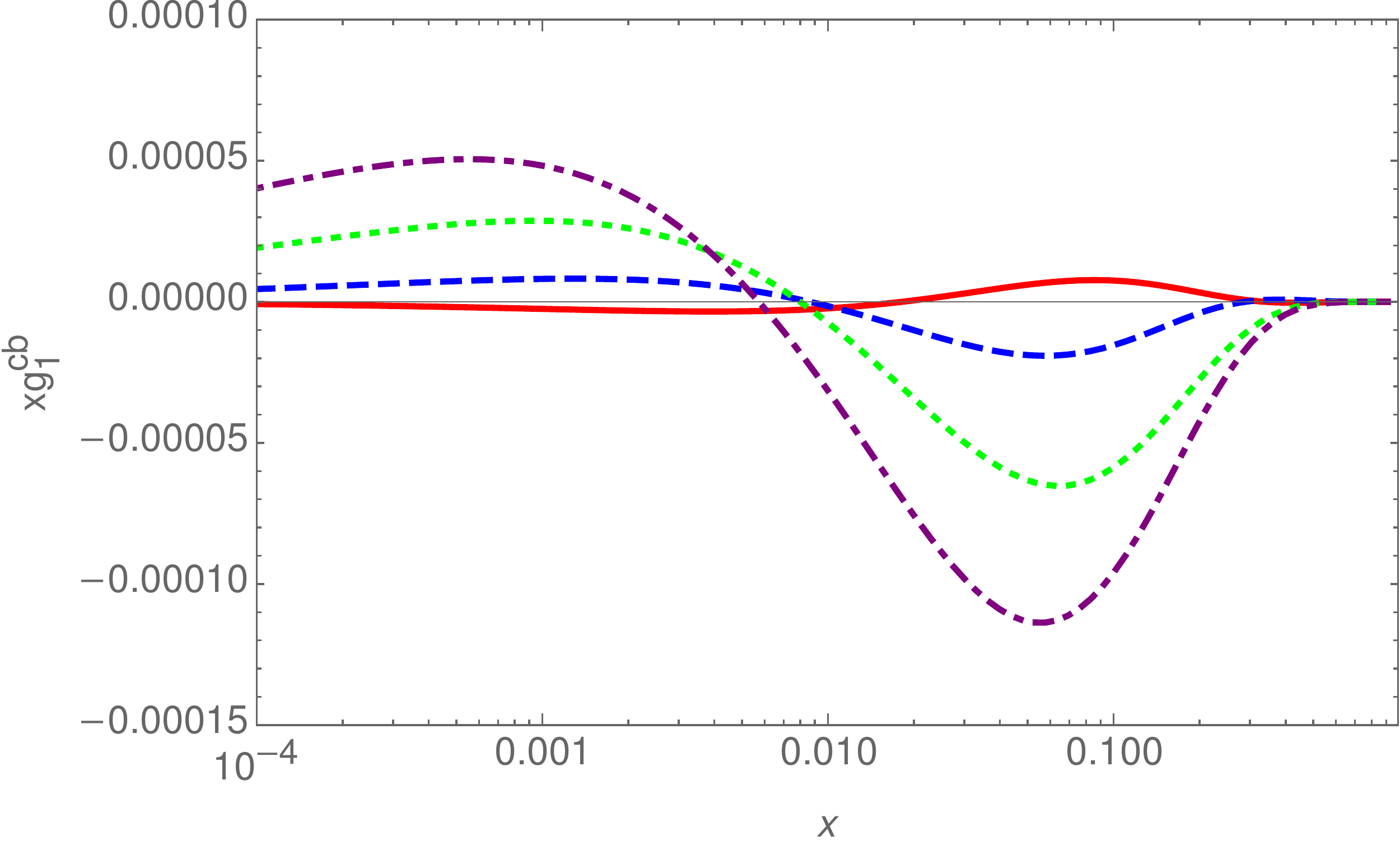} 
\caption[]{\small \sf The $O(a_s^2)$ two--mass contribution the polarized structure function $xg_1(x,Q^2)$ 
as a function of $x$ for $Q^2 = 10~\GeV^2$ (full line); $100~\GeV^2$ (dashed line); $1000~\GeV^2$ 
(dotted line), and  $10000~\GeV^2$ (dash--dotted line) for $m_c = 1.59~\GeV$ and $m_b = 4.78~\GeV$ 
and the parton distribution functions \cite{Blumlein:2010rn}. 
}
\label{fig:FIG11} 
\end{figure}
\begin{figure}[H]
\centering
\includegraphics[width=0.8 \linewidth]{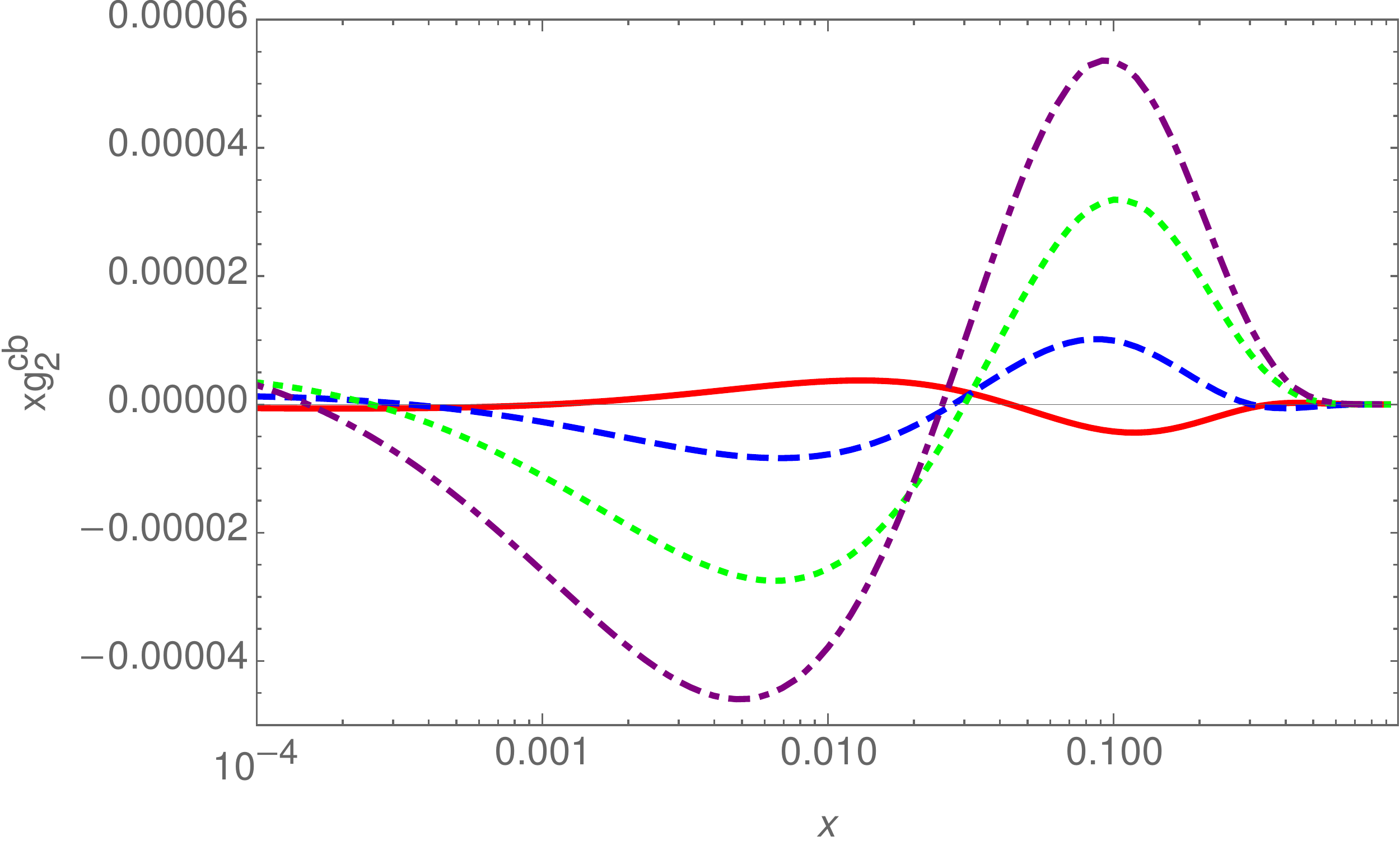}
\caption[]{\small \sf The $O(a_s^2)$ two--mass contribution the polarized structure function $xg_2(x,Q^2)$ 
as a function of $x$ for $Q^2 = 10~\GeV^2$ (full line); $100~\GeV^2$ (dashed line); $1000~\GeV^2$ 
(dotted line), and  $10000~\GeV^2$ (dash--dotted line) for $m_c = 1.59~\GeV$ and $m_b = 4.78~\GeV$ 
and the parton distribution functions \cite{Blumlein:2010rn}. 
}
\label{fig:FIG12} 
\end{figure}
Figure~\ref{fig:FIG4} shows the corresponding contributions for the structure 
functions $xg_{2}(x,Q^2)$. The numerical integrals have been performed using the {\tt Fortran} code 
{\tt AIND} \cite{PIESSENS}.
The contributions to $xg_1$ turn out to be two to three times larger than to $xg_2$. 

In 
Figures~\ref{fig:FIG5} and \ref{fig:FIG6} the corresponding contributions due to bottom quarks are 
shown. They are suppressed by a factor of $\sim 8$ compared with the $O(a_s)$ terms due to charm 
quarks. Comparing Figures~\ref{fig:FIG1} and \ref{fig:FIG3}, the $O(a_s)$ charm contribution is suppressed by 
about one order of magnitude compared to the massless case for the structure function $xg_1(x,Q^2)$ and 
similarly for the structure function $xg_2(x,Q^2)$. Yet for future precision measurements, contributions of 
this kind become important.

We now turn  to the single mass $O(a_s^2)$ contributions. They are shown in Figures~\ref{FIG7} and 
\ref{FIG9} for the charm contributions to $xg_{1(2)}(x,Q^2)$ and for those from the bottom quark contributions to 
in Figures~\ref{FIG8} and \ref{FIG10}. Here we show the combination of the non--singlet and 
different singlet contributions. In the large $x$ region the non--singlet contribution dominates, while the
singlet contributions dominate in the lower $x$ region. Towards large values of $x$ the Wandzura--Wilczek 
relation implies $g_2(x,Q^2) \simeq - g_1(x,Q^2)$. The bottom quark corrections turn out to be 
about a factor of 1.5 to 2 smaller than the charm quark contributions. The corrections of $O(a_s^2)$ are a factor
2 to 3 smaller than the $O(a_s)$ corrections in the case of charm, and similarly for bottom. Concerning the present 
illustrations, the corrections for bottom
quarks can be trusted only in the higher $Q^2$ region. For the lower range of $Q^2$ one would need to consider
also power corrections, which we did not do in the present analysis.

At $O(a_s^2)$ there are also contributions with two heavy quark lines in single graphs, due to heavy
quark polarization insertions in the external gluon line. Their contributions are illustrated in
Figures~\ref{fig:FIG11} and \ref{fig:FIG12}. They are smaller in size by factors of 10--20 than the
$O(a_s)$ charm contributions.

\section{The gluonic OMEs for the variable flavor number scheme}
\label{sec:7}

\vspace{1mm}\noindent
The matching between parton densities at large scales $Q^2 \gg m^2$ can be performed by using
the variable flavor number scheme, cf. e.g.~\cite{Blumlein:2018jfm}. Due to the similar size 
of $m_c$ and $m_b$ one often has to decouple both masses at the same time, see 
Eqs.~(\ref{eq:FI}--\ref{eq:ma2VNS})  and (\ref{eq:tm1}, \ref{eq:tm2}), cf.~\cite{Ablinger:2017err}. 
Besides the OMEs given in Section~\ref{sec:4} already the polarized gluonic OMEs contribute which 
we calculate in the following. For the unrenormalized operator matrix element $\Delta A_{gq,Q}^{(2)}$ 
one obtains 
\begin{eqnarray} 
\label{eq:Agq1}
\Delta A_{gq,Q}^{(2)} &=&	S_\ep^2 
\left( \frac{m^2}{\mu^2} \right)^\ep  
(N+2) \biggl\{
      \frac{1}{\ep^2} \frac{32}{3 N (1+N)}
      +\frac{16}{\ep} \biggl[
            \frac{(2+5 N)}{9 N (N+1)^2}
            -\frac{1}{3 N (N+1)} S_1
      \biggr]
\nonumber\\ &&
      +\frac{8 \big(22+41 N+28 N^2\big)}{27 N (N+1)^3}
      -\frac{8 (2+5 N)}{9 N (N+1)^2}  S_1
      +\frac{4}{3 N (N+1)} \left[S_1^2 + S_2 + 2 \zeta_2\right]
\nonumber \\ &&
      +\ep 
      \biggl[
            \frac{4 \big(98+369 N+408 N^2+164 N^3\big)}{81 N (N+1)^4}
            -\biggl(
                   \frac{4 \big(22+41 N+28 N^2\big)}{27 N (N+1)^3}
                  +\frac{2}{3 N (N+1)} S_2
            \biggr) 
\nonumber \\ && \times S_1
            +\frac{2(2+5 N)}{9 N (N+1)^2} S_1^2
            -\frac{2}{9 N (N+1)} S_1^3
            +\frac{2 (2+5 N)}{9 N (N+1)^2} S_2
            -\frac{4}{9 N (N+1)} S_3
\nonumber \\ &&
            +\biggl(
                  \frac{4 (2+5 N)}{9 N (N+1)^2}
                  -\frac{4}{3 N (N+1)} S_1
            \biggr) \zeta_2
            +\frac{8}{9 N (N+1)} \zeta_3
      \biggr]
\biggr\} +O(\ep^2).
\label{eq:Agq2}
\end{eqnarray}
The structure of $\Delta A_{gq,Q}^{(2), \rm L}$ is predicted, cf.~\cite{Bierenbaum:2009mv}, by
\begin{eqnarray}
\Delta \hat{\hat{A}}_{gq,Q}^{(2), \rm L} = \left(\frac{m^2}{\mu^2}\right)^\ep S_\ep^2 \left[- \frac{2 
\beta_{0,Q}}{\ep^2} 
\Delta P_{gq}^{(0)} -  \frac{1}{2\ep} \Delta \hat{P}_{gq}^{(1)} + a_{gq,Q}^{(2)} + \ep 
\overline{a}_{gq,Q}^{(2)} \right] + O(\ep^2),
\end{eqnarray}
\begin{eqnarray}
\Delta A_{gq,Q}^{(2), \rm L} &=& - \frac{\beta_{0,Q}}{2} \Delta P_{gq}^{(0)} 
\ln^2\left(\frac{m^2}{\mu^2}\right)
- \frac{1}{2} \Delta \hat{P}_{gq}^{(1)} \ln\left(\frac{m^2}{\mu^2}\right)
+ a_{gq,Q}^{(2)} + \frac{\beta_{0,Q} \zeta_2}{2} \Delta P_{gq}^{(0)}.
\\
&=& \textcolor{black}{C_F T_F} \Biggl\{\frac{8}{3} \frac{N+2}{N(N+1)} \ln^2\left(\frac{\mu^2}{m^2}\right) 
+ 16 (N+2) \Biggl[\frac{S_1}{3N(N+1)} - \frac{(2+5N)}{9N(N+1)^2}\Biggr]
\ln\left(\frac{\mu^2}{m^2}\right) \nonumber\\ &&
+
(N+2) \Biggl[
      \frac{8 \big(22+41 N+28 N^2\big)}{27 N (N+1)^3}
      -\frac{8 (2+5 N)}{9 N (N+1)^2}  S_1
      +\frac{4}{3 N (N+1)} \left[S_1^2 + S_2\right]\Biggr]
\Biggr\}.
\nonumber\\
\end{eqnarray}
The unrenormalized OMEs $\Delta \hat{\hat{A}}_{gg,Q}^{(1,(2)), \rm L}$ are given by\footnote{Please note 
that (\ref{eq:hhAggQ2L})
replaces Eq.~(280) of \cite{Hasselhuhn:2013swa}, which contained typographical errors.}
\begin{eqnarray}
\label{eq:AggQ1}
 \Delta \hat{\hat{A}}_{gg,Q}^{(1)}
 &={}&
  \left(\frac{m^2}{\mu^2}\right)^{\ep/2} 
  S_{\ep} \left(- \frac{2 \beta_{0,Q}}{\ep}\right) \exp\left[\sum_{i=2}^\infty \frac{\zeta_i}{i} 
\left(\frac{\ep}{2}\right)^i\right],
\\
 \Delta \hat{\hat{A}}_{gg,Q}^{(2), L} &=& \left( \frac{m^2}{\mu^2} \right)^{\ep} S_\ep^2 
\Biggl\{
\frac{1}{\ep^2} \Biggl[ \textcolor{black}{C_F T_F}
        \frac{16 (N-1) (2+N) }{N^2 (1+N)^2}
        +\textcolor{black}{T_F^2} \frac{64 }{9}
        +\textcolor{black}{C_A T_F} \Biggl(
                \frac{64}{3 N (1+N)}
\nonumber\\ &&               
 -\frac{32}{3} S_1
        \Biggr)
\Biggr]
+\frac{1}{\ep} \Biggl[
        - \textcolor{black}{C_F T_F} \frac{4 R_2}{N^3 (1+N)^3}
        +\textcolor{black}{C_A T_F} \Biggl(
                +\frac{16 R_1}{9 N^2 (1+N)^2}
                -\frac{80}{9} S_1
        \Biggr)
\Biggr]
\nonumber\\ &&
+\textcolor{black}{C_F T_F} \Biggl(
        -\frac{R_5}{3 N^4 (1+N)^4}
        +\frac{4 (N-1) (2+N) \zeta_2}{N^2 (1+N)^2}
\Biggr)
+\textcolor{black}{C_A T_F} \Biggl(
        \frac{2 R_3}{27 N^3 (1+N)^3}
\nonumber\\ &&
        -\frac{4 (47+56 N)}{27 (1+N)} S_1
        +\frac{16 \zeta_2}{3 N (1+N)}
        -\frac{8}{3} S_1 \zeta_2
\Biggr)
+ \textcolor{black}{T_F^2} \frac{16}{9} \zeta_2
+\ep \Biggl[
        \textcolor{black}{C_F T_F} \Biggl(
                -\frac{\zeta_2 R_7}{N^3 (1+N)^3}
\nonumber\\ &&              
  -\frac{R_6}{12 N^5 (1+N)^5}
                +\frac{4 (N-1) (2+N) \zeta_3}{3 N^2 (1+N)^2}
        \Biggr)
        +\textcolor{black}{C_A T_F} \Biggl(
                \frac{4 \zeta_2 R_1}{9 N^2 (1+N)^2}
\nonumber\\ &&            
    +\frac{R_4}{81 N^4 (1+N)^4}
                -\frac{2 \big(
                        283+584 N+328 N^2\big)}{81 (1+N)^2} S_1
                -\frac{S_1^2}{3 (1+N)}
                +\frac{(1+2 N)}{3 (1+N)} S_2
\nonumber\\ &&               
 -\frac{20}{9}\zeta_2 S_1
                +\frac{16 \zeta_3}{9 N (1+N)}
                -\frac{8}{9} S_1 \zeta_3
        \Biggr)
        + \textcolor{black}{T_F^2} \frac{8}{27}  \zeta_3
\Biggr],
\label{eq:hhAggQ2L}
\end{eqnarray}
with the polynomials
\begin{eqnarray}
R_1 &=& 3 N^4+6 N^3+16 N^2+13 N-3,
\\
R_2 &=& 3 N^6+9 N^5+7 N^4+3 N^3+8 N^2-2 N-4,
\\
R_3 &=& 15 N^6+45 N^5+374 N^4+601 N^3+161 N^2-24 N+36,
\\
R_4 &=& 3 N^8+12 N^7+2080 N^6+5568 N^5+4602 N^4+1138 N^3-3 N^2-36 N-108,
\\
R_5 &=& 13 N^8+52 N^7+54 N^6+4 N^5+13 N^4+12 N^2+36 N+24,
\\
R_6 &=& 35 N^{10}+175 N^9+254 N^8+62 N^7+55 N^6+347 N^5+384 N^4+72 N^3-96 N^2
\nonumber\\ &&
-120 N-48,
\\
R_7 &=& N^6 + 3 N^5 + 5 N^4 + N^3 - 8 N^2 + 2 N  + 4. 
\end{eqnarray}
In Eqs.~(\ref{eq:Agq2}, \ref{eq:hhAggQ2L}) we also present the terms of $O(\ep)$ which are needed in the 
calculation of the NNLO contributions, cf.~\cite{Bierenbaum:2009mv}. Furthermore, one has
\begin{eqnarray}
\label{eq:Agg2un}
\Delta \hat{\hat{A}}_{gg,Q}^{(2)} &=& \left(\frac{\hat{m}^2}{\mu^2}\right)^\ep S_\ep^2 \Biggl[
\frac{1}{2\ep^2} \Biggl\{ \Delta P_{gq}^{(0)} \Delta \hat{P}_{qg}^{(0)}
+ 2 \beta_{0,Q} \left( - \Delta P_{gg}^{(0)} + 2 \beta_0 + 4 \beta_{0,Q}\right)\Biggr\}
\nonumber\\ &&
+ \frac{1}{2\ep} \left[- \Delta \hat{P}^{(1)}_{gg} + 4 \delta m_1^{(-1)} \beta_{0,Q}\right]
+ a_{gg,Q}^{(2)} + 2 \delta m_1^{(0)} \beta_{0,Q} + \beta_{0,Q}^2 \zeta_2
\nonumber\\ &&
+ \ep \left[ \overline{a}_{gg,Q}^{(2)} + 2 \delta m_1^{(1)} \beta_{0,Q}  + \frac{1}{6} \beta_{0,Q} 
\zeta_3\right]
\Biggr].
\end{eqnarray}
The renormalized OME $\Delta A_{gg,Q}^{(2)}$ is then given by
\begin{eqnarray}
\Delta A_{gg,Q}^{(1)} &=& -\beta_{0,Q} \ln\left(\frac{m^2}{\mu^2}\right),
\\
\Delta A_{gg,Q}^{(2)} &=& \frac{1}{8} \left\{2 \beta_{0,Q}\left(-\Delta P_{gg}^{(0)} + 2 \beta_0\right)
+ \Delta P_{gq}^{(0)} \Delta P_{qg}^{(0)} + 8 \beta_{0,Q}^2 \right\} \ln^2\left(\frac{m^2}{\mu^2}\right)
\nonumber\\ &&
- \frac{1}{2} \Delta \hat{P}_{gg}^{(1)} \ln\left(\frac{m^2}{\mu^2}\right)
- \frac{\zeta_2}{8} \left[2 \beta_{0,Q}\left( - \Delta P_{gg}^{(0)} + 2 \beta_0\right)
+ \Delta P_{gq}^{(0)} \Delta P_{qg}^{(0)} \right] + a_{gg,Q}^{(2)}
\\
&=&
\Biggl[
        \textcolor{black}{C_F T_F} \frac{4 (N-1) (2+N)}{N^2 (1+N)^2}
        + \textcolor{black}{T_F^2}
\frac{16 }{9}
        +\textcolor{black}{C_A T_F}  \Biggl(
                \frac{16}{3 N (1+N)}
                -\frac{8}{3} S_1
        \Biggr)
\Biggr] \ln^2\left(\frac{\mu^2}{m^2}\right)
\nonumber\\ &&
+ \Biggl[
        -\textcolor{black}{C_F T_F} \frac{4 R_7}{N^3 (1+N)^3}
        +\textcolor{black}{C_A T_F} \Biggl(
                -\frac{16 R_1}{9 N^2 (1+N)^2}
                +\frac{80}{9} S_1
        \Biggr)
\Biggr] \ln\left(\frac{\mu^2}{m^2}\right)
\nonumber\\ &&
+\textcolor{black}{C_A T_F} \Biggl(
        \frac{2 R_3}{27 N^3 (1+N)^3}
        -\frac{4 (47+56 N)}{27 (1+N)} S_1
\Biggr)
+ \textcolor{black}{C_F T_F} \frac{R_8}{N^4 (1+N)^4},
\end{eqnarray}
with
\begin{eqnarray}
R_8 &=& -15 N^8-60 N^7-82 N^6-44 N^5-15 N^4-4 N^2-12 N-8.
\end{eqnarray}

The following transition rules hold in the two--flavor VFNS, 
cf.~\cite{Blumlein:2018jfm}, to next--to--leading order in Mellin $N$ space
\begin{eqnarray}
\label{eq:FI}
\Delta f_{{\sf NS},i}(N_F+2,\mu^2) &=& \Biggl\{1 + a_s^2(\mu^2) \left[
 \Delta A_{qq,Q}^{{\sf NS},(2,c)}
+ \Delta A_{qq,Q}^{{\sf NS},(2,b)}\right] \Biggr\} \Delta f_{{\sf NS},i}(N_F,\mu^2),
\\
\label{eq:FIS}
\Delta \Sigma(N_F+2,\mu^2) &=& \Biggl\{1 + a_s^2(\mu^2)\Bigl[
 \Delta A_{qq,Q}^{{\sf NS},(2,c)} + \Delta A_{qq,Q}^{{\sf PS},(2,c)}
+ \Delta A_{qq,Q}^{{\sf NS},(2,b)} + \Delta A_{qq,Q}^{{\sf PS},(2,b)} \Bigr] \Biggr\}
\nonumber\\ && \times
\Delta \Sigma(N_F,\mu^2)
\nonumber\\ &&
\hspace*{-3mm}
+ \Biggl\{a_s(\mu^2) \Bigl[
 \Delta A_{Qg}^{(1,c)}
+ \Delta A_{Qg}^{(1,b)}\Bigr] 
+a_s^2(\mu^2) 
\Bigl[
 \Delta A_{Qg}^{(2,c)}
+ \Delta A_{Qg}^{(2,b)} 
+ \Delta A_{Qg}^{(2,cb)}\Bigr]\Biggr\} 
\nonumber\\ && 
\times
\Delta G(N_F,\mu^2),
\\
\label{eq:GNF2}
\Delta G(N_F+2,\mu^2) &=& 
 \Biggl\{1 + a_s(\mu^2) \Bigl[
 \Delta A_{gg,Q}^{(1,c)}
+ \Delta A_{gg,Q}^{(1,b)}\Bigr] 
+a_s^2(\mu^2) 
\Bigl[
 \Delta A_{gg,Q}^{(2,c)}
+ \Delta A_{gg,Q}^{(2,b)} 
\nonumber\\
&&
+ \Delta A_{gg,Q}^{(2,cb)}\Bigr]\Biggr\} 
\Delta G(N_F,\mu^2)
+a_s^2(\mu^2)\Bigl[
  \Delta A_{gq,Q}^{(2,c)}
+ \Delta A_{gq,Q}^{(2,b)} \Bigr] \Delta \Sigma(N_F,\mu^2), \nonumber\\ &&
\\
\label{eq:cc}
\lefteqn{\hspace*{-2.5cm}\Bigl[\Delta f_c+ \Delta f_{\bar{c}}\Bigr](N_F+2,\mu^2) = a_s^2(\mu^2)
  \Delta A_{Qq}^{{\sf PS},(2,c)}
 \Delta \Sigma(N_F,\mu^2) 
+ \Biggl\{a_s(\mu^2) 
\Delta A_{Qg}^{(1,c)}
} 
\nonumber\\
&&~~~~~~
+a_s^2(\mu^2) 
\Bigl[
 \Delta A_{Qg}^{(2,c)}
+ \frac{1}{2} \Delta A_{Qg}^{(2,cb)}\Bigr]\Biggr\} \Delta G(N_F,\mu^2),
\\
\label{eq:ma2VNS}
\lefteqn{\hspace*{-2.5cm}\Bigl[\Delta f_b+ \Delta f_{\bar{b}}\Bigr](N_F+2,\mu^2) = a_s^2(\mu^2)
  \Delta A_{Qq}^{{\sf PS},(2,b)}
 \Delta \Sigma(N_F,\mu^2) + \Biggl\{a_s(\mu^2) 
\Delta A_{Qg}^{(1,b)}
} 
\nonumber\\
&&~~~~~~
+a_s^2(\mu^2) 
\Bigl[
 \Delta A_{Qg}^{(2,b)}
+ \frac{1}{2} \Delta A_{Qg}^{(2,cb)}\Bigr]\Biggr\} \Delta G(N_F,\mu^2),
\end{eqnarray}
and 
\begin{eqnarray}
\label{eq:NS1}
\Delta f_{{\sf NS},i}(N_F,\mu^2) &=& \Delta q_i(\mu^2) + \Delta \bar{q}_i(\mu^2).
\end{eqnarray}
The two--mass OMEs read
\begin{eqnarray}
\label{eq:tm1}
\Delta A_{Qg}^{(2,cb)}   &=& 
-\beta_{0,Q} \Delta \hat{\gamma}_{qg}^{(0)}  
\ln\left(\frac{\mu^2}{m_c^2}\right)
\ln\left(\frac{\mu^2}{m_b^2}\right),
\\
\label{eq:tm2}
\Delta A_{gg,Q}^{(2,cb)}   &=& 
2 \beta_{0,Q}^2 
\ln\left(\frac{\mu^2}{m_c^2}\right)
\ln\left(\frac{\mu^2}{m_b^2}\right).
\end{eqnarray}
Very recently the three--loop corrections to $\Delta A_{gg,Q}^{(3)}$
have been completed \cite{Ablinger:2022wbb}. 
\section{Conclusions}
\label{sec:8}

\vspace{1mm}\noindent
We calculated the two--loop single and double mass corrections to the polarized twist--2 
structure function $g_1(x,Q^2)$ in the asymptotic range  $Q^2 \gg m^2$ in analytic form. 
Those to $g_2(x,Q^2)$ are related by the Wandzura--Wilczek relation. The corrections include  
all but the power contributions $\propto (m^2/Q^2)^k,~k \in \mathbb{N}, k \geq 1$. Parts of 
the results in Ref. \cite{BUZA2} were confirmed, and other  parts were corrected. In \cite{BUZA2} 
a series of contributions to the Wilson coefficients of the structure function $g_1(x,Q^2)$, like
additional terms contributing to the non--singlet Wilson coefficient, 
$\Delta H_g^{(2)}$, and $\Delta L_g^{(2)}$, were 
left out. Also the two--mass corrections were 
not considered there. We perform the calculation of the Feynman diagrams using the hypergeometric 
method \cite{GHF} for general values of the dimensional parameter $\ep$ in the Larin scheme and 
transform then to the {\sf M} scheme and do not use IBP reduction. In Mellin space one obtains 
more compact results than in momentum fraction space. In Ref.~\cite{BUZA2}, 24 Nielsen integrals 
\cite{NIELSEN} were needed, whereas the $N$--space result depends only on two functions using 
also structural relations \cite{STRUCT}. In the small $x$ region the heavy flavor contributions  
are suppressed by at least one power of $\ln(x)$ if compared to the expected leading logarithmic 
behaviour of $O((a_s \ln^2(x))^k)$ in the massless case. We illustrated the different contributions
to two--loop order for the structure functions $xg_1(x,Q^2)$ and $xg_2(x,Q^2)$ in a wide kinematic
range for planning future experiments and possible re--analysis of the existing data.

The contributions calculated in the present paper are of importance for precision measurements of the 
structure functions $g_1(x,Q^2)$ and $g_2(x,Q^2)$ in future high luminosity measurements, e.g. at 
the EIC \cite{Boer:2011fh}, and associated precision measurements of the strong coupling constant 
$a_s$~\cite{alphas} and the charm quark mass \cite{Alekhin:2012vu}. We also presented the polarized 
NLO expansion coefficients in the 2--heavy flavor variable flavor number scheme and the next order 
terms $O(\ep)$ needed in the calculation of the $O(a_s^3)$ massive OMEs.
\appendix 
\section{Results for the Individual Diagrams}
\label{sec:A}

\vspace*{1mm}
\noindent
In this appendix we list the results for the individual diagrams to $O(\varepsilon)$, prior 
to renormalization. The calculation was performed in Feynman gauge. 
We suppress the argument in $S_{\vec{a}}(N) \equiv S_{\vec{a}}$ and
the factor 	
 \begin{eqnarray}
   i a_s^2 S_{\ep}^2 \Biggl(\frac{m^2}{\mu^2}\Biggr)^{\ep/2}\frac{1-(-1)^N}{2}. 
   \N
 \end{eqnarray} 
The notation follows Ref.~\cite{BBK1}, where also the individual diagrams are depicted. 
\begin{eqnarray} 
%
%
%
 \Delta A^{Qg}_a &=&T_FC_F\Biggl\{
            \frac{1}{\ep^2}\Biggl[
                               \frac{-16(N-1)}
                                    {N^2(N+1)^2}
                           \Biggr]
             +\frac{1}{\ep}\Biggl[
                             8\frac{(N-1)(2N+1)}
                                    {N^3(N+1)^3}
                           \Biggr]
                            -\frac{4(N-1)}
                                   {N^2(N+1)^2}
                                \Bigl(
                                      2S_2
                                     +\zeta_2
                                \Bigr) \N\\
&&                          +\frac{4\hat{P}_1}
                                   {N^4(N+1)^4}
             +\ep          \Biggl[
                            \frac{-4(N-1)}{3N^2(N+1)^2}
                                \Bigl(
                                      3S_3
                                     +\zeta_3
                                \Bigr)
                            +2\frac{(N-1)(2N+1)}
                                   {N^3(N+1)^3}
                                \Bigl(
                                      2S_2
                                     +\zeta_2
                                \Bigr) \N\\ &&
                            - \frac{2\hat{P}_2}
                                   {N^5(N+1)^5}
                           \Biggr]
                   \Biggr\}~, \\
\hat{P}_1&=&(N-1)(3N^4+2N^3-2N^2+N+1)~, \\
\hat{P}_2&=&2N^7+10N^6+21N^5+7N^4-7N^3-3N^2+N+1
                   \label{resA}~. \\\N\\
%
%
 \Delta A^{Qg}_b &=&T_FC_F\Biggl\{
            \frac{1}{\ep^2}\Biggl[
                               \frac{32(N-1)}
                                    {N(N+1)}
                                  \Bigl(
                                        S_1
                                        -1
                                  \Bigr)
                           \Biggr]
             +\frac{1}{\ep}\Biggl[
                             \frac{8(N-1)}
                                   {N(N+1)}
                                 \Bigl(
                                   S_1^2
                                  -3S_2
                                 \Bigr)
                            - \frac{16(N^2+1)}
                                    {N(N+1)^2}S_1 \N\\ &&
                              +\frac{32N}{(N+1)^2}
                           \Biggr]
                            +\frac{4(N-1)}
                                  {3N(N+1)}
                                 \Bigl(
                                        12S_{2,1}
                                       -10S_3
                                       +3S_1S_2
                                       +S_1^3
                                       +6S_1\zeta_2
                                       -6\zeta_2
                                 \Bigr)
\N\\ &&
                            -\frac{4(N^2-7)}
                                   {N(N+1)^2}S_2  
                            -\frac{4(N^2+1)}
                                   {N(N+1)^2}S_1^2
                            +16\frac{N^4+2N^3+N^2-2N+1}
                                    {N^2(N+1)^3}S_1
\N\\ &&
                            -16\frac{2N^3+2N^2+1}
                                    {N(N+1)^3}  
\N\\ &&
             +\ep          \Biggl[
                            \frac{N-1}
                                  {N(N+1)}
                                 \Bigl(
                                   -8S_{2,1,1}
                                   +8S_{3,1}
                                   -11S_4
                                   +8S_{2,1}S_1
                                   +\frac{4}{3}S_3S_1
                                   -\frac{7}{2}S_2^2
                                   +S_2S_1^2
                                   +\frac{1}{6}S_1^4  \N\\ &&
                                   -\frac{8}{3}\zeta_3
                                   -6\zeta_2S_2
                                   +2\zeta_2S_1^2
                                   +\frac{8}{3}S_1\zeta_3
                                 \Bigr)
                            +\frac{N^2+1}
                                  {N(N+1)^2}
                                 \Bigl(
                                   -8S_{2,1}
                                   -2S_2S_1
                                   -\frac{2}{3}S_1^3
                                   -4S_1\zeta_2
                                 \Bigr)  \N\\ &&
                            -\frac{4}{3}\frac{N^2-11}
                                             {N(N+1)^2}S_3
                            +4\frac{N^4+2N^3+N^2+2N+1}
                                   {N^2(N+1)^3}S_2
                            +\frac{8N\zeta_2}
                                   {(N+1)^2}
                            -\frac{8\hat{P}_3S_1}
                                   {N^2(N+1)^4} \N\\ &&
                            +4\frac{N^4+2N^3+N^2-2N+1}
                                   {N^2(N+1)^3}S_1^2
                            +8\frac{4N^4+8N^3+4N^2+2N+3}
                                   {N(N+1)^4}
                           \Biggr]
                    \Biggr\}~, \\
\hat{P}_3&=&2N^5+6N^4+6N^3+3N^2+8N+2 
                   \label{resB}~. \\\N\\
%
%
 \Delta A^{Qg}_c &=&T_FC_F\Biggl\{
            \frac{1}{\ep^2}\Biggl[
                                \frac{8(N-1)}{N(N+1)}
                           \Biggr]
             +\frac{1}{\ep}\Biggl[
                              -4\frac{13N^4+14N^3+2N^2+5N+2}
                                     {N^2(N+1)^2(N+2)}
                           \Biggr] \N\\
&&                            -\frac{2(N-1)}
                                     {N(N+1)}
                                    \Bigl(
                                           10S_2
                                          -\zeta_2
                                    \Bigr)
                              +\frac{2\hat{P}_4}
                                    {N^3(N+1)^3(N+2)}
             +\ep          \Biggl[
                              -\frac{2(N-1)}
                                    {3N(N+1)}
                                    \Bigl(
                                           15S_3
                                          -\zeta_3
                                    \Bigr) 
\N
\end{eqnarray}
\begin{eqnarray}
&&                            -\frac{13N^4+14N^3+2N^2+5N+2}
                                     {N^2(N+1)^2(N+2)}\zeta_2
                              +2\frac{7N^3-10N-1}
                                     {N^2(N+1)^2}S_2
                              -\frac{\hat{P}_5}
                                     {N^4(N+1)^4(N+2)}
                           \Biggr]
                   \Biggr\}~, 
\nonumber\\
\\
\hat{P}_4&=&16N^6+32N^5-4N^4-44N^3-11N^2-7N-2 ~, \\
\hat{P}_5&=&32N^8+96N^7+65N^6-45N^5-24N^4+19N^3+18N^2+9N+2
                   \label{resC}~.
\\
%
%
 \Delta A^{Qg}_d &=&T_FC_F\Biggl\{
            \frac{1}{\ep^2}\Biggl[
                               \frac{16(N-1)}{N(N+1)}
                           \Biggr]
             +\frac{1}{\ep}\Biggl[
                               \frac{8(N-1)}{N(N+1)}S_1
                               -8\frac{N^3+8N^2-5N-10}
                                      {N(N+1)^2(N+2)}
                           \Biggr]\N\\
&&                       +\frac{2(N-1)}
                               {N(N+1)}
                             \Bigl(
                                 S_2
                                +S_1^2
                                +2\zeta_2
                             \Bigr)
                         -4\frac{N^3+6N^2-11N+2}
                                {N^2(N+1)^2}S_1
                         +8\frac{N^4+7N^3-9N+6}
                                {N(N+1)^3(N+2)} \N\\
&&           +\ep          \Biggl[
                         \frac{1}{3}\frac{N-1}
                               {N(N+1)}
                             \Bigl(
                                 12S_{2,1}
                                +2S_3
                                +3S_2S_1
                                +S_1^3
                                +4\zeta_3
                                +6\zeta_2S_1
                             \Bigr) \N\\
&&                       -\frac{N^3+6N^2-11N+2}
                               {N^2(N+1)^2}\Bigl(S_2+S_1^2\Bigr)
                         -2\frac{N^3+8N^2-5N-10}
                                {N(N+1)^2(N+2)}\zeta_2 \N\\
&&                       +4\frac{N^4+6N^3-5N^2-2N+1}
                                {N^2(N+1)^3}S_1
                         -\frac{4\hat{P}_6}
                                {N(N+1)^4(N+2)} 
                           \Biggr]
                   \Biggr\} ~, \\
\hat{P}_6&=&2N^5+16N^4+14N^3-21N^2-22N-8
                   \label{resD}~. \\\N\\
%
%
 \Delta A^{Qg}_e &=&T_F\Bigl(C_F-\frac{C_A}{2}\Bigr)\Biggl\{
            \frac{1}{\ep^2}\Biggl[
                              \frac{-16(N-1)}{N(N+1)}
                           \Biggr]
             +\frac{1}{\ep}\Biggl[
                               \frac{8S_1}
                                     {N}
                              +\frac{8(N-1)}
                                     {N^2(N+1)^2}
                           \Biggr]
                              +\frac{2(N-2)}
                                     {N(N+2)}S_1^2  \N\\ &&
                              +2\frac{9N^2+7N-10}
                                     {N(N+1)(N+2)}S_2 
                              -4\frac{4N^4+7N^3+9N^2+14N-8}
                                     {N^2(N+1)^2(N+2)}S_1
                              -\frac{4\hat{P}_7}
                                     {N^3(N+1)^3(N+2)}\N\\ &&
                              -\frac{4(N-1)}
                                     {N(1+N)}\zeta_2
            +\ep          \Biggl[
                              2\frac{2S_{2,1}+S_1\zeta_2}
                                     {N}
                              +\frac{N-2}
                                    {3N(N+2)} 
                                        \Bigl(3S_2S_1+S_1^3\Bigr)
                              -\frac{4}{3}\frac{N-1}
                                               {N(N+1)}\zeta_3  \N\\ &&
                              +\frac{2}{3}\frac{13N^2+11N-14}
                                               {N(N+1)(N+2)}S_3
                              -\frac{20N^3+39N^2+13N+2}
                                    {N(N+1)^2(N+2)}S_2
                              +\frac{2\hat{P}_8S_1}
                                     {N^2(N+1)^3(N+2)} \N\\ &&
                              -\frac{4N^4+3N^3+5N^2+14N-8}
                                    {N^2(N+1)^2(N+2)}S_1^2 
                              +\frac{2(N-1)}
                                     {N^2(N+1)^2}\zeta_2 
                              +\frac{2\hat{P}_9}
                                     {N^4(N+1)^4(N+2)}
                           \Biggr]
                   \Biggr\}~, \\
\hat{P}_7&=&2N^5-13N^4-28N^3+4N^2+N+2~, \\
\hat{P}_8&=&8N^5+22N^4+35N^3+35N^2-18N-16~, \\
\hat{P}_9&=&8N^7-4N^6-26N^5+34N^4+38N^3+3N^2-3N-2
                   \label{resE}~. \\\N\\
%
%
 \Delta A^{Qg}_f &=&T_F\Bigl(C_F-\frac{C_A}{2}\Bigr)\Biggl\{
              \frac{1}{\ep}\Biggl[
                               \frac{16(N-1)}
                                    {N(N+1)}S_2
                            -16\frac{(2N+1)(N-1)}
                                    {N^2(N+1)^2}S_1
                           \Biggr]
                            -\frac{8(N-1)}
                                  {N(N+1)}
                                      \Bigl(
                                         2S_{2,1}
                                        -S_3
                                      \Bigr) \N\\ &&
                            -4\frac{(2N+3)(7N+1)}
                                   {N^2(N+1)^2}S_2
                            -4\frac{(2N+1)(N-1)}
                                   {N^2(N+1)^2}S_1^2
                            +8\frac{2N^3+11N^2+21N+6}
                                    {N^2(N+1)^3}S_1 \N\\ &&
             +\ep          \Biggl[
                             \frac{4(N-1)}
                                  {N(N+1)}
                                   \Bigl(
                                      -2S_{2,1,1}
                                      +2S_4
                                      +S_2^2
                                      +S_2\zeta_2
                                   \Bigr)
                            +\frac{8(3N+1)}
                                  {N^2(N+1)^2}S_{2,1}
                            -\frac{20}{3}\frac{4N^2+7N+1}
                                              {N^2(N+1)^2}S_3 \N
\end{eqnarray} \begin{eqnarray}
&&
                            +2\frac{(2N+1)(N-1)}
                                  {3N^2(N+1)^2}
                                   \Bigl(
                                      -3S_2S_1
                                      -S_1^3
                                      -6S_1\zeta_2
                                   \Bigr) 
                            -2\frac{10N^3+17N^2+11N+2}
                                   {N^2(N+1)^3}S_2 \N\\ &&
                            +2\frac{2N^3+11N^2+21N+6}
                                   {N^2(N+1)^3}S_1^2
                            -4\frac{4N^4+4N^3-13N^2-33N-10}
                                   {N^2(N+1)^4}S_1
                           \Biggr]
                   \Biggr\}
                   \label{resF}~.
\\
%
%
 \Delta A^{Qg}_j &=&T_FC_A\Biggl\{
            \frac{1}{\ep^2}\Biggl[
                              16\frac{(N+4)(N-1)}
                                     {N^2(N+1)^2}
                           \Biggr]
             +\frac{1}{\ep}\Biggl[
                             -8\frac{(N+4)(N^3+2N+1)}
                                     {N^3(N+1)^3}
                           \Biggr]   
\nonumber\\ &&
                             +4\frac{(N+4)(N-1)}
                                   {N^2(N+1)^2}
                                    \Bigl(
                                       2S_2
                                      +\zeta_2
                                    \Bigr)
                             +4\frac{(N+4)(4N^3-4N^2-3N-1)}
                                    {(N+1)^4N^4}
\nonumber\\ &&
             +\ep          \Biggl[
                              -\frac{2\hat{P}_{10}}
                                     {N^5(N+1)^5}  
                              +4\frac{(N+4)(N-1)}
                                   {3N^2(N+1)^2}
                                    \Bigl(
                                       3S_3
                                      +\zeta_3
                                    \Bigr)
\N\\ &&
                              -2\frac{(N+4)(N^3+2N+1)}
                                   {N^3(N+1)^3}
                                    \Bigl(
                                      2S_2
                                      +\zeta_2
                                    \Bigr)
                           \Biggr]
                   \Biggr\}~,\\
\hat{P}_{10}&=&(N+4)(N^5-7N^4+6N^3+7N^2+4N+1)
                   \label{resJ}~. \\\N\\
%
%
 \Delta A^{Qg}_l &=&T_FC_A\Biggl\{
            \frac{1}{\ep^2}\Biggl[
                              -\frac{8(2N-1)}
                                    {N(N+1)}S_1
                              -16\frac{N^3+N^2-2N-1}
                                      {(N+1)^2N^2}
                           \Biggr]
             +\frac{1}{\ep}\Biggl[
                              \frac{-2(2N-1)}
                                   {N(N+1)}
                                     \Bigl(
                                       S_2+S_1^2
                                     \Bigr) \N\\ &&
                              +\frac{12S_1}
                                    {N(N+1)^2}
                              +\frac{8\hat{P}_{11}}
                                    {N^3(N+1)^3}
                           \Biggr]
                              -\frac{2N-1}
                                    {3N(N+1)}
                                     \Bigl(
                                        12S_{2,1}
                                       +2S_3
                                       +3S_2S_1
                                       +S_1^3
                                       +6S_1\zeta_2
                                     \Bigr) \N\\ &&
                              +\frac{3S_1^2}
                                    {N(N+1)^2}
                              +2\frac{7N^3-2N^2+4N+4}
                                     {N^2(N+1)^3}S_1
                              +\frac{8N^3-N+8}
                                    {N^2(N+1)^2}S_2
\N\\ &&
                              -4\frac{N^3+N^2-2N-1}
                                     {N^2(N+1)^2}\zeta_2 
                              -\frac{4\hat{P}_{12}}
                                     {N^4(N+1)^4}
             +\ep          \Biggl[
                               \frac{2N-1}
                                    {N(N+1)}
                                     \Bigl(
                                        2S_{2,1,1}
                                       -2S_{3,1}
                                       -\frac{5}{4}S_4
\nonumber\\ & &                                       
-2S_{2,1}S_1
                                       -\frac{1}{3}S_3S_1
                                       -\frac{9}{8}S_2^2
                                       -\frac{1}{4}S_2S_1^2 
                                       -\frac{1}{24}S_1^4
                                       -\frac{1}{2}S_2\zeta_2
                                       -\frac{1}{2}S_1^2\zeta_2
                                       -\frac{2}{3}S_1\zeta_3
                                     \Bigr)
\N\\ &&
                              +\frac{
                                      12S_{2,1}
                                     +3S_2S_1
                                     +S_1^3
                                     +6\zeta_2S_1}
                                    {2N(N+1)^2}
                              -\frac{\hat{P}_{13}S_2}
                                    {2N^3(N+1)^3} \N\\ &&
                              +\frac{7N^3-2N^2+4N+4}
                                    {2N^2(N+1)^3}S_1^2
                              -\frac{14N^4+13N^3+12N^2+6N+8}
                                    {N^2(N+1)^4}S_1
                              +\frac{2\hat{P}_{14}\zeta_2}
                                    {N^3(N+1)^3} \N\\ &&
                              -\frac{4}{3}\frac{N^3+N^2-2N-1}
                                               {N^2(N+1)^2}\zeta_3
                              +\frac{4N^3-N+4}
                                    {N^2(N+1)^2}S_3
                              +\frac{2\hat{P}_{15}}
                                    {N^5(N+1)^5}
                           \Biggr]
                   \Biggr\}~,\\
\hat{P}_{11} &=&(2N+1)(N^4+N^3+2N+1)~, \\
\hat{P}_{12}&=&4N^7+12N^6+5N^5+4N^4-2N^3-10N^2-5N-1~, \\
\hat{P}_{13}&=&16N^5+21N^4+2N^3-36N^2-36N-8~, \\
\hat{P}_{14}&=&(2N+1)(N^4+N^3+2N+1)~, \\
\hat{P}_{15}&=&(2N+1)(4N^8+14N^7+9N^6+N^5-N^4+6N^3+7N^2+4N+1)
                   \label{resL}~. 
\end{eqnarray}

\begin{eqnarray}
%
%
 \Delta A^{Qg}_m &=&T_FC_A\Biggl\{
            \frac{1}{\ep^2}\Biggl[
                              -\frac{16(N-1)}
                                    {N^2(N+1)^2}
                           \Biggr]
+\frac{1}{\ep}\Biggl[
                               4\frac{N^4+7N^3+3N^2+3N+2}
                                     {N^3(N+1)^3}
                           \Biggr]
\N
\end{eqnarray}\begin{eqnarray}
&&
                              -\frac{4(N-1)} 
                                    {N^2(N+1)^2}
                                      \Bigl(
                                         2S_2
                                        +\zeta_2
                                      \Bigr) 
                              -\frac{2\hat{P}_{16}}
                                    {N^4(N+1)^4}
             +\ep          \Biggl[
                               \frac{-4(N-1)} 
                                    {3N^2(N+1)^2}
                                      \Bigl(
                                          3S_3
                                         +\zeta_3
                                      \Bigr)
\nonumber\\ &&                              
+\frac{N^4+7N^3+3N^2+3N+2} 
                                    {N^3(N+1)^3}
                                      \Bigl(
                                          2S_2
                                         +\zeta_2
                                      \Bigr) 
                              -\frac{\hat{P}_{17}}
                                    {N^5(N+1)^5}
                           \Biggr]
                   \Biggr\}~,\\
\hat{P}_{16}&=&6N^5+5N^4+N^3-13N^2-5N-2~, \\
\hat{P}_{17}&=&3N^6-10N^5-9N^4-29N^3-18N^2-7N-2
                   \label{resM}~.\\
%
%
 \Delta A^{Qg}_n &=&T_FC_A\Biggl\{
            \frac{1}{\ep^2}\Biggl[
                               -\frac{8(2N-3)}
                                     {N(N+1)}S_1
                               +8\frac{N^3+N-4}
                                      {N^2(N+1)^2}
                           \Biggr]
             +\frac{1}{\ep}\Biggl[
                               -\frac{16(N-1)}
                                     {N(N+1)}S_{-2}
                               +\frac{2(2N-1)}
                                     {N(N+1)}S_2 \N\\ &&
                               -\frac{2(2N-3)}
                                     {N(N+1)}S_1^2
                               +4\frac{N^3-2N^2+8N+2}
                                      {N^2(N+1)^2}S_1
                               -4\frac{2N^4+3N^3+8N^2+N+8}
                                      {N^2(N+1)^3}
                           \Biggr] \N\\ &&
                               +\frac{8(N-1)}
                                     {N(N+1)}
                                       \Bigl(
                                          2S_{-2,1}
                                         -S_{-3}
                                         -2S_{-2}S_1
                                       \Bigr)
                               +\frac{4S_{2,1}}
                                     {N(N+1)}
                               -\frac{2N-3}
                                     {3N(N+1)}
                                       \Bigl(
                                          S_1^3
                                         +6S_1\zeta_2
                                       \Bigr) \N\\ &&
                               -\frac{2(8N-9)}{3N(N+1)}S_3
                               -\frac{10N-11}
                                     {N(N+1)}S_2S_1
                               -\frac{16(N-1)}
                                     {N(N+1)^2}S_{-2}
                               +\frac{N^3-10N^2-20N-22}
                                     {N^2(N+1)^2}S_2 \N\\ &&
                               +\frac{N^3-2N^2+8N+2}
                                     {N^2(N+1)^2}S_1^2
                               +2\frac{N^3+N-4}
                                      {N^2(N+1)^2}\zeta_2
                               -2\frac{2N^4+2N^3-4N^2-35N-12}
                                      {N^2(N+1)^3}S_1 \N\\ &&
                               +\frac{2\hat{P}_{18}}
                                     {N^2(N+1)^4}
             +\ep          \Biggl[
                               \frac{4(N-1)}
                                    {N(N+1)}
                                      \Bigl(
                                        -4S_{-2,1,1}
                                        +2S_{-3,1}
                                        +2S_{-2,2}
                                        -S_{-4}
                                        +4S_{-2,1}S_1
\N\\ &&
                                        -2S_{-3}S_1 
                                        -2S_{-2}S_2
                                        -2S_{-2}S_1^2
                                        -S_{-2}\zeta_2
                                      \Bigr)
                               +\frac{8(N-1)}
                                    {N(N+1)^2}
                                      \Bigl(
                                         2S_{-2,1}
                                        -S_{-3}
                                        -2S_{-2}S_1
                                      \Bigr)
\nonumber\\ &&
                               +\frac{2(4N-3)}
                                     {N(N+1)}S_{3,1} 
                               -\frac{2N-1}
                                     {2N(N+1)}
                                      \Bigl(
                                         4S_{2,1,1}
                                        -4S_{2,1}S_1
                                        -S_2\zeta_2
                                      \Bigr)
                               -\frac{2N-3}
                                    {24N(N+1)}
                                      \Bigl(
                                         30S_4
\nonumber\\ &&
                                        +S_1^4
                                        +12S_1^2\zeta_2
                                        +16S_1\zeta_3
                                      \Bigr) \N\\ &&
                               -\frac{38N-39}
                                     {3N(N+1)}S_3S_1
                               +\frac{3(10N-7)}
                                     {8N(N+1)}S_2^2
                               -\frac{18N-19}
                                     {4N(N+1)}S_2S_1^2
                               +2\frac{N^3+2N^2+4N+2}
                                      {N^2(N+1)^2}S_{2,1} \N\\ &&
                               +\frac{N^3-38N^2-10N-34}
                                     {3N^2(N+1)^2}S_3
                               +\frac{N^3-2N^2+8N+2}
                                     {6N^2(N+1)^2}S_1^3
                               +\frac{N^3-18N^2+24N+2}
                                     {2N^2(N+1)^2}S_2S_1 \N\\ &&
                               +\frac{N^3-2N^2+8N+2}
                                     {N^2(N+1)^2}S_1\zeta_2
                               +\frac{2}{3}\frac{N^3+N-4}
                                                {N^2(N+1)^2}\zeta_3
                               +\frac{8(N^2+3)}
                                     {N(N+1)^3}S_{-2} \N\\ &&
                               -\frac{2N^4-18N^3-20N^2-3N+36}
                                     {2N^2(N+1)^3}S_2
                               -\frac{2N^4+2N^3-4N^2-35N-12}
                                     {2N^2(N+1)^3}S_1^2 \N\\ &&
                               -\frac{2N^4+3N^3+8N^2+N+8}
                                     {N^2(N+1)^3}\zeta_2
                               +\frac{\hat{P}_{19}}
                                     {N^2(N+1)^4}S_1
                               -\frac{\hat{P}_{20}}
                                     {N^2(N+1)^5}
                           \Biggr]
                   \Biggr\}~,
\end{eqnarray}
\begin{eqnarray}
\hat{P}_{18}&=&4N^5+10N^4+11N^3-16N^2-19N-24~, \\
\hat{P}_{19}&=&4N^5+8N^4-10N^2+51N+20~, \\
\hat{P}_{20}&=&8N^6+28N^5+42N^4+19N^3+24N^2+13N+40
                   \label{resN}~. \\\N
\end{eqnarray}

Furthermore, one has
\begin{eqnarray}
%
%
%
\Delta  A^{Qg}_g & = & \Delta A^{Qg}_h = \Delta A^{Qg}_i 
                   =   \Delta A^{Qg}_k =   \Delta A^{Qg}_o 
                   =   \Delta A^{Qg}_p = \Delta A^{Qg}_q 
                   =   \Delta A^{Qg}_r = \Delta A^{Qg}_{r'} \N\\
                 & = & \Delta A^{Qg}_s = \Delta A^{Qg}_t = 0~.
                         \label{res}
\end{eqnarray}
In Table~\ref{TFIG1} we show, for comparison, numerical values for some moments 
of the diagrams calculated above.

\vspace{2mm}\noindent
The non--singlet diagrams are the same as in the unpolarized case, cf.
\cite{BBK1,BBKS8a}. These sums were calculated both with the help of integral 
representations and by applying the package {\tt Sigma},~\cite{Schneider:2007a,Schneider:2013a}. 


\begin{table}[H]
\begin{center}
\includegraphics[angle=0, width=7.7cm]{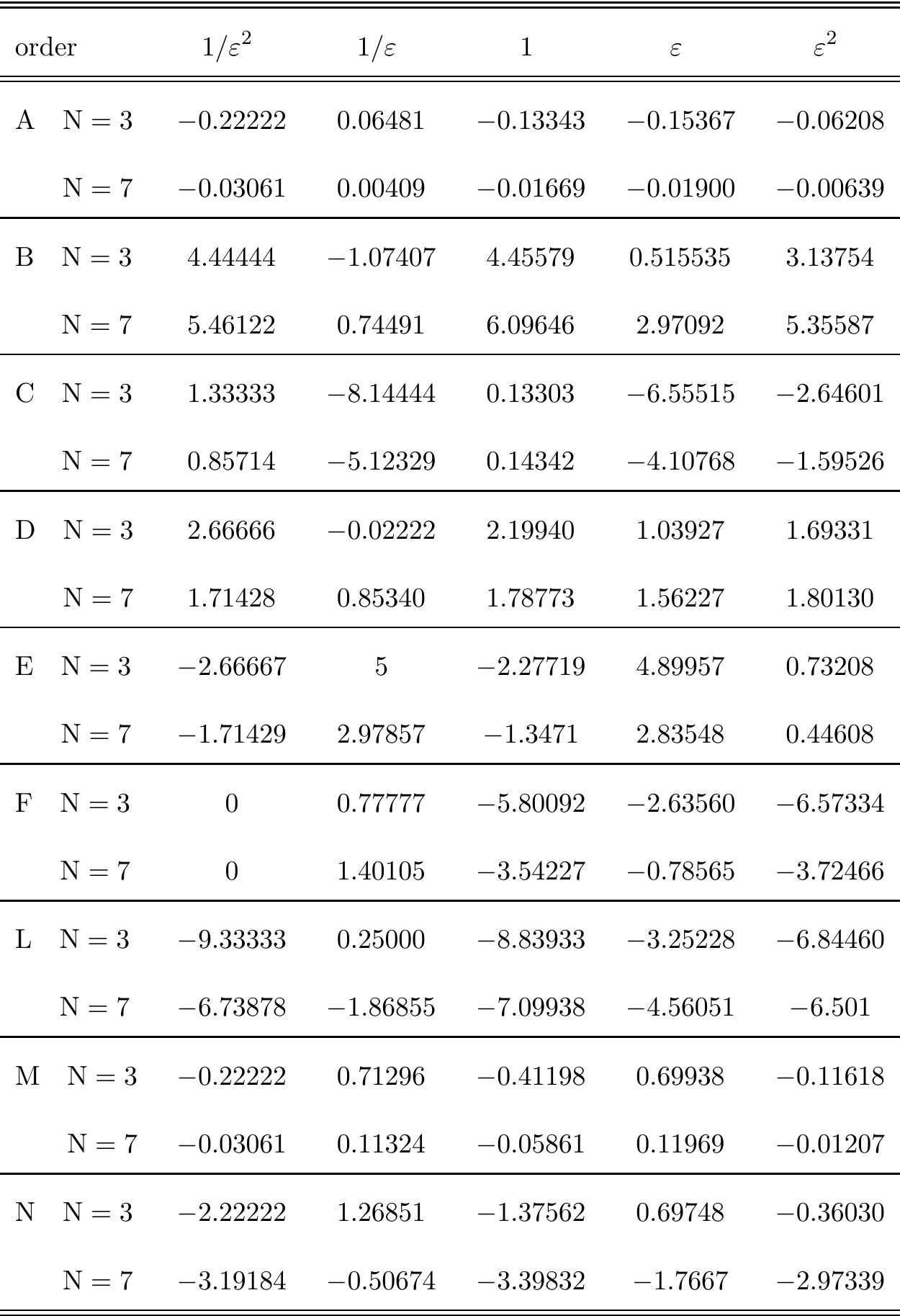}
\end{center}
\caption[]{\label{TFIG1}
\sf \small Numerical moments for $N = 3,7$ for the different contributions in the dimensional parameter
$\ep$ to the diagrams $A-N$ of Ref.~\cite{BUZA2}.} 
\end{table}

\section{Representations through generalized hypergeometric series}
\label{sec:B}

\vspace{1mm}\noindent
      In the present calculation the Feynman diagrams were evaluated without
      using the integration-by-parts method. As an example, we describe in the following
      the evaluation of  a $5$--propagator integral emerging in diagram~$f$, 
      see Figure~3, Ref.~\cite{BBK1}.
\begin{eqnarray}
       I_f(N)&=&\sum_{i=0}^{N-2}\int_0^1 \! \int_0^1 \!\frac{d^Dq~d^Dk}{(2\pi)^D(2\pi)^D}
          \frac{(\Delta q)^{i}(\Delta k)^{N-2-i}}{
          (\!q^2-\!m^2)((\!q-\!p)^2-\!m^2)(\!k^2-\!m^2)((\!k-\!p)^2-\!m^2)^2(\!k-\!q)^2}~. 
\nonumber\\
\end{eqnarray}
The diagram has a 4--dimensional Feynman parameterization over the generalized unit-cube.
After the momentum integrals are carried out one obtains 
\begin{eqnarray}
I_f(N)       &=&\frac{(\Delta p)^{N-2}\Gamma(1-\ep)}{(4\pi)^{4+\ep}(m^2)^{1-\ep}}
          \int_0^1 
          \int_0^1 \!
          \int_0^1 \!
          \int_0^1 \!
dudzdydx 
          \frac{(1-u)^{-\ep/2}z^{-\ep/2}(1-z)^{\ep/2-1}}{(1-u+uz)^{1-\ep}(x-y)}
          \N\\
        &&\Biggl[ \Bigl(zyu+x(1-zu)\Bigr)^{N-1}
                -\Bigl((1-u)x+uy\Bigr)^{N-1}\Biggr]~.
\end{eqnarray}
It is very useful to apply the following transformations of variables 
given in Ref.~\cite{HAMB}, 
\begin{eqnarray}
         x'&:=&xy~, \quad \hspace{16.5mm} y':=\frac{x(1-y)}{1-xy}~, \N\\
         x&=&x'+y'-x'y'~,\quad y=\frac{x'}{y'+x'-x'y'}~,\N\\
         &&\frac{\partial(x,y)}{\partial(x',y')}=\frac{1-x'}{x'+y'-x'y'}~,
         \label{trafo1}
\end{eqnarray}
which yields
\begin{eqnarray}
        \int_0^1 \! \int_0^1 \! dx dy f(x,y)(xy)^N =
        \int_0^1 \! \int_0^1 \! dx' dy'\frac{(1-x')(x')^N}{x'+y'-x'y'}
                                   f\Bigl(y'+x'-x'y',
                                  \frac{x'}{x'+y'-x'y'}\Bigr)~.
\end{eqnarray}
Similarly, terms of the form $(x-y)^N$ can be combined using  
       \begin{alignat}{3}
          \underline{x>y}:&                & \qquad     \underline{x<y}:&    \N\\
                   x'&:=x-y~,             & \qquad               x'&:=y-x~,    \N\\
                   y'&:=\frac{y}{1-x+y}~, & \qquad               y'&:=\frac{1-y}{1+x-y}~, \N\\
                    x&=x'+y'-x'y'~,        & \qquad                x&=(1-x')(1-y')~,\N\\
                    y&=(1-x')y'~,         & \qquad                y&=1-(1-x')y'~,\N\\
         \frac{\partial(x,y)}{\partial(x',y')}&=1-x'~. 
      & \qquad   \frac{\partial(x,y)}{\partial(x',y')}&=1-x'~, \label{trafo2}
       \end{alignat} \\ 
leading to, cf.~Ref.~\cite{HAMB},
       \begin{eqnarray}
        \int_0^1 \! \int_0^1 \! dx dy f(x,y)(x-y)^N\! &=&\!
        \int_0^1 \! \int_0^1 \! dx' dy'{x'}^N(1-x')\Bigl(
                                f(y'+x'-x'y',(1-x')y')\N\\
                                &&+(-1)^Nf((1-y')(1-x'),1-(1-x')y')\Bigr)~.
       \end{eqnarray}
The substitution (\ref{trafo1})  
$u':=uz$ and shifting
$z' \rightarrow 1-z~, u'\rightarrow 1-u$  afterwards,  yields
     \begin{eqnarray}
      I_f(N) &=& \frac{(\Delta p)^{N-2}\Gamma(1-\ep)}{(4\pi)^{4+\ep}(m^2)^{1-\ep}}
         \int_0^1 \!
         \int_0^1 \!
         \int_0^1 \!
         \int_0^1 \! dudzdydx 
         \frac{(1-u)^{-\ep/2}z^{-\ep/2}(1-z)^{\ep/2-1}}{(1-u+uz)^{1-\ep}(x-y)} 
         \N\\
       &&\Biggl[ \Bigl((1-u)y+ux\Bigr)^{N-1}
               -\Bigl((1-u)x+uy\Bigr)^{N-1}\Biggr]~.
     \end{eqnarray}
Now transformation (\ref{trafo2}) is used 
by setting $x':=\pm(x-y)$. Thus 
     \begin{eqnarray}
      I_f(N) &=& \frac{(\Delta p)^{N-2}\Gamma(1-\ep)}{(4\pi)^{4+\ep}(m^2)^{1-\ep}}
         \int_0^1 \!
         \int_0^1 \!
         \int_0^1 \!
         \int_0^1 \! dudzdy'dx' 
         \frac{(1-u)^{-\ep/2}z^{-\ep/2}(1-z)^{\ep/2-1}}{(1-u+uz)^{1-\ep}}
         \frac{1-x'}{x'}
         \N\\
       &&\Biggl[ \Bigl(y'(1-x')+ux'\Bigr)^{N-1}
                -\Bigl(y'(1-x')+x'(1-u)\Bigr)^{N-1}
                -\Bigl(1-ux'-y'(1-x')\Bigr)^{N-1} \N\\
       &&       +\Bigl(1-x'+ux'-y'(1-x')\Bigr)^{N-1}\Biggr]~.
     \end{eqnarray}
This form allows to perform the $y'$-integration. Further we set $x=x'$, 
$u\rightarrow 1-u$, giving
     \begin{eqnarray}
I_f(N)
     &=&\frac{2(\Delta p)^{N-2}\Gamma(1-\ep)}{(4\pi)^{4+\ep}(m^2)^{1-\ep}N}
        \int_0^1 \! \int_0^1 \! \int_0^1 \! dudzdx 
        \frac{u^{-\ep/2}z^{-\ep/2}(1-z)^{\ep/2-1}}{(u+z-uz)^{1-\ep}x}
        \Biggl[ x^{N}u^N-x^N(1-u)^N\N\\
      &&    +(1-ux)^N-(1-x(1-u))^N
         \Biggr]~\N\\
     &=&\frac{2(\Delta p)^{N-2}\Gamma(1-\ep)}{(4\pi)^{4+\ep}(m^2)^{1-\ep}N}
        \Biggl[
        \frac{1}{N}\int_0^1 \! \int_0^1 \! dudz 
        \frac{u^{-\ep/2}z^{-\ep/2}(1-z)^{\ep/2-1}}{(u+z-uz)^{1-\ep}}
        \Bigl[u^N-(1-u)^N\Bigr]~\N\\
      &&+\sum_{i=1}^N\binom{N}{i}\frac{(-1)^i}{i}
        \frac{u^{-\ep/2}z^{-\ep/2}(1-z)^{\ep/2-1}}{(u+z-uz)^{1-\ep}}
        \Bigl[u^i-(1-u)^i \Bigr]~
         \Biggr] \N\\
     &=&\frac{2(\Delta p)^{N-2}\Gamma(1-\ep)}{(4\pi)^{4+\ep}(m^2)^{1-\ep}N}
        \sum_{i=1}^N\Biggl\{\binom{N}{i}(-1)^i+\delta_{i,N}\Biggr\}
        \frac{1}{i}\int_0^1 \! \int_0^1 \! dudz 
        \frac{u^{-\ep/2}z^{-\ep/2}(1-z)^{\ep/2-1}}{(u+z-uz)^{1-\ep}}\N\\
      &&  \times \Bigl[u^i-(1-u)^i \Bigr]~.
     \end{eqnarray}
Here also the $x$--integral was carried out.
The latter expression can now be rewritten in terms of a generalized
hypergeometric series \cite{GHF} by applying 
\begin{eqnarray}
I_1 &=& \int_0^1 \! \int_0^1 \! dy dw (1-w)^a w^b (1-y)^c y^d [1-y(1-w)]^{-e} \nonumber\\ 
    &=& B(d+1,c+1) \sum_{k=0}^{\infty} \frac{(e)_k (d+1)_k}{k! (2+d+c)_k} \int_0^1 \! dw (1-w)^{a+k} w^b
        \nonumber\\
    &=& B(d+1,c+1) B(a+1,b+1) _3F_2 \Biggl[\begin{array}{c} e, d+1, a+1 \\ 
                                                             2+ d +c, 2 + a + b  \end{array} ;1\Biggr]~.
\end{eqnarray}
Here $(a)_k = \prod_{l=0}^k (a+l)$ denotes the Pochhammer--Appell symbol and $B(a,b)$
is Euler's Beta-function.
One thus obtains
     \begin{eqnarray}
I_f(N)     &=&\frac{S_{\varepsilon}^2(\Delta p)^{N-2}}{(4\pi)^4(m^2)^{1-\varepsilon}}
        \exp\Biggl\{\sum_{l=2}^{\infty}\frac{\zeta_l}{l}\varepsilon^l\Biggr\}
        \frac{2\pi}{N\sin(\frac{\pi}{2}\varepsilon)}
        \sum_{j=1}^{N}\Biggl\{\binom{N}{j}(-1)^j+\delta_{j,N}\Biggr\}
        \nonumber \\
     && \times\Biggl\{
        \frac{\Gamma(j)\Gamma(j+1-\frac{\varepsilon}{2})}
        {\Gamma(j+2-\varepsilon)\Gamma(j+1+\frac{\varepsilon}{2})}
       -\frac{B(1-\frac{\varepsilon}{2},1+j)}{j}~ 
        _3F_2 \Biggl[ \begin{array}{c} 1-\varepsilon,\frac{\varepsilon}{2},j+1 \\ 1,j+2
        -\frac{\varepsilon}{2} \end{array} ;1\Biggr]
        \Biggr\}. \label{ffinal}
\nonumber\\
     \end{eqnarray}
     Note that although (\ref{ffinal}) is a double sum, the 
     summation parameters and the variable $N$ are not nested.
     This expression can be expanded in $\ep$ and calculated
     using the sums given in Appendix~B of Ref.~\cite{BBK1}.

     The same kind of transformation was performed to 
     obtain a result for the $5$--propagator integral of diagram $n$. Although
     a little more work is needed, it could be treated in a 
     quite similar manner as diagram $f$. One of the most important aspects 
     is to write all sums which have to be introduced in such a way
     that there is {\sf no nesting} of summation indexes with $N$. 
     Note that in the case of only $3$ massive propagators, analytic
     results for fixed values of $N$ can be obtained quite easily 
     by choosing the momentum flow in such a way that one momentum 
     follows the massive propagators. Thus no denominator structure 
     emerges in the parameter integral. One obtains
     \begin{eqnarray}
      I_n(N) &=&\sum_{j=0}^{N-2}\int\!\!\frac{d^Dq}{(2\pi)^{D}}\int\!\!\frac{d^Dk}{(2\pi)^D}
            \frac{(\Delta q)^{j}(\Delta q -\Delta k)^{N-2-j}}
            {(q^2\!-\!m^2)((q\!-\!p)^2\!-\!m^2)((k-q)^2\!-\!m^2)
            (\!k-\!p)^2 k^2}
           \nonumber \\  
      &=&\frac{\Gamma(1-\ep)(\Delta p)^{N-2}}{(4\pi)^{4+\ep}(m^2)^{1-\ep}}
         \int_0^1 \! \int_0^1 \! \int_0^1 \! \int_0^1 \! dwdydvdz \sum_{i=0}^{N-2}
         (1-w)w^{\ep/2-1}(1-y)^{1-\ep/2}y^{-\ep/2}(1-y)^j \N\\ 
       && \times (1-w)^j(z-v)^j(y((1-w)v+wz)+(1-y)z)^{N-j-2}~.\label{in8}
     \end{eqnarray}
     Calculating (\ref{in8}) for arbitrary values of $N$ analytically involves 
     some work, while for fixed values of 
     $N$, (\ref{in8}) decomposes into a finite sum of Beta-functions,
     which can be handled by {\tt MAPLE}. The general $N$ expression reads
%
%

 \begin{eqnarray}
  I_n&=&\int\!\!\frac{dq}{(2\pi)^{D}}\int\!\!\frac{dk}{(2\pi)^D}
       \frac{(\Delta q)^{N-1}\!-\!(\Delta k)^{N-1}}{(\Delta q\!-\!\Delta k)}
       \frac{1}{(q^2\!-\!m^2)((q\!-\!p)^2\!-\!m^2)(k^2\!-\!m^2)
       (q\!-\!k-\!p)^2\!(k\!-\!q)^2}
       \nonumber \\
     &=&\frac{(\Delta p)^{N-2}\Gamma(1-\ep)}{(4\pi)^{4+\ep}(m^2)^{1-\ep}}
        \Biggl\{
        B(N+1-\ep/2,1-\ep/2)\Biggl(
             \frac{(-1)^N-1}{N^2}B(N+1,\ep/2) \N\\
           && -\frac{2(N+1)+\ep}{N(N+1)\ep}B(N+2,\ep/2)
            +(-1)^N\frac{B(N+1,\ep/2+1)}{N(N+1)}
            ~_3F_2\Biggr[\begin{array}{c} \ep/2+1,N+1,1 \\ N+2,N+2+\ep/2 \end{array};1
                         \Biggr] 
                            \Biggr) \N\\
       &&-\sum_{l=1}^{N-1}\binom{N-1}{l}(-1)^lB(l+1-\ep/2,2-\ep/2)
           \sum_{k=0}^{l-1}(-1)^k\frac{B(k+2,\ep/2)B(k+2,N-1-k)}{k+1} \N
\\
   &&-\frac{1}{N}\sum_{l=1}^{N-1}\binom{N-1}{l}(-1)^lB(l+1-\ep/2,2-\ep/2)
           \sum_{k=0}^{l-1}\frac{B(k+2,\ep/2)}{k+1} \N
\\
       &&+\Biggl[\sum_{l=0}^{N-1}\binom{N-1}{l}(-1)^l
                B(l+1-\ep/2,2-\ep/2)\Biggr] \N\\
       &&  \times
           \int_0^1 \int_0^1 dxdv~(1-x)^{\ep/2-1}v^{N-1}
           \left(\ln(v(1-x)+x)-\ln(v(1-x))\right) 
        \Biggr\} \label{1} \\
     &=&\frac{S_{\ep}^2(\Delta p)^{N-2}}{(4\pi)^4(m^2)^{1-\ep}}
        \exp\Biggl\{\sum_{l=2}^{\infty}\frac{\zeta_l}{l}\varepsilon^l\Biggr\}
        \frac{1}{N(N+1)}\Biggl\{
          \frac{1}{\ep} \Biggl[
                        4S_1(N)
                        +2\frac{(-1)^N-1}{N}
                         \Biggr] \N\\
      &&                +2S_{-2}(N)-S_2(N)
                        +S^2_1(N)
                        +\frac {2(3N+1)S_1(N)}{N(N+1)}
                        +2\frac{(-1)^N-1}{N(N+1)}
+ O(\ep)
                        \Biggr\} 
\label{2}~,
 \end{eqnarray}
where
 \begin{eqnarray}
  &&\int_0^1 \int_0^1
dxdv~(1-x)^{\ep/2-1}v^{N-1}\left(\ln(v(1-x)+x)-\ln(v(1-x))\right)\N\\
  &=&\frac{4}{\ep^2 N}+\frac{2}{\ep N^2}
     -\frac{B(1+\ep/2,1)}{N(N+1)}~
       _3F_2\Biggl[\begin{array}{c} 1,1+\ep/2,1 \\ N+2,2+\ep/2 \end{array};1\Biggr]~.
       \label{4} 
 \end{eqnarray}

\section{\boldmath  \hspace*{-5mm} The heavy quark $O(a_s^2)$ Wilson 
coefficients for \boldmath{$Q^2 \gg m^2$}}
\label{sec:C}

\vspace{1mm}\noindent
In the following we give the representation of the heavy quark two--loop polarized Wilson 
coefficients for $Q^2 \gg m^2$ in $N$-- and $z$ space and correct some errors in 
\cite{BUZA2}.
Furthermore, for the case of the inclusive heavy flavor corrections, further conceptual changes
 w.r.t. \cite{BUZA2} are necessary. The structure of the Wilson coefficients has been given in 
Eqs.~(\ref{eqH2g}--\ref{eqL2g}). In Ref.~\cite{BUZA2} the Wilson coefficient $\Delta 
L_{g,g_1}^{(2)}$ has not been considered.

In the following we set both the factorization and renormalization scales $\mu^2 = Q^2$. Thus
the asymptotic heavy flavor Wilson coefficients depend on the logarithms $\ln(Q^2/m_Q^2)$ 
only, with $m_Q = m_c$ or $m_b$.

The massive asymptotic polarized flavor non--singlet Wilson coefficient \cite{BUZA2}, Eq.~(B.4), 
reads
\begin{eqnarray}
\label{eq:NSWVN}
\Delta L^{{\rm NS}, (2), {\sf [1]}}\left(z, \frac{Q^2}{m^2}\right)
&=&
T_F C_F \Biggl\{ 
\frac{4}{3}\, \frac{1+{z}^{2}}{1-z} \ln^2\left(\frac{Q^2}{m^2}\right)
\nonumber\\ & & 
+
\Biggl[  \frac{1+{z}^{2}}{1-z}  \Biggl( 
\frac{8}{3} \ln(1-z)
-\frac{16}{3} \,\ln  \left( z \right) -{
\frac {58}{9}} \Biggr) 
-2+6\,z  \Biggr] \ln\left(\frac{Q^2}{m^2}\right)
\nonumber\\ & & 
+ 
 \frac{ 1+{z}^{2}}{1-z}  \Biggl[ -\frac{8}{3}\, \Li_2\left(1-z 
\right) 
-\frac{8}{3}\,\zeta_2 -
\frac{16}{3}\,\ln ( z ) \ln ( 1-z ) +\frac{4}{3}\, 
\ln^2( 1-z ) 
\nonumber\\ & &
+ 4\, \ln^2( z )
- \frac {58}{9}\,\ln ( 1-z) + \frac 
{134}{9}\,
\ln  ( z ) +\frac {359}{27} \Biggr] - \left( 2-6\,z \right) \ln
 \left( 1-z \right) 
\nonumber\\ & &
+ \left( \frac{10}{3}-10\,z \right) \ln  \left( z \right) 
+{\frac {19}{3}}-19\,z \Biggr\}.
\end{eqnarray}
In Ref.~\cite{BUZA2} the higher functions are expressed in terms of
polylogarithms \cite{LEWIN} and Nielsen-integrals \cite{NIELSEN}. They obey the following integral representations 
\begin{eqnarray}
{\rm S}_{n,p}(x) &=& \frac{(-1)^{n+p-1}}{(n-1)! p!} \int_0^1 \frac{dz}{z} \ln^{n-1}(z) \ln^p(1-xz)
\\
\Li_{n}(x) &=& {\rm S}_{n-1,1}(x)~.
\end{eqnarray}
In Eq.~(\ref{eq:NSWVN})  the variable $z$ obeys $z \in [0, Q^2/(m^2+Q^2) < 1]$, 
since only real heavy quark production has been considered in \cite{BUZA2}. Approaching the region 
$z \sim 1$ the $+$-prescription has to be used, and $z \in [0,1]$. 
Furthermore, a soft and virtual term has to be added, cf.~\cite{BUZA2}. This is not the complete 
result, however, since a term containing other virtual initial state corrections with massless 
quark final 
states is yet missing \cite{Behring:2015zaa,Blumlein:2016xcy} and the foregoing result still
violates the polarized Bjorken sum rule, since there is a logarithmic correction in the limit $Q^2 
\gg m^2$. To obtain the correct expression one has to consider the inclusive heavy flavor Wilson 
coefficient for a structure function, cf.~\cite{Behring:2015zaa,Blumlein:2016xcy}.\footnote{It is 
very well possible that different analysis programmes still refer to the results of Ref.~\cite{BUZA2}, 
which has to be corrected.}  In particular, its first moment does not yield the correct result 
for the Bjorken sum rule. Relations of this kind are used in the tagged flavor case, see also \cite{Hekhorn:2018ywm}.
They do not apply to the structure functions. The polarized flavor non--singlet Wilson coefficient is given in the 
$\overline{\sf MS}$ scheme due to the Ward--Takahashi \cite{WT} identity, which can be used since the 
local operator is located on the massless fermion line. 
\begin{eqnarray}
\label{eq:HNScorr}
\Delta L^{{\rm NS}, (2), \overline{\sf MS}}\left(N, \frac{Q^2}{m^2}\right)
&=& \textcolor{black}{C_F T_F} \Biggl\{
        -\frac{4}{3} \ln^2\left(\frac{Q^2}{m^2}\right)
\Bigl(-\frac{2+3 N+3 N^2}{2 N (1+N)}
                +2 S_1
        \Bigr)
        + \Biggl(
                \frac{2 T_1}{9 N^2 (1+N)^2}
\nonumber\\ && 
                +\frac{80}{9} S_1
                -\frac{16}{3} S_2
        \Biggr) \ln\left(\frac{Q^2}{m^2}\right)
        +\frac{T_4}{27 N^3 (1+N)^3}
        +\Biggl(
                -\frac{2 T_3}{27 N^2 (1+N)^2}
\nonumber\\ &&
  +\frac{8}{3} S_2
        \Biggr) S_1
        -\frac{2 \big(
                -6+29 N+29 N^2\big)}{9 N (1+N)} S_1^2
        +\frac{2 \big(
                -2+35 N+35 N^2\big)}{3 N (1+N)} S_2
\nonumber\\ &&
        -\frac{8}{9} S_1^3
        -\frac{112}{9} S_3
        +\frac{16}{3} S_{2,1} \Biggr\},
\end{eqnarray}
with 
\begin{eqnarray}
T_1 &=&-3 N^4-6 N^3-47 N^2-20 N+12,
\\
T_2 &=& 57 N^4+96 N^3+65 N^2-10 N-24,
\\
T_3 &=& 359 N^4+772 N^3+335 N^2+30 N+72,
\\
T_4 &=& 795 N^6+2043 N^5+2075 N^4+517 N^3-298 N^2+156 N+216.
\end{eqnarray}
The first moment of (\ref{eq:HNScorr}) yields $8 a_s^2 \textcolor{black}{C_F T_F}$ for $\mu^2 = Q^2$ in accordance 
with the polarized Bjorken sum rule \cite{Bjorken:1969mm} for a single quark
flavor, shifting $N_F \rightarrow N_F +1$ in the limit $Q^2 \gg m^2$. For a detailed discussion
of the finite mass effects see Ref.~\cite{Blumlein:2016xcy}.

The corresponding expression in $z$ space are represented using 
harmonic polylogarithms \cite{Remiddi:1999ew}, which are 
the iterative integrals over the alphabet
\begin{eqnarray}
\mathfrak{A} = \left\{f_0 = \frac{1}{z}, f_1 = \frac{1}{1-z}, f_{-1} = \frac{1}{1+z} \right\},
\end{eqnarray}
and are given by
\begin{eqnarray}
H_{b,\vec{a}}(z) = \int_0^z \frac{dy}{y} f_b(y) H_{\vec{a}}(y),~~~~H_\emptyset = 1,~~~b, a_i \in 
\{0,1,-1\}.
\end{eqnarray}
In the following we use the shorthand notation $H_{\vec{a}}(z) \equiv H_{\vec{a}}$.

One obtains
\begin{eqnarray}
\label{eq:NS1}
\Delta L^{{\rm NS}, (2), \overline{\sf MS}}\left(z, \frac{Q^2}{m^2}\right)
&=& \textcolor{black}{C_F T_F} \Biggl\{
\Biggl[
\frac{1}{1-z} \Biggl[
        \frac{8}{3} \ln^2\left(\frac{Q^2}{m^2}\right)
        + \frac{718}{27}
        +\frac{16}{9} \big(
                5+3 \HA_0\big) \ln\left(\frac{Q^2}{m^2}\right)
\nonumber\\ &&
        +\frac{268}{9} H_0
        +8 H_0^2
        +\frac{116}{9} H_1
        +\frac{16}{3} H_0 H_1
        +\frac{8}{3} H_1^2
        +\frac{16}{3} H_{0,1}
        -\frac{32}{3} \zeta_2
\Biggr]
\Biggr]_+
\nonumber\\ &&
+ \Biggl(
        2 \ln^2\left(\frac{Q^2}{m^2}\right)  +\frac{2}{3} \ln\left(\frac{Q^2}{m^2}\right)
+        \frac{265}{9}
\Biggr) 
\delta(1-z)
\nonumber\\ &&
        -\frac{4}{3} (1+z) \ln^2\left(\frac{Q^2}{m^2}\right)
        -\frac{4}{27} (47+218 z)
        + \Biggl(
                -\frac{8}{9} (11 z-1)
\nonumber\\ &&
                -\frac{8}{3} (1+z) H_0
        \Biggr) \ln\left(\frac{Q^2}{m^2}\right)
        -\frac{8}{9} (13+28 z) H_0
        -4 (1+z) H_0^2
\nonumber\\ &&
        +\Biggl(
                -\frac{8}{9} (5+14 z)
                -\frac{8}{3} (1+z) H_0
        \Biggr) H_1
        -\frac{4}{3} (1+z) \left[H_1^2
        +2 H_{0,1}
        -4 \zeta_2 \right]
\Biggr\}.
\nonumber\\
\end{eqnarray}
The corresponding expression in the Larin scheme is given in \cite{LOGPOL}.
Here the $+$-distribution is defined by
\begin{eqnarray}
\int_0^1 dz [f(z)]_+ g(z) = \int_0^1 dz [f(z)-f(1)]g(z)~.
\end{eqnarray}

The Mellin convolution of the different contributions in (\ref{eq:NS1}) are defined by~\cite{Blumlein:1989gk}
\begin{eqnarray}
\left(\frac{f(z)}{1-z}\right)_+ \otimes g(z) &=& \int_z^1 dy \frac{f(y)}{1-y} \left[ 
\frac{1}{y} g\left(\frac{z}{y}\right) - g(z) \right] - g(z) \int_0^z dy \frac{f(y)}{1-y}
\\
\delta(1-z) \otimes g(z) &=& g(z)
\\
f(z) \otimes g(z) &=& \int_z^1 \frac{dy}{y} f\left(\frac{z}{y}\right) g(y).
\end{eqnarray}

In the pure singlet case one obtains in Mellin space for the massive Wilson coefficient in the
Larin scheme
\begin{eqnarray}
\Delta H^{{\rm PS}, (2), L}\left(N, \frac{Q^2}{m^2}\right)
&=& 
\textcolor{black}{C_F T_F} \Biggl\{
        -\frac{4 (N-1) (2+N)}{N^2 (1+N)^2} \ln^2\left(\frac{Q^2}{m^2}\right)
        -\frac{8 (2+N) \big(
                1+2 N+N^3\big)}{N^3 (1+N)^3} 
\nonumber\\ && \times
\ln\left(\frac{Q^2}{m^2}\right)
        +\frac{8 T_5}{(N-1) N^4 (1+N)^4 (2+N)}
        +\frac{4 (N-1) (2+N)}{N^2 (1+N)^2} 
\nonumber\\ &&  
\times [S_1^2 - 3 S_2]
        +\frac{8 (2+N) \big(
              2+N-N^2+2 N^3\big) }{N^3 (1+N)^3} S_1
\nonumber\\ &&  
        -\frac{64}{(N-1) N (1+N) (2+N)} S_{-2}
\Biggr\},
\\
T_5 &=& 3 N^8+10 N^7-N^6-22 N^5-14 N^4-18 N^3-30 N^2+8.
\end{eqnarray}
The corresponding result in $z$ space reads
\begin{eqnarray}
\Delta H^{{\rm PS}, (2), L}\left(z, \frac{Q^2}{m^2}\right)
&=&
\textcolor{black}{C_F T_F} \Biggl\{ 
        - \left[
                20 (1-z)
                + 8 (1+z) H_0
        \right] \ln^2\left(\frac{Q^2}{m^2}\right)
        + 8 [
                (1-z)
                - (1- 3 z) H_0
\nonumber\\ &&        
        - (1+z) H_0^2
        ] \ln\left(\frac{Q^2}{m^2}\right)
        +\frac{592}{3} (1-z)
        +\Biggl(
                \frac{256}{3} (2-z)
                -32 (1+z) \zeta_2
        \Biggr) H_0
\nonumber\\ && 
        -\frac{32 (1+z)^3}{3z} H_{-1} H_0
        +\frac{8}{3} \big(
                21+2 z^2\big) H_0^2
        +\frac{16}{3} (1+z) H_0^3
        + 8 (1-z) 
\nonumber\\ && \times 
(11 + 10 H_0) H_1
        +20 (1-z) H_1^2
        -16 \big[
                (1-3 z)
                -2 (1+z) H_0
        \big] H_{0,1}
\nonumber\\ && 
        +\frac{32 (1+z)^3}{3 z} H_{0,-1}
        -16(1+z) ( 2 H_{0,0,1}
        - (1+z) H_{0,1,1})
\nonumber\\ &&
        -\frac{32}{3} \big(
                9-3 z+z^2\big) \zeta_2
        +16 (1+z) \zeta_3
\Biggr\},
\end{eqnarray}
which agrees with  Eq.~(B.3) of Ref.~\cite{BUZA2}.
For the pure singlet Wilson coefficient $H_{Qq}^{\rm PS,(2)}$  there is no finite transformation to 
the {\sf M} scheme at $O(a_s^2)$ since the correction to the massive OME and the massless Wilson
coefficient compensate each other.

The gluonic Wilson coefficient $\Delta H_{Qg}^{{\rm S},(2)}$ is given by
\begin{eqnarray}
\label{eq:HQgN}
\lefteqn{
\Delta H_{Qg}^{{\rm S},(2)}\left(N, \frac{Q^2}{m^2}\right) =}
\nonumber\\
&& \Biggl\{
        -
        \textcolor{black}{T_F^2} \frac{16 (N-1)}{3 N (1+N)}
        +\textcolor{black}{C_F T_F} \Biggl[
                \frac{2 (N-1) \big(
                        2+3 N+3 N^2\big)}{N^2 (1+N)^2}
                -\frac{8 (N-1) S_1}{N (1+N)}
        \Biggr]
\nonumber\\ && 
        +\textcolor{black}{C_A T_F} \Biggl[
                -\frac{16 (N-1)}{N^2 (1+N)^2}
                +\frac{8 (N-1) S_1}{N (1+N)}
        \Biggr]
\Biggr\} \ln^2\left(\frac{Q^2}{m^2}\right)
+
\Biggl\{
        \textcolor{black}{C_F T_F} \Biggl[
                -\frac{4 (N-1) T_8}{N^3 (1+N)^3}
\nonumber\\ &&                
 +\frac{(N-1) \big(
                        16 S_1^2
                        -16 S_2
                \big)}{N (1+N)}
                +\frac{4 (N-1) \big(
                        -6-N+3 N^2\big) S_1}{N^2 (1+N)^2}
        \Biggr] 
+        \textcolor{black}{T_F^2} \Biggl[
                \frac{16 (N-1)^2}{3 N^2 (1+N)}
\nonumber\\ &&                
 +\frac{16 (N-1) S_1}{3 N (1+N)}
        \Biggr]
       +  \textcolor{black}{C_A T_F} \Biggl[
                \frac{8 T_{11}}{N^3 (1+N)^3}
                -\frac{(N-1) \big(
                        8 S_1^2
                        +8 S_2
                        +16 S_{-2}
                \big)}{N (1+N)}
\nonumber\\ &&             
    +\frac{32 S_1}{N (1+N)^2}
        \Biggr]
\Biggr\} \ln\left(\frac{Q^2}{m^2}\right)
+\textcolor{black}{C_A T_F} \Biggl[
        \frac{4 S_2 T_7}{N^2 (1+N)^2 (2+N)}
        -\frac{16 S_1 T_{13}}{N^3 (1+N)^3 (2+N)}
\nonumber\\ &&     
    -\frac{8 T_{14}}{(N-1) N^4 (1+N)^4 (2+N)^2}
        +\frac{(N-1) \big(
                32 S_1 S_2
                -16 S_{2,1}
        \big)}{N (1+N)}
\nonumber\\ &&        
 -\frac{4 \big(
                12-N+N^2+2 N^3\big) S_1^2}{N^2 (1+N) (2+N)}
        +\frac{8 \big(
                -2+3 N+3 N^2\big) S_3}{N (1+N) (2+N)}
\nonumber\\ &&         
+\Biggl(
                \frac{16 T_6}{(N-1) N (1+N) (2+N)^2}
                +\frac{32 S_1}{2+N}
        \Biggr) S_{-2}
        -\frac{8 (N-2) (3+N) S_{-3}}{N (1+N) (2+N)}
\nonumber\\ &&         
        -\frac{16 \big(
                2+N+N^2\big) S_{-2,1}}{N (1+N) (2+N)}
-\frac{24 \big(
                2+N+N^2\big)}{N (1+N) (2+N)} \zeta_3
\Biggr]
\nonumber\\ && 
+\textcolor{black}{C_F T_F} \Biggl[
        -\frac{2 S_1^2 T_9}{N^2 (1+N)^2 (2+N)}
        +\frac{2 S_2 T_{10}}{N^2 (1+N)^2 (2+N)}
        +\frac{4 S_1 T_{12}}{N^3 (1+N)^3 (2+N)}
\nonumber\\ &&         
+\frac{4 T_{15}}{(N-1) N^4 (1+N)^4 (2+N)^2}
        +\frac{N-1}{N (1+N)} \big(
                -8 S_1^3
                +8 S_1 S_2
                +16 S_{2,1}
        \big)
\nonumber\\ && 
        -\frac{16 \big(
                2+N+N^2\big) S_3}{N (1+N) (2+N)}
        +\Biggl(
                \frac{16 \big(
                        10+N+N^2\big)}{(N-1) (2+N)^2}
                -\frac{128 S_1}{N (1+N) (2+N)}
        \Biggr) S_{-2}
\nonumber\\ &&         
-\frac{64 S_{-3}}{N (1+N) (2+N)}
        +\frac{128 S_{-2,1}}{N (1+N) (2+N)}
        +\frac{48 \big(
                2+N+N^2\big)}{N (1+N) (2+N)} \zeta_3
\Biggr],
\end{eqnarray}
with
\begin{eqnarray}
T_6 &=& N^4+3 N^3-4 N^2-8 N-4,
\\
T_7 &=& 2 N^4-N^3-24 N^2-17 N+28,
\\
T_8 &=& 4 N^4+5 N^3+3 N^2-4 N-4,
\\
T_9 &=& 9 N^4+6 N^3-35 N^2-16 N+20,
\\
T_{10} &=& 11 N^4+42 N^3+47 N^2+32 N+12,
\\
T_{11} &=& N^5+N^4-4 N^3+3 N^2-7 N-2,
\\
T_{12} &=& 2 N^6+5 N^5-22 N^4-95 N^3-114 N^2-24 N+16,
\\
T_{13} &=& 2 N^6+5 N^5-3 N^4-7 N^3+2 N^2-11 N-8,
\\
T_{14} &=& N^{10}+3 N^9-15 N^8-56 N^7-8 N^6+90 N^5+60 N^4+67 N^3+86 N^2
\nonumber\\ &&
        -12 N-24,
\\
T_{15} &=& 5 N^{10}+23 N^9+31 N^8-N^7+54 N^6+268 N^5+342 N^4+98 
N^3-60 N^2
\nonumber\\ &&
-8 N+16.
\end{eqnarray}
The rightmost pole in (\ref{eq:HQgN}) is located at $N=1$, as expected.
In $z$ space it reads
\begin{eqnarray}
\label{HQqz}
\lefteqn{
\Delta H_{Qg}^{{\rm S},(2)}\left(z, \frac{Q^2}{m^2}\right) =}
\nonumber\\
&&
\Biggl\{
        -\textcolor{blue}{T_F^2} \frac{16}{3} (-1+2 z)
        +\textcolor{blue}{C_A T_F}\big[
                48 (-1+z)
                -16 (1+z) H_0
                +8 (-1+2 z) H_1
        \big]
\nonumber\\ && 
        +\textcolor{blue}{C_F T_F} \big(
                6
                +(-1+2 z) \big(
                        -4 H_0
                        -8 H_1
                \big)
\Biggr\} \ln\left(\frac{Q^2}{m^2}\right)
+ \Biggl\{
        \textcolor{blue}{T_F^2} \Biggl[
                \frac{16}{3} (-3+4 z)
\nonumber\\ &&                
 +(-1+2 z) \Biggl(
                        \frac{16}{3} H_0
                        +\frac{16}{3} H_1
                \Biggr)
        \Biggr]
        +\textcolor{blue}{C_A T_F} \Biggl[
                -8 (-12+11 z)
                +(1+2 z) 
\nonumber\\ &&  \times
\big(
                        -16 H_{-1} H_0
                        -8 H_0^2
                        +16 H_{0,-1}
                \big)
                +8 (1+8 z) H_0
                -32 (-1+z) H_1
\nonumber\\ &&                
 -8 (-1+2 z) H_1^2
                -16 \zeta_2
        \Biggr]
        +\textcolor{blue}{C_F T_F} \Biggl[
                4 (-17+13 z)
                +(-1+2 z) \big(
                        8 H_0^2
                        +32 H_0 H_1
\nonumber\\ &&                
         +16 H_1^2
                        -8 H_{0,1}
                \big)
                +16 (-3+2 z) H_0
                +4 (-17+20 z) H_1
                -24 (-1+2 z) \zeta_2
        \Biggr]
\Biggr\} 
\nonumber\\ &&  \times
\ln\left(\frac{Q^2}{m^2}\right)
+\textcolor{blue}{C_A T_F} \Biggl\{
        -\frac{8}{3} (-101+104 z)
        +(1+2 z) \big(
                16 H_{-1} H_{0,1}
                -16 H_{0,1,-1}
\nonumber\\ &&               
 -16 H_{0,-1,1}
        \big)
        +(-1+2 z) \big(
                -16 H_0 H_1^2
                +16 H_1 H_{0,1}
        \big)
        +\frac{4}{3} \big(
                194-163 z+6 z^2\big) H_0
\nonumber\\ &&      
   -\frac{16 \big(
                2+3 z+9 z^2+11 z^3\big) H_{-1} H_0}{3 z}
        +16 z^2 H_{-1}^2 H_0
        +\Biggl(
                \frac{2}{3} \big(
                        126-48 z+41 z^2\big)
\nonumber\\ &&                
 -4 \big(
                        -3-6 z+2 z^2\big) H_{-1}
        \Biggr) H_0^2
        +\frac{8}{3} (3+4 z) H_0^3
        +4 \big(
                43-53 z+2 z^2\big) H_1
\nonumber\\ &&     
    -4 \big(
                -53+56 z+z^2\big) H_0 H_1
        +4 \big(
                3-6 z+2 z^2\big) H_0^2 H_1
        +2 \big(
                19-24 z+z^2\big) H_1^2
\nonumber\\ &&    
     +4 \big(
                -19+28 z+2 z^2\big) H_{0,1}
        -8 \big(
                -7-10 z+2 z^2\big) H_0 H_{0,1}
       +\Biggl(
                -32 z^2 H_{-1}
\nonumber\\ &&  
+                \frac{16 \big(
                        2+3 z+9 z^2+11 z^3\big)}{3 z}
                +8 \big(
                        -3-6 z+2 z^2\big) H_0
        \Biggr) H_{0,-1}
        +8 \big(
                -9-10 z
\nonumber\\ && 
+2 z^2\big) H_{0,0,1}
        -8 \big(
                -3-6 z+2 z^2\big) H_{0,0,-1}
        +48 H_{0,1,1}
        +32 z^2 H_{0,-1,-1}
\nonumber\\ &&  
       +\Biggl(
                -\frac{4}{3} \big(
                        114-84 z+47 z^2\big)
                -32 (2+z) H_0
                -16 (-1+z)^2 H_1
                +16 \big(
                        -1-2 z
\nonumber\\ && 
+z^2\big) H_{-1}
        \Biggr) \zeta_2
        -8 \big(
                -1-10 z+4 z^2\big) \zeta_3
\Biggr\}
+\textcolor{blue}{C_F T_F} \Biggl\{
        -\frac{20}{3} (-20+17 z)
\nonumber\\ && 
        +(-1+2 z) \big(
                -\frac{8}{3} H_0^3
                -16 H_0 H_1^2
                -8 H_1^3
                -16 H_1 H_{0,1}
                +24 H_{0,1,1}
        \big)
        -\frac{8}{3} \big(
                -46
\nonumber\\ && 
+53 z+6 z^2\big) H_0
        +\frac{16 \big(
                4+12 z^2+13 z^3\big)}{3 z} H_{-1} H_0
        -32 (1+z)^2 H_{-1}^2 H_0
\nonumber\\ &&       
  +\Biggl(
                -\frac{4}{3} \big(
                        -27+6 z+23 z^2\big)
                +16 (1+z)^2 H_{-1}
        \Biggr) H_0^2
        -4 \big(
                -47+41 z+4 z^2\big) H_1
\nonumber\\ && 
       +8 \big(
                8-14 z+z^2\big) H_0 H_1
        -16 z^2 H_0^2 H_1
        -2 \big(
                -33+40 z+2 z^2\big) H_1^2
        -16 \big(
                -6+z^2\big) 
\nonumber\\ &&  \times
H_{0,1}
        +32 (-1+z)^2 H_0 H_{0,1}
        +\Biggl(
                -\frac{16 \big(
                        4+12 z^2+13 z^3\big)}{3 z}
                -32 (-1+z)^2 H_0
\nonumber\\ && 
                +64 (1+z)^2 H_{-1}
        \Biggr) 
H_{0,-1}
        -32 (-1+z)^2 H_{0,0,1}
        +32 \big(
                1-6 z+z^2\big) H_{0,0,-1}
\nonumber\\ &&   
      -64 (1+z)^2 H_{0,-1,-1}
        +\Biggl(
                \frac{8}{3} \big(
                        -60+42 z+29 z^2\big)
                +(-1+2 z) \big(
                        32 H_0
                        +16 H_1
                \big)
\nonumber\\ &&                
 +32 z^2 H_1
                -32 (1+z)^2 H_{-1}
        \Biggr) \zeta_2
        +8 \big(
                1+14 z+8 z^2\big) \zeta_3
\Biggr\}.
\end{eqnarray}

To compare with the representation of $H_{Qg}^{{\rm S},(2)}\left(z\right)$ in \cite{BUZA2}, Eq.~(B.2), we use the 
relation
\begin{eqnarray}
\Li_3\left[\frac{1-z}{1+z}\right] - \Li_3\left[-\frac{1-z}{1+z}\right] &=&
-\frac{1}{2} H_{-1}^2 H_0
-H_{-1} H_0 H_1
-\frac{1}{2} H_0 H_1^2
+ \big( H_1 + H_{-1} \big) 
\nonumber\\ && \times
\left(H_{0,1} + H_{0,-1} - \frac{3}{2} \zeta_2\right)
-H_{0,1,1}(z)
-H_{0,1,-1}(z)
\nonumber\\ &&
-H_{0,-1,1}(z)
-H_{0,-1,-1}(z)
+\frac{7}{4} \zeta_3.
\end{eqnarray}

The expressions corresponding to Eqs.~(\ref{eq:HQgN}, \ref{HQqz}) in Ref.~\cite{BUZA2} do not agree 
with our results. Our result differs by 
\begin{eqnarray}
&& \frac{16}{3} \frac{(N-1)}{N(N+1)}\Biggl\{\frac{3 C_F T_F}{N} - T_F^2 \Biggl[
 \ln^2\left(\frac{Q^2}{m^2}\right)
- \left(\frac{N-1}{N} + S_1
\right) \ln\left(\frac{Q^2}{m^2}\right) \Biggr] \Biggr\}
\end{eqnarray}
from that in Eq.~(B.2) of Ref.~\cite{BUZA2}. The renormalization formulae in \cite{BUZA2} 
are different from 
those in \cite{Bierenbaum:2009mv}, which fully refer to  the $\overline{\sf MS}$ scheme for charge 
renormalization, being related to the $O(T_F^2)$ terms.

Finally, the Wilson coefficient $\Delta  L_g^{(2)}$ reads
\begin{eqnarray}
\Delta  L_g^{(2)}\left(N, \frac{Q^2}{m^2}\right) = \frac{16}{3} \frac{N-1}{N(N+1)} T_F^2  
\left[\frac{N-1}{N} + S_1\right] 
\ln\left(\frac{Q^2}{m^2}\right).
\end{eqnarray}
It is a gluonic single heavy quark correction to virtual photon--gluon fusion with $N_F$ massless 
final state quarks. In $z$ space one has
\begin{eqnarray}
\Delta  L_g^{(2)}\left(z,\frac{Q^2}{m^2}\right) = \frac{16}{3} T_F^2  \left[
 4 z -3 + (2 z-1) (H_0 + H_1)\right]
\ln\left(\frac{Q^2}{m^2}\right).
\end{eqnarray}
Furthermore, the two--mass corrections (\ref{eqTWOm}) contribute. Both the latter contributions 
have not been considered in \cite{BUZA2}.

\vspace{5mm}\noindent
{\bf Acknowledgments.}~~We would like to thank A.~Behring, E.~Reya, M.~Saragnese, C.~Schneider, J.~Smith, 
D.~St\"ockinger, and J.~Vermaseren for useful discussions and A.~Vogt for providing the massless two--loop Wilson 
coefficients of \cite{Vogt:2008yw} for comparison, which are in agreement with the earlier {\tt FORTRAN} 
code by W.L.~van~Neerven \cite{FORTRAN} and the recent results in Ref.~\cite{Blumlein:2022gpp}. This work was 
supported in part by Studienstiftung des Deutschen Volkes, EU TMR network SAGEX agreement No.~764850 (Marie 
Sk\l{}odowska-Curie), from the European Research
Council (ERC) under the European Union's Horizon 2020 research and innovation programme
grant agreement 101019620 (ERC Advanced Grant TOPUP).


\begin{thebibliography}{99}
%
\bibitem{BUZA2}
  M.~Buza, Y.~Matiounine, J.~Smith and W.L.~van Neerven,
  Nucl.\ Phys.\ B {\bf 485} (1997) 420--456
  [arXiv:hep-ph/9608342].
%
\bibitem{SPUZ}
  M.J.~Alguard {\it et al.} (SLAC),
  Phys.\ Rev.\ Lett.\  {\bf 37} (1976) 1261--1265;
{\bf 41} (1978) 70--73;\\
  G.~Baum {\it et al.},
  Phys.\ Rev.\ Lett.\  {\bf 51} (1983) 1135--1138;\\
  J.~Ashman {\it et al.}  [European Muon Collaboration],
  Phys.\ Lett.\  B {\bf 206} (1988) 364--370;
  Nucl.\ Phys.\  B {\bf 328} (1989) 1--35.
%
\bibitem{EXP1}
  B.~Adeva {\it et al.}  [Spin Muon Collaboration],
  Phys.\ Lett.\  B {\bf 302} (1993) 533--539;\\
  P.L.~Anthony {\it et al.}  [E142 Collaboration],
  Phys.\ Rev.\  D {\bf 54} (1996) 6620--6650
  [arXiv:hep-ex/9610007];\\
  K.~Ackerstaff {\it et al.}  [HERMES Collaboration],
  Phys.\ Lett.\  B {\bf 404} (1997) 383--389
  [arXiv:hep-ex/9703005];\\
  K.~Abe {\it et al.}  [E154 Collaboration],
  Phys.\ Rev.\ Lett.\  {\bf 79} (1997) 26--30
  [arXiv:hep-ex/9705012];\\
  B.~Adeva {\it et al.}  [Spin Muon Collaboration],
  Phys.\ Rev.\  D {\bf 58} (1998) 112001;\\
  K.~Abe {\it et al.}  [E143 collaboration],
  Phys.\ Rev.\  D {\bf 58} (1998) 112003
  [arXiv:hep-ph/9802357];\\
  A.~Airapetian {\it et al.}  [HERMES Collaboration],
  Phys.\ Lett.\  B {\bf 442} (1998) 484--492
  [arXiv:hep-ex/9807015];\\
  P.L.~Anthony {\it et al.}  [E155 Collaboration],
  Phys.\ Lett.\  B {\bf 463} (1999) 339--345
  [arXiv:hep-ex/9904002];
  Phys.\ Lett.\  B {\bf 493} (2000) 19--28
  [arXiv:hep-ph/0007248];\\
  X.~Zheng {\it et al.}  [Jefferson Lab Hall A Collaboration],
  Phys.\ Rev.\ Lett.\  {\bf 92} (2004) 012004
  [arXiv:nucl-ex/0308011];\\
  A.~Airapetian {\it et al.}  [HERMES Collaboration],
  Phys.\ Rev.\  D {\bf 71} (2005) 012003
  [arXiv:hep-ex/0407032];\\
  E.S.~Ageev {\it et al.}  [COMPASS Collaboration],
  Phys.\ Lett.\  B {\bf 612} (2005) 154--164
  [arXiv:hep-ex/0501073];
  Phys.\ Lett.\  B {\bf 647} (2007) 330--340
  [arXiv:hep-ex/0701014];\\
  A.~Airapetian {\it et al.}  [HERMES Collaboration],
  Phys.\ Rev.\  D {\bf 75} (2007) 012007  [hep-ex/0609039];\\
  C.~Adolph {\it et al.} [COMPASS Collaboration],
  Phys.\ Rev.\ D {\bf 87} (2013) no.5,  052018
  [arXiv: 1211.6849 [hep-ex]];\\
  C.~Adolph {\it et al.} [COMPASS Collaboration],
  Phys.\ Lett.\ B {\bf 769} (2017) 34--41
  [arXiv: 1612.00620 [hep-ex]];\\
  M.~Aghasyan {\it et al.} [COMPASS Collaboration],
  Phys.\ Lett.\ B {\bf 781} (2018) 464--472
  [arXiv: 1710.01014 [hep-ex]];\\
  K.P.~Adhikari {\it et al.} [CLAS Collaboration],
  Phys.\ Rev.\ Lett.\  {\bf 120} (2018) no.6,  062501
  [arXiv:1711.01974 [nucl-ex]];\\
  A.~Korzenev,
  Eur.\ Phys.\ J.\ A {\bf 31} (2007) 606--609.
%
\bibitem{LR}
  B.~Lampe and E.~Reya,
  Phys.\ Rept.\  {\bf 332} (2000) 1--163
  [arXiv:hep-ph/9810270].
%
\bibitem{Blumlein:2012bf}
  J.~Bl\"umlein,
  Prog.\ Part.\ Nucl.\ Phys.\  {\bf 69} (2013) 28--84
  [arXiv:1208.6087 [hep-ph]].
%
\bibitem{Deur:2018roz}
  A.~Deur, S.J.~Brodsky and G.F.~De T\'eramond,
  Rep. Prog. Phys. {\bf 82} (2019) 076201
  [arXiv:1807.05250 [hep-ph]].
%
\bibitem{Boer:2011fh}
  D.~Boer, M.~Diehl, R.~Milner, R.~Venugopalan, W.~Vogelsang, D.~Kaplan, H.~Montgomery and S.~Vigdor {\it et al.},
  {\it Gluons and the quark sea at high energies: Distributions, polarization, tomography},
  arXiv:1108.1713 [nucl-th].
%
\bibitem{PROC}
J. Bl\"umlein and W.D. Nowak, {\sf  Workshop on the Prospects of Spin Physics at HERA},
DESY, August 1995, DESY 95--200;\\
  J.~Bl\"umlein, A.~De Roeck, T.~Gehrmann and W.~D.~Nowak, eds.,
  {\sf Deep inelastic scattering off polarized targets: Theory meets experiment.
  Physics with polarized protons at HERA. Proceedings, Workshops, SPIN'97 }
  Zeuthen, Germany, September 1-5, 1997 and Hamburg, Germany, March-September
  1997, DESY 97-200;\\
  J.~Bl\"umlein,
  {\it On the measurability of the structure function $g_1(x,Q^2)$ in $ep$
  collisions at HERA},
  arXiv:hep-ph/9508387.
%
\bibitem{UHEAV1}
  E.~Laenen, S.~Riemersma, J.~Smith and W.L.~van Neerven,
  Nucl.\ Phys.\  B {\bf 392} (1993) 162--228;\\
  S.~Riemersma, J.~Smith and W.L.~van Neerven,
  Phys.\ Lett.\  B {\bf 347} (1995) 143--151
  [arXiv:hep-ph/9411431].
%
\bibitem{BUZA1}
  M.~Buza, Y.~Matiounine, J.~Smith, R.~Migneron and W.L.~van Neerven,
  Nucl.\ Phys.\  B {\bf 472} (1996) 611--658
  [arXiv:hep-ph/9601302].
%
\bibitem{BBK1}
  I.~Bierenbaum, J.~Bl\"umlein and S.~Klein,
  Nucl.\ Phys.\  B {\bf 780} (2007) 40--75
  [arXiv:hep-ph/0703285].
%
\bibitem{BBK1a}
  I.~Bierenbaum, J.~Bl\"umlein and S.~Klein,
  Phys.\ Lett.\  B {\bf 648} (2007) 195--200
  [arXiv:hep-ph/0702265];
  Nucl.\ Phys.\ Proc.\ Suppl.\  {\bf 160} (2006) 85--90
  [arXiv:hep-ph/0607300];
  Acta Phys.\ Polon.\  B {\bf 38} (2007) 3543--3550
  [arXiv:0710.3348 [hep-ph]];
  Acta Phys.\ Polon.\  B {\bf 39} (2008) 1531--1538
  [arXiv:0806.0451 [hep-ph]].
%
\bibitem{BBK2}
  I.~Bierenbaum, J.~Bl\"umlein and S.~Klein,
  Phys. Lett. B {\bf 672} (2009) 401--406 [arXiv:0901.0669 [hep-ph]].
%
\bibitem{Bierenbaum:2009mv}
  I.~Bierenbaum, J.~Bl\"umlein and S.~Klein,
  Nucl.\ Phys.\  B {\bf 820} (2009) 417--482
  [arXiv:0904.3563 [hep-ph]].
%
\bibitem{BBK-III}
I.~Bierenbaum, J.~Bl\"umlein and S.~Klein,
  Nucl.\ Phys.\ Proc.\ Suppl.\  {\bf 183} (2008) 162--167
  [arXiv:0806.4613 [hep-ph]].
%
\bibitem{BBKS8a}
  I.~Bierenbaum, J.~Bl\"umlein, S.~Klein and C.~Schneider,
  Nucl.\ Phys.\  B {\bf 803} (2008) 1--41
  [arXiv:0803.0273 [hep-ph]].
%
\bibitem{FL}
  J.~Bl\"umlein, A.~De Freitas, W.L.~van Neerven and S.~Klein,
  Nucl.\ Phys.\  B {\bf 755} (2006) 272--285
  [arXiv:hep-ph/0608024].
%
\bibitem{WIL1a}
  A.D.~Watson,
  Z.\ Phys.\  C {\bf 12} (1982) 123--125;\\
  M.~Gl\"uck, E.~Reya and W.~Vogelsang,
  Nucl.\ Phys.\  B {\bf 351} (1991) 579--592.
%
\bibitem{WIL1b}
  W.~Vogelsang,
  Z.\ Phys.\  C {\bf 50} (1991) 275--284.
%
\bibitem{AB}
  S.I.~Alekhin and J.~Bl\"umlein,
  Phys.\ Lett.\  B {\bf 594} (2004) 299--307
  [arXiv:hep-ph/0404034].
%
\bibitem{PHOTO}
  I.~Bojak and M.~Stratmann,
  Nucl.\ Phys.\ B {\bf 540} (1999) 345--381
   Erratum: [Nucl.\ Phys.\ B {\bf 569} (2000) 694]
  [hep-ph/9807405].
%
\bibitem{Hekhorn:2018ywm}
  F.~Hekhorn and M.~Stratmann,
  Phys.\ Rev.\ D {\bf 98} (2018) no.1,  014018
  [arXiv:1805.09026 [hep-ph]].
%
\bibitem{COMPASS}
  K.~Kurek,
  {\it $\Delta(G)$ from COMPASS},
  arXiv:hep-ex/0607061;\\
  G.~Brona  [COMPASS Collaboration],
  {\it Measurement of the gluon polarisation at COMPASS},
  arXiv:0705.2372 [hep-ex].
%
\bibitem{BRN1}
  J.~Bl\"umlein, V.~Ravindran and W.L.~van Neerven,
  Phys.\ Rev.\  D {\bf 68} (2003) 114004
  [arXiv:hep-ph/0304292].
%
\bibitem{BB}
  J.~Bl\"umlein and H.~B\"ottcher,
  Nucl.\ Phys.\  B {\bf 636} (2002) 225--263
  [arXiv:hep-ph/0203155].
%
\bibitem{OPDF}
M.~Gl\"uck, E.~Reya, M.~Stratmann and W.~Vogelsang,
Phys. Rev. D \textbf{63} (2001) 094005
[arXiv:hep-ph/0011215 [hep-ph]];\\
M.~Hirai, S.~Kumano and N.~Saito,
Phys. Rev. D \textbf{74} (2006) 014015
[arXiv:hep-ph/0603213 [hep-ph]];\\
D.~de Florian, R.~Sassot, M.~Stratmann and W.~Vogelsang,
Phys. Rev. D \textbf{80} (2009) 034030
[arXiv:0904.3821 [hep-ph]];\\
E.~Leader, A.V.~Sidorov and D.B.~Stamenov,
Phys. Rev. D \textbf{82} (2010) 114018
[arXiv: 1010.0574 [hep-ph]];\\
P.~Jimenez-Delgado, W.~Melnitchouk and J.F.~Owens,
J. Phys. G \textbf{40} (2013) 093102
[arXiv: 1306.6515 [hep-ph]];\\
E.R.~Nocera \textit{et al.} [NNPDF],
Nucl. Phys. B \textbf{887} (2014) 276--308
[arXiv:1406.5539 [hep-ph]];\\
M.~Salimi-Amiri, A.~Khorramian, H.~Abdolmaleki and F.~I.~Olness,
Phys. Rev. D \textbf{98} (2018) no.5, 056020
[arXiv:1805.02613 [hep-ph]].
%
\bibitem{Blumlein:2010rn}
  J.~Bl\"umlein and H.~B\"ottcher,
  Nucl.\ Phys.\ B {\bf 841} (2010) 205--230
  [arXiv:1005.3113 [hep-ph]].
%
\bibitem{Blumlein:2019zux}
J.~Bl\"umlein, C.G.~Raab and K.~Sch\"onwald,
Nucl. Phys. B \textbf{948} (2019) 114736
[arXiv: 1904.08911 [hep-ph]].
%
\bibitem{HQ3N}
  J.~Ablinger, A.~Behring, J.~Bl\"umlein, A.~De Freitas, A.~von Manteuffel and C.~Schneider,
  Nucl.\ Phys.\ B {\bf 890} (2014) 48--151
  [arXiv:1409.1135 [hep-ph]];\\
  J.~Ablinger, J.~Bl\"umlein, A.~De Freitas, A.~Hasselhuhn, A.~von Manteuffel, M.~Round, C.~Schneider and 
  F.~Wi\ss{}brock,
  Nucl.\ Phys.\ B {\bf 882} (2014) 263--288
  [arXiv:1402.0359 [hep-ph]].
%
\bibitem{Blumlein:2014zxa}
  J.~Bl\"umlein, A.~DeFreitas and C.~Schneider,
  Nucl.\ Part.\ Phys.\ Proc.\  {\bf 261-262} (2015) 185--201
  [arXiv:1411.5669 [hep-ph]].
%
\bibitem{Ablinger:2014vwa}
  J.~Ablinger, A.~Behring, J.~Bl\"umlein, A.~De Freitas, A.~Hasselhuhn, A.~von Manteuffel, M.~Round, C.~Schneider and
  F.~Wi\ss{}brock
  Nucl.\ Phys.\ B {\bf 886} (2014) 733--823
  [arXiv:1406.4654 [hep-ph]].
%
\bibitem{ZN}
  E.B.~Zijlstra and W.L.~van Neerven,
  Nucl.\ Phys.\  B {\bf 417} (1994) 61--100
  [Errata-ibid.\  B {\bf 426} (1994) 245; B {\bf 773} (2007) 105--106;
   B {\bf 501} (1997) 599].
%
\bibitem{MAT1}
  Y.~Matiounine, J.~Smith and W.L.~van Neerven,
  Phys.\ Rev.\  D {\bf 58} (1998) 076002
  [arXiv:hep-ph/9803439].
%
\bibitem{Jegerlehner:2000dz}%
  J.G.~K\"orner, D.~Kreimer and K.~Schilcher,
  Z.\ Phys.\ C {\bf 54} (1992) 503--512;\\
  F.~Jegerlehner,
  Eur.\ Phys.\ J.\  C {\bf 18} (2001) 673--679
  [arXiv:hep-th/0005255].
%
\bibitem{SP_PS}
  R.~Mertig and W.L.~van Neerven,
  Z.\ Phys.\  C {\bf 70} (1996) 637--654 [hep-ph/9506451].
%
\bibitem{Weinzierl:1999xb}
  S.~Weinzierl,
  {\it Equivariant dimensional regularization},
  hep-ph/9903380.
%
\bibitem{WW}
  S.~Wandzura and F.~Wilczek,
  Phys.\ Lett.\  B {\bf 72} (1977) 195--198.
%
\bibitem{G2A}
  J.D.~Jackson, G.G.~Ross and R.G.~Roberts,
  Phys.\ Lett.\  B {\bf 226} (1989) 159--166;\\
  R.G.~Roberts and G.G.~Ross,
  Phys.\ Lett.\  B {\bf 373} (1996) 235--245
  [arXiv:hep-ph/9601235].
%
\bibitem{G2B}
  J.~Bl\"umlein and N.~Kochelev,
  Nucl.\ Phys.\  B {\bf 498} (1997) 285--309
  [arXiv:hep-ph/9612318];
  Phys.\ Lett.\  B {\bf 381} (1996) 296--309
  [arXiv:hep-ph/9603397].
%
\bibitem{Landshoff:1971xb}
  P.V.~Landshoff and J.C.~Polkinghorne,
  Phys.\ Rept.\  {\bf 5} (1972) 1--55.
%
\bibitem{DIFFR}
  J.~Bl\"umlein and D.~Robaschik,
  Phys.\ Rev.\  D {\bf 65} (2002) 096002
  [arXiv:hep-ph/0202077].
%
\bibitem{BT}
  J.~Bl\"umlein and A.~Tkabladze,
  Nucl.\ Phys.\  B {\bf 553} (1999) 427--464
  [arXiv:hep-ph/9812478];
  Nucl.\ Phys.\ Proc.\ Suppl.\  {\bf 79} (1999) 541--544
  [arXiv:hep-ph/9905524].
%
\bibitem{TM}
  J.~Bl\"umlein, B.~Geyer and D.~Robaschik,
  Nucl.\ Phys.\  B {\bf 755} (2006) 112--136
  [arXiv:hep-ph/0605310];
  Eur.\ Phys.\ J.\  C {\bf 61} (2009) 279--298
  [arXiv:0812.1899 [hep-ph]].
%
\bibitem{BR1}
  J.~Bl\"umlein and D.~Robaschik,
  Nucl.\ Phys.\  B {\bf 581} (2000) 449--473
  [arXiv:hep-ph/0002071].
%
\bibitem{Blumlein:2016xcy}
  J.~Bl\"umlein, G.~Falcioni and A.~De Freitas,
  Nucl.\ Phys.\ B {\bf 910} (2016) 568--617
  [arXiv: 1605.05541 [hep-ph]].
%
\bibitem{Blumlein:2019qze}
J.~Bl\"umlein, A.~De Freitas, C.G.~Raab and K.~Sch\"onwald,
Nucl. Phys. B \textbf{945} (2019) 114659
[arXiv:1903.06155 [hep-ph]].
%
\bibitem{Blumlein:2011mi}
J.~Bl\"umlein, A.~De Freitas and W.~van Neerven,
Nucl. Phys. B \textbf{855} (2012) 508--569
[arXiv:1107.4638 [hep-ph]].
%
\bibitem{Blumlein:2020jrf}
J.~Bl\"umlein, A.~De Freitas, C.~Raab and K.~Sch\"onwald,
Nucl. Phys. B \textbf{956} (2020) 115055
[arXiv:2003.14289 [hep-ph]].
%
\bibitem{GAM5}
  S.A.~Larin,
  Phys.\ Lett.\  B {\bf 303} (1993) 113--118
  [arXiv:hep-ph/9302240].
%
\bibitem{TRANSV}
  J.~Bl\"umlein, S.~Klein and B.~T\"odtli,
  Phys.\ Rev.\ D {\bf 80} (2009) 094010
  [arXiv:0909.1547 [hep-ph]].
%
\bibitem{Moch:2014sna}
  S.~Moch, J.A.M.~Vermaseren and A.~Vogt,
  Nucl.\ Phys.\ B {\bf 889} (2014) 351--400
  [arXiv: 1409.5131 [hep-ph]].
%
\bibitem{Behring:2019tus}
A.~Behring, J.~Bl\"umlein, A.~De Freitas, A.~Goedicke, S.~Klein, A.~von Manteuffel, C.~Schneider and K.~Sch\"onwald,
Nucl. Phys. B \textbf{948} (2019) 114753
[arXiv:1908.03779 [hep-ph]].
%
\bibitem{Blumlein:2021ryt}
J.~Bl\"umlein, P.~Marquard, C.~Schneider and K.~Sch\"onwald,
JHEP \textbf{01} (2022) 193
[arXiv:2111.12401 [hep-ph]].
%
\bibitem{Blumlein:2022ndg}
J.~Bl\"umlein, P.~Marquard, C.~Schneider and K.~Sch\"onwald,
Nucl. Phys. B \textbf{980} (2022) 115794
[arXiv:2202.03216 [hep-ph]].
%
\bibitem{IBP}
J. Lagrange, {\sf Nouvelles recherches sur la nature et la propagation
du son}, Miscellanea Taurinensis, t. II, 1760-61; Oeuvres t. I, p. 263;\\
C.F. Gauss, {Theoria attractionis corporum sphaeroidicorum ellipticorum
homogeneorum methodo novo tractate}, Commentationes societas scientiarum
Gottingensis recentiores, Vol III, 1813, Werke Bd. {\bf V} pp. 5-7;\\
G. Green, {\sf Essay on the Mathematical Theory of Electricity and
Magnetism}, Nottingham, 1828 [Green Papers, pp. 1-115];\\
M.~Ostrogradsky (presented: November 5, 1828 ; published: 1831) 
M\'emoires de l'Acad\'emie imp\'eriale des sciences de St.~P\'etersbourg,
series {\bf 6} 1: 129--133;\\
  K.G.~Chetyrkin and F.V.~Tkachov,
  Nucl.\ Phys.\ B {\bf 192} (1981) 159--204.
%
\bibitem{Vermaseren:1998uu}
  J.A.M.~Vermaseren,
  Int.\ J.\ Mod.\ Phys.\  A {\bf 14} (1999) 2037--2076
  [arXiv:hep-ph/9806280].
%
\bibitem{Blumlein:1998if}
  J.~Bl\"umlein and S.~Kurth,
  Phys.\ Rev.\  D {\bf 60} (1999) 014018
  [arXiv:hep-ph/9810241].
%
\bibitem{BW}
  I.~Bierenbaum and S.~Weinzierl,
  Eur.\ Phys.\ J.\  C {\bf 32} (2003) 67--78
  [arXiv:hep-ph/0308311].
%
\bibitem{GHF}
N. Bailey, {\sf Generalized Hypergeometric Series},
(Cambridge University Press, Cambridge, 1935);\\
L.J. Slater, {\sf Generalized Hypergeometric Functions},
(Cambridge University Press, Cambridge, 1966).
%
\bibitem{Blumlein:2018cms}
J.~Bl\"umlein and C.~Schneider,
Int. J. Mod. Phys. A \textbf{33} (2018) no.17, 1830015
[arXiv:1809.02889 [hep-ph]];\\
J.~Bl\"umlein,
{\it Analytic Integration Methods in Quantum Field Theory: An Introduction},
in: {\sf Anti-Differentiation and the calculation of Feynman diagrams}, eds. J.~Bl\"umlein 
and C.~Schneider, (Springer, Berlin, 2021) 1--33,
[arXiv:2103.10652 [hep-th]].
%
\bibitem{Behring:2015zaa}
  A.~Behring, J.~Bl\"umlein, A.~De Freitas, A.~von Manteuffel and C.~Schneider,
  Nucl.\ Phys.\ B {\bf 897} (2015) 612--644
  [arXiv:1504.08217 [hep-ph]].
%
\bibitem{Ablinger:2019etw}
J.~Ablinger, A.~Behring, J.~Bl\"umlein, A.~De Freitas, A.~von Manteuffel, C.~Schneider and K.~Sch\"onwald,
Nucl. Phys. B \textbf{953} (2020) 114945
[arXiv:1912.02536 [hep-ph]].
%
\bibitem{Behring:2021asx}
A.~Behring, J.~Bl\"umlein, A.~De Freitas, A.~von Manteuffel, K.~Sch\"onwald and C.~Schneider,
Nucl. Phys. B \textbf{964} (2021) 115331
[arXiv:2101.05733 [hep-ph]].
%
\bibitem{LOGPOL}
J.~Bl\"umlein, A.~De Freitas, M.~Saragnese, C.~Schneider and K.~Sch\"onwald,
Phys. Rev. D \textbf{104} (2021) no.3, 034030
[arXiv:2105.09572 [hep-ph]].
%
\bibitem{Ablinger:2017err}
J.~Ablinger, J.~Bl\"umlein, A.~De Freitas, A.~Hasselhuhn, C.~Schneider and F.~Wi\ss{}brock,
Nucl. Phys. B \textbf{921} (2017) 585--688
[arXiv:1705.07030 [hep-ph]].
%
\bibitem{Ablinger:2020snj}
J.~Ablinger, J.~Bl\"umlein, A.~De Freitas, A.~Goedicke, M.~Saragnese, C.~Schneider and K.~Sch\"onwald,
Nucl. Phys. B \textbf{955} (2020) 115059
[arXiv:2004.08916 [hep-ph]].
%
\bibitem{Ablinger:2019gpu}
J.~Ablinger, J.~Bl\"umlein, A.~De Freitas, M.~Saragnese, C.~Schneider and K.~Sch\"onwald,
Nucl. Phys. B \textbf{952} (2020) 114916
[arXiv:1911.11630 [hep-ph]].
%
\bibitem{Schonwald:2019gmn}
K.~Sch\"onwald,
{\it Massive two- and three-loop calculations in QED and QCD},
PhD Thesis, TU Dortmund, October 2019.
%
\bibitem{HVBM}
  G.~'t Hooft and M.J.G.~Veltman,
  Nucl.\ Phys.\  B {\bf 44} (1972) 189--213;\\
  D.A.~Akyeampong and R.~Delbourgo,
  Nuovo Cim.\  A {\bf 17} (1973) 578--586;
A {\bf 18} (1973) 94--104;
A {\bf 19} (1974) 219--224;\\
  P.~Breitenlohner and D.~Maison,
  Commun.\ Math.\ Phys.\  {\bf 52} (1977) 11--38.
%
\bibitem{Blumlein:2022gpp}
J.~Bl\"umlein, P.~Marquard, C.~Schneider and K.~Sch\"onwald,
{\it The massless three-loop Wilson coefficients for the deep-inelastic structure functions $F_2, F_L, xF_3$ and 
$g_1$}, JHEP in print
[arXiv:2208.14325 [hep-ph]].
%
\bibitem{Blumlein:2018jfm}
J.~Bl\"umlein, A.~De Freitas, C.~Schneider and K.~Sch\"onwald,
Phys. Lett. B \textbf{782} (2018) 362--366
[arXiv:1804.03129 [hep-ph]].
%
\bibitem{Bardin:1996ch}
  D.Y.~Bardin, J.~Bl\"umlein, P.~Christova and L.~Kalinovskaya,
  Nucl.\ Phys.\ B {\bf 506} (1997) 295--328
  [hep-ph/9612435].
%
\bibitem{Arbuzov:1995id}
  A.~Arbuzov, D.Y.~Bardin, J.~Bl\"umlein, L.~Kalinovskaya and T.~Riemann,
  Comput.\ Phys.\ Commun.\  {\bf 94} (1996) 128--184
  [hep-ph/9511434].
%
\bibitem{Buza:1996wv}
M.~Buza, Y.~Matiounine, J.~Smith and W.~L.~van Neerven,
Eur. Phys. J. C \textbf{1} (1998) 301--320
[arXiv:hep-ph/9612398 [hep-ph]].
%
\bibitem{Moch:2008fj}
  S.~Moch, J.A.M.~Vermaseren and A.~Vogt,
  Nucl.\ Phys.\ B {\bf 813} (2009) 220--258
  [arXiv: 0812.4168 [hep-ph]].
%
\bibitem{Zijlstra:1992kj}
  E.B.~Zijlstra and W.L.~van Neerven,
  Phys.\ Lett.\ B {\bf 297} (1992) 377--384.
%
\bibitem{Moch:1999eb}
  S.~Moch and J.A.M.~Vermaseren,
  Nucl.\ Phys.\ B {\bf 573} (2000) 853--907
  [hep-ph/9912355].
%
\bibitem{Vogt:2008yw}
  A.~Vogt, S.~Moch, M.~Rogal and J.A.M.~Vermaseren,
  Nucl.\ Phys.\ Proc.\ Suppl.\  {\bf 183} (2008) 155--161
  [arXiv:0807.1238 [hep-ph]].
%
\bibitem{FORTRAN}
W.L.~van Neerven, {\tt FORTRAN}-code for the massless polarized 2-loop
Wilson coefficient (2003).
%
\bibitem{Zijlstra:1992qd}
  E.B.~Zijlstra and W.L.~van Neerven,
  Nucl.\ Phys.\ B {\bf 383} (1992) 525--574.
%
\bibitem{Vogelsang:1995vh}
  W.~Vogelsang,
  Phys.\ Rev.\  D {\bf 54} (1996) 2023--2029
  [arXiv:hep-ph/9512218].
%
\bibitem{Vogelsang:1996im}
  W.~Vogelsang,
  Nucl.\ Phys.\  B {\bf 475} (1996) 47--72
  [arXiv:hep-ph/9603366].
%
\bibitem{WIL_p1}
  G.T.~Bodwin and J.W.~Qiu,
  Phys.\ Rev.\  D {\bf 41} (1990) 2755--2766.
%
\bibitem{FP}
  W.~Furmanski and R.~Petronzio,
  Z.\ Phys.\ C {\bf 11} (1982) 293--314.
%
\bibitem{Bjorken:1969mm}
  J.D.~Bjorken,
  Phys.\ Rev.\ D {\bf 1} (1970) 1376--1379.
%
\bibitem{Kodaira:1978sh}
  J.~Kodaira, S.~Matsuda, T.~Muta, K.~Sasaki and T.~Uematsu,
  Phys.\ Rev.\ D {\bf 20} (1979) 627--629.
%
\bibitem{Blumlein:2004xs}
  J.~Bl\"umlein and A.~Guffanti,
  Nucl.\ Phys.\ Proc.\ Suppl.\  {\bf 152} (2006) 87--91
  [hep-ph/0411110].
%
\bibitem{Blumlein:2021lmf}
J.~Bl\"umlein and M.~Saragnese,
Phys. Lett. B \textbf{820} (2021) 136589
[arXiv:2107.01293 [hep-ph]].
%
\bibitem{Blumlein:2000wh}
  J.~Bl\"umlein, V.~Ravindran and W.L.~van Neerven,
  Nucl.\ Phys.\ B {\bf 586} (2000) 349--381
  [hep-ph/0004172].
%
\bibitem{Curci:1980uw}
  G.~Curci, W.~Furmanski and R.~Petronzio,
  Nucl.\ Phys.\ B {\bf 175} (1980) 27--92.
%
\bibitem{zuber}
C.~Itzykson~and~J.~Zuber,~{\sf~Quantum~Field~Theory},~(McGraw-Hill~Inc.,
NewYork,~1980).
%
\bibitem{FORM}
  J.A.M.~Vermaseren,
  {\it New features of FORM},
  arXiv:math-ph/0010025.
%
\bibitem{STRUCT}
  J.~Bl\"umlein,
  Nucl.\ Phys.\ Proc.\ Suppl.\  {\bf 135} (2004) 225--231
  [arXiv:hep-ph/0407044];
  Comput.\ Phys.\ Commun.\  {\bf 180} (2009) 2218--2249
  [arXiv:0901.3106 [hep-ph]];
  Clay Math.\ Proc.\  {\bf 12} (2010) 167--188
  [arXiv:0901.0837 [math-ph]].
%
\bibitem{MATH}
  J.~Bl\"umlein and V.~Ravindran,
  Nucl.\ Phys.\  B {\bf 716} (2005) 128--172
  [arXiv:hep-ph/0501178];
  Nucl.\ Phys.\  B {\bf 749} (2006) 1--24
  [arXiv:hep-ph/0604019].
%
\bibitem{MVV2}
  J.A.M.~Vermaseren, A.~Vogt and S.~Moch,
  Nucl.\ Phys.\  B {\bf 724} (2005) 3--182
  [arXiv:hep-ph/0504242].
%
\bibitem{Moch:2004pa}
S.~Moch, J.A.M.~Vermaseren and A.~Vogt,
Nucl. Phys. B \textbf{688} (2004) 101--134
[arXiv:hep-ph/0403192 [hep-ph]].
%
\bibitem{Blumlein:2021enk}
J.~Bl\"umlein, P.~Marquard, C.~Schneider and K.~Sch\"onwald,
Nucl. Phys. B \textbf{971} (2021) 115542
[arXiv:2107.06267 [hep-ph]].
%
\bibitem{Ablinger:2017tan}
J.~Ablinger, A.~Behring, J.~Bl\"umlein, A.~De Freitas, A.~von Manteuffel and C.~Schneider,
Nucl. Phys. B \textbf{922} (2017) 1--40
[arXiv:1705.01508 [hep-ph]].
%
\bibitem{KLEINphd}
S.~Klein, {\it Charm production in deep--inelastic scattering: Mellin moments of 
heavy flavor contributions to $F_2(x,Q^2)$ at NNLO}, PhD Thesis, TU Dortmund, 2009,
Springer Theses, (Springer, Berlin, 2012). 
%
\bibitem{PHO_G1}
  I.~Antoniadis and C.~Kounnas,
  Phys.\ Rev.\  D {\bf 24} (1981) 505--525;\\
  A.V.~Efremov and O.V.~Teryaev,
  Phys.\ Lett.\  B {\bf 240} (1990) 200--202;\\
  S.D.~Bass,
  Int.\ J.\ Mod.\ Phys.\  A {\bf 7} (1992) 6039--6052;\\
  S.~Narison, G.M.~Shore and G.~Veneziano,
  Nucl.\ Phys.\  B {\bf 391} (1993) 69--99;\\
  A.~Freund and L.M.~Sehgal,
  Phys.\ Lett.\  B {\bf 341} (1994) 90--94;\\
  S.D.~Bass, S.J.~Brodsky and I.~Schmidt,
  Phys.\ Lett.\  B {\bf 437} (1998) 417--424
  [arXiv:hep-ph/9805316];\\
  T.~Ueda, T.~Uematsu and K.~Sasaki,
  Phys.\ Lett.\  B {\bf 640} (2006) 188--195
  [arXiv:hep-ph/0606267];
  Acta Phys.\ Polon.\  B {\bf 37} (2006) 1103--109.
%
\bibitem{Kirschner:1983di}
  R.~Kirschner and L.~Lipatov,
  Nucl.\ Phys.\  B {\bf 213} (1983) 122--148.
%
\bibitem{Blumlein:1995jp}
  J.~Bl\"umlein and A.~Vogt,
  Phys.\ Lett.\  B {\bf 370} (1996) 149--155
  [arXiv:hep-ph/9510410].
%
\bibitem{Blumlein:1996hb}
  J.~Bl\"umlein and A.~Vogt,
  Phys.\ Lett.\  B {\bf 386} (1996) 350--358
  [arXiv:hep-ph/9606254].
%
\bibitem{Bartels:1996wc}
  J.~Bartels, B.I.~Ermolaev and M.G.~Ryskin,
  Z.\ Phys.\  C {\bf 72} (1996) 627--635
  [arXiv:hep-ph/9603204].
%
\bibitem{SX2}
  J.~Bl\"umlein and A.~Vogt,
  Acta Phys.\ Polon.\  B {\bf 27} (1996) 1309--1322
  [arXiv:hep-ph/9603450].
%
\bibitem{Bartels:1995iu}
  J.~Bartels, B.I.~Ermolaev and M.G.~Ryskin,
  Z.\ Phys.\ C {\bf 70} (1996) 273--280
  [hep-ph/9507271].
%
\bibitem{Alekhin:2012vu}
S.~Alekhin, J.~Bl\"umlein, K.~Daum, K.~Lipka and S.~Moch,
Phys. Lett. B \textbf{720} (2013) 172--176
[arXiv:1212.2355 [hep-ph]].
%
\bibitem{PDG}
K.A.~Olive et al. [Particle Data Group], Chin. Phys. C {\bf 38} (2014) 090001.
%
\bibitem{PIESSENS}
R.~Piessens, Angew. Informatik {\bf 9} (1973) 399--401.
%
\bibitem{Hasselhuhn:2013swa}
A.~Hasselhuhn, PhD Thesis, TU Dortmund,
{\it 3-Loop Contributions to Heavy Flavor Wilson Coefficients of Neutral and Charged Current DIS},
DESY-THESIS-2013-050.
%
\bibitem{Ablinger:2022wbb}
J.~Ablinger, A.~Behring, J.~Bl\"umlein, A.~De Freitas, A.~Goedicke, A.~von 
Manteuffel, C.~Schneider and 
K.~Sch\"onwald,
{\it The Unpolarized and Polarized Single-Mass Three-Loop Heavy Flavor Operator Matrix 
Elements 
$A_{gg,Q}$ and $\Delta A_{gg,Q}$}.
[arXiv:2211.05462 [hep-ph]].
%
\bibitem{NIELSEN}
N.~Nielsen {Nova Acta Leopold.} {\bf XC} (1909) Nr. 3, 125--211;\\
K.S.~K{\"o}lbig, J.A.~Mignoco, and E.~Remiddi, {BIT} {\bf 10} (1970) 38--74;\\
K.S. K{\"o}lbig, { SIAM J. Math. Anal.} {\bf 17} (1986)
1232--1258;\\
A.~Devoto and D.W.~Duke, { Riv. Nuovo Cim.} {\bf 7N6} (1984)
1--39.
%
\bibitem{alphas}
  S.~Bethke {\it et al.},
  {\it Workshop on Precision Measurements of $\alpha_s$},
  arXiv:1110.0016 [hep-ph];\\
  S.~Moch {\it et al.},
  {\it High precision fundamental constants at the TeV scale},
  arXiv:1405.4781 [hep-ph];\\
  S.~Alekhin, J.~Bl\"umlein and S.O.~Moch,
  Mod.\ Phys.\ Lett.\ A {\bf 31} (2016) no.25,  1630023;\\
D.~d'Enterria et al., {\it The strong coupling constant: State of the art and the decade ahead},
[arXiv:2203.08271 [hep-ph]].
%
\bibitem{Schneider:2007a}
C.~Schneider, { S\'em.~Lothar. Combin.\/} {\bf 56} (2007) 1--36,
{article B56b}.
%
\bibitem{Schneider:2013a}
 C.~Schneider, in: {\sf Computer Algebra in Quantum Field Theory: Integration,
  Summation and Special Functions}, Texts and Monographs in Symbolic
  Computation eds. C.~Schneider and J.~Bl{\"u}mlein (Springer, Wien, 2013)
  325--360 [arXiv:1304.4134 [cs.SC]].
%
\bibitem{HAMB}
R. Hamberg, {\it Second order gluonic contributions to physical quantities}, PhD Thesis,
(Leiden, 1991). 
%
\bibitem{LEWIN}
 L.~Lewin, {\sf Dilogarithms and associated functions}, (Macdonald, London,
  1958);
{\sf Polylogarithms and associated functions}, (North Holland, New
  York, 1981).
%
\bibitem{WT}
  J.C.~Ward,
  Phys.\ Rev.\  {\bf 78} (1950) 182;\\
  Y.~Takahashi,
  Nuovo Cim.\  {\bf 6} (1957) 371--375.
%
\bibitem{Remiddi:1999ew}
E.~Remiddi and J.A.M.~Vermaseren,
Int. J. Mod. Phys. A \textbf{15} (2000) 725--754
[arXiv:hep-ph/9905237 [hep-ph]].
%
\bibitem{Blumlein:1989gk}
  J.~Bl\"umlein,
  Z.\ Phys.\ C {\bf 47} (1990) 89--94.
\end{thebibliography}
\end{document}